# Rippled graphene pores as fluidic memristive devices with synaptic and neuromorphic functionalities

Wenzhe Zhou[1]#, Dongjiao Ge[2]#*, Ao Zhang[3]#, Jincheng Xu[1], Yu Ji[1], Yiran Gong[1], Wenchang Zhang[1], Jidong Li[4], Li Lin[5], Zhiping Xu[3]*, Pengzhan Sun[1]*
[1]Institute of Applied Physics and Materials Engineering, University of Macau, Macau SAR, China
[2]Faculty of Data Science, City University of Macau, Macau SAR, China
[3]Applied Mechanics Laboratory, Department of Engineering Mechanics and Center for Nano and Micro Mechanics, Tsinghua University, Beijing, China
[4]Institute for Frontier Science, Nanjing University of Aeronautics and Astronautics, Nanjing, China
[5]School of Materials Science and Engineering, Peking University, Beijing, China
These authors contributed equally: Wenzhe Zhou, Dongjiao Ge, Ao Zhang.
Corresponding emails: djge@cityu.edu.mo; xuzp@tsinghua.edu.cn; pengzhansun@um.edu.mo

## Abstract

Nanofluidic memristive devices work with nanoscale pores (or channels) and ions dissolved in water, which harness the ionic memory effect aiming to store and process information. These devices share the same charge carriers as biological systems and bring hope for better emulating the neural functions and developing ionic circuits for neuromorphic applications. Specially, theory and experiments suggest that nanoconfinement is essential for inducing a memory effect, which places limit on the pore size to nm-scale or smaller. Such devices are, however, difficult to scale up with precision and operate with long-term stability. Here, we show that a micrometer size pore, generally expected to exhibit a linear ion transport, can display a pronounced memory effect, if its rim is wrapped by strongly curved and tightly stacked graphene. We attribute the observation to slow ion dynamics confined in the rippled graphene edges. The devices are easy to scale up and integrate into fluidic circuits. The memory effect is ion-selective and, crucially, exhibits long endurance, which enables reversible modification of the conductance states using programmable voltage spikes and various electrolytes over a long time, akin to biological synaptic plasticity. Thanks to this plasticity, our devices and their integrated circuits enable storing, transmitting and processing information with exceptionally high reliability, fidelity and accuracy, as evidenced by the effective processing of largescale and event-based information in the identification of both greyscale and color images, and in the real-time analysis of emulated neural signals. Our results highlight nanoscale morphology of the pore wall as an important parameter regulating ion transport, in addition to the well-documented pore's charge and size, and indicate that the stringent nanoconfinement for ionic memory can be lifted from restricting the pore size to designing its rim structure. The devices and their integrated circuits may find use in ionic neuromorphic applications.

As a basic element of biological neural networks, the synapse serves as a junction connecting two neurons and transmitting neuronal impulses (voltage spikes)[1,2]. The connection strength can be continuously adjusted in response to external stimuli (so-called synaptic plasticity)[1,2], for example, by sequentially transmitted voltage spikes. This synaptic plasticity is believed to be important for learning and memory[1,2]. Similar to the working function of synapse, a resistive switching device, or memristor[3,4], changes its conductance state $G$ depending on the history of applied voltage $V$. This feature enables mimicking the synaptic plasticity and hence, the device has attracted considerable attention due to its potential to store and process information in a single device and for use as building blocks of artificial neuromorphic systems[5,6]. The well-documented memristors are mostly solid-state devices[5,6]. In those devices, the observed memory effects are typically digital-type with well-defined high- (HRS) and low- (LRS) resistance states, and arise from coupled dynamics of electrons with a specific type of ions in solids. The fast electronic dynamics results in high operation frequencies[7] $f$ of the order of $10^5$ – $10^8$ Hz, whereby the HRS and LRS can be used to emulate the ones and zeros of the binary code for digital computing. In contrast, the biological neural networks rely on different ions dissolved in water to process information, in particular the nonlinear response arising from selective ion transport through the voltage-gated synaptic channels[1,2]. In stark contrast to the solid-state memristors, the synaptic channels typically display an analog-type memory characteristic of continuum change in resistance states[8], and process physiological voltage spikes of a lower $f$ = 10 – 100 Hz[9,10]. Thanks to the advancement of nanomaterials and nanofabrication techniques[11], one can fabricate nanofluidic memristors[12-23] based on artificial pores or channels and manipulating different ions in an aqueous media, to better mimic those biological features and further, develop bio-inspired and ion-based technologies such as neuromorphic integration and interfacing with biological systems.

Recently, the ionic memory effect has been reported for some nanofluidic devices, to list a few, nanocapillaries by van der Waals assembly of layered crystals and nanoscale spacers[16,18,23], glass nanopipettes and nanopores with designer surface functionalization[17,20], and porous crystals[19,21,22]. In those devices, the core was nanoscale pores or channels and many had asymmetric entrances. The asymmetric channel geometry led to nonlinear and rectified ion transport[16,18], whereas nanoscale confinement slowed down the diffusion dynamics of ions, particularly those along the solid surfaces[15-17]. These two effects are believed to be important for inducing a memory effect with conductance hysteresis $G(V)$ so that information can be stored and processed as different $G$[16]. To make these nanofluidic devices, complex and multiple fabrication procedures are typically involved. For this reason, the resulting devices are difficult to scale up and integrate on a circuit level. Also, they inevitably display large device-to-device variations, significant evolutions of memory effect over time and insufficient stability for long-term operation especially in concentrated solutions[11,23]. This is because fabricating pores or channels at nm- or Å-scale with precision is challenging, and nanofluidic devices are susceptible to clogging or device integrity issues in electrolytes[11]. To overcome these difficulties, a straightforward strategy is perhaps to work with pores notably larger than few nm and fabricate them in a more scalable and controllable way. However, the memory effect is expected to degrade rapidly with increasing the pore size where fast and bulk ion transport gradually dominates[15,16]. Furthermore, to mimic the biological functions of synapses and use the devices for neuromorphic integration, one of the prerequisite is long endurance so that the programmed and event-based spike signals can be reliably processed over a long time[5-7]. Taking the synaptic proteins in biological systems as an example, which require regular replacement to maintain function and prevent damage, the lifetime typically ranges from days to months[24,25]. By contrast, the reported endurance for fluidic memristors, which can be characterized by the time span of continuous spike responses, is typically in the range of a few minutes up to about 2 h in response to $10^1$ – $10^4$ spikes having timescales from tens of ms to s[12-14,16,17,19,20,22]. This is, however, far from the lifetime of biological counterparts.

### *Observing an ionic memory effect in micropores*

Devices used in this study are micrometer size pores with corrugated graphene edges (Figs. 1a, S1). In brief, solid-state pores having diameters $D$ of 0.5 – 8 μm were microfabricated in silicon nitride ($SiN_x$) chips of 100 – 500 nm thick ($L$), as reported previously[26-28] and detailed in Methods ("Device fabrication"). Monolayer graphene films of A3 or A4 size were prepared by chemical vapor deposition (CVD) ("Synthesis of graphene" in Methods). The latter were transferred sealing the $SiN_x$ pores using polymethylmethacrylate (PMMA) as support. After dissolving the polymer by solvents, the whole assembly was blow-dried and then annealed to remove polymer residues. Because of surface tension and rapid evaporation of solvents, the freestanding graphene membranes over the $SiN_x$ pores tend to break, leaving the broken segments contracting toward the pore periphery and folding tightly to form a rippled structure[29,30] (Figs. 1a, S1). The $SiN_x$ chips with controllable pore size and number can be made on a wafer scale using the standard microfabrication techniques. Large-area graphene films can be continuously prepared by CVD. Only one transfer step is involved and could potentially be automated with robotics[31]. Therefore, the described device fabrication procedures are promising to scale up.

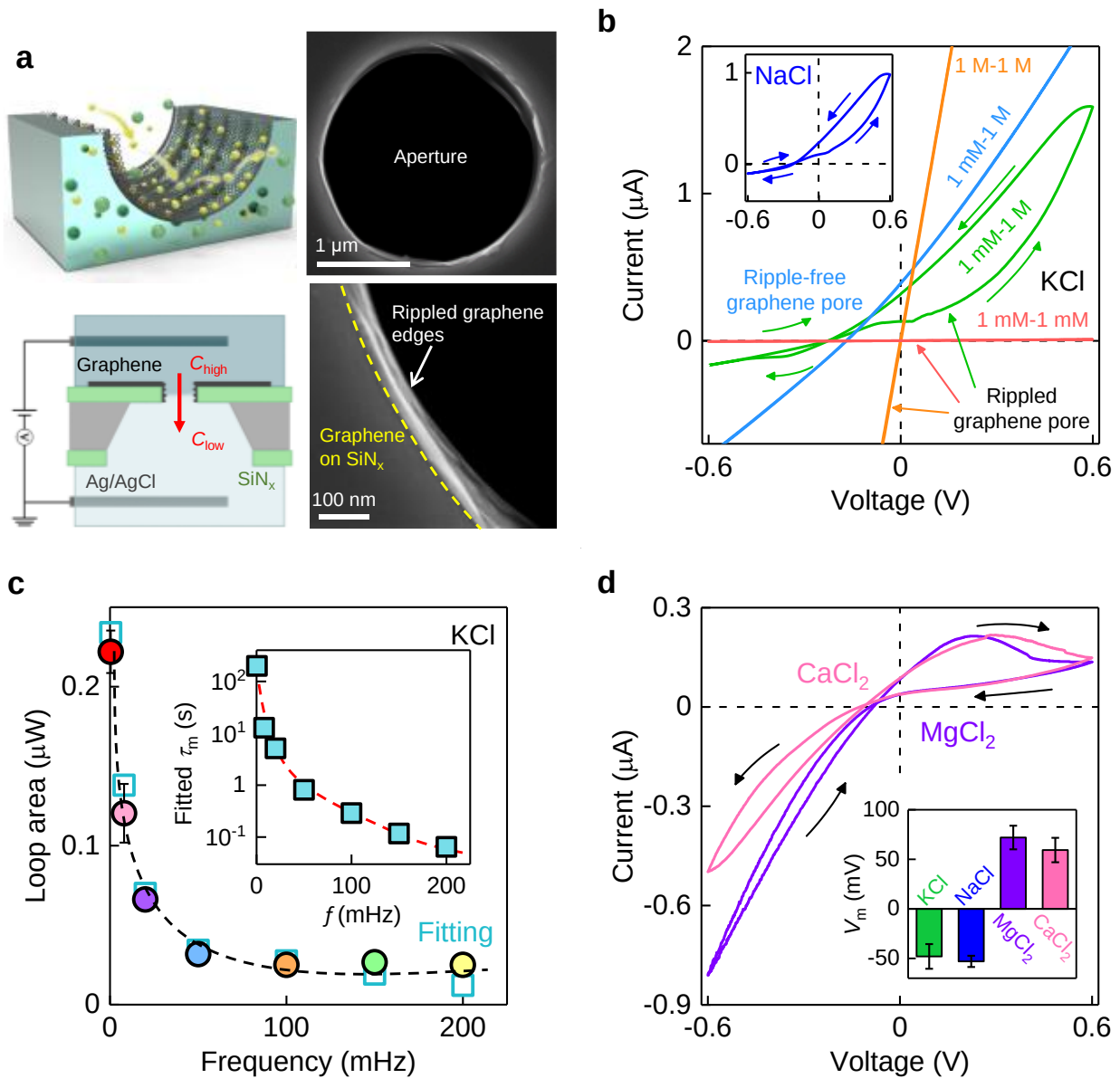

**Fig. 1: Experimental device and ionic memory effect.** **a** Schematics for our micropores with rippled graphene edges (top left) and the measurement setup (bottom left). Electron micrographs for one of our devices (top right) and zoom-in showing the pore edges (bottom right). **b** Representative *I*-*V* curves acquired in KCl solutions under different $C$ (color coded). Inset, same device but in NaCl solutions of $C_{high}$ = 1 M and $C_{low}$ = 1 mM. For comparison, the blue curve in the main figure is for a control device without rippled graphene edges (Fig. S1). **c** Frequency $f$ dependence of $A_{loop}$ in KCl solutions. Solid circles, experimental data averaged over at least 3 devices with error bars (shown only if larger than the symbols) indicating SD. Empty squares, fitted $A_{loop}$ using the circuit model detailed in the main text and SI ("Equivalent circuit model"). Inset, fitted memory timescale $\tau_m$ using the same model. Dashed curves, guides to the eyes. **d** Typical *I*-*V* curves in $MgCl_2$ and $CaCl_2$ solutions (color coded). Inset, $V_m$ for different electrolytes, same color coding as in (b, d). Error bars, SD based on 3 devices. In (c, d), $C_{high}$ = 1 M, $C_{low}$ = 1 mM. In (b, d), voltage sweep rate, $r_s$ = 1 mV/s; arrows indicate the directions of hysteresis loops and dashed black lines mark zero $I$ and $V$. For all our devices, $D \approx 2$ μm, $L \approx 500$ nm.

To measure ion transport through the fabricated pore, we employed a fluidic cell where the device separated two reservoirs containing electrolyte solutions of concentration $C$ (Fig. 1a, "Electrical measurements" in Methods). The time-dependent ion current $I(t)$ was recorded using a pair of Ag/AgCl electrodes under a periodic voltage sweep $V(t)$. Unless otherwise specified, all the measured devices had a single pore and the measurements were performed with devices of $D \approx 2$ μm, $L \approx 500$ nm and in KCl solutions of 1 mM ($C_{low}$) and 1 M ($C_{high}$). Under a triangular voltage wave from -0.6 V to +0.6 V and at a constant sweep rate $r_s$ = 1 mV/s, a nonlinear and rectified *I*-*V* characteristic was observed in the first few cycles (Fig. S2). The rectified characteristic then evolved into a pronounced anticlockwise hysteresis loop in the first quadrant that pinched on the x-axis and extended into the third quadrant as a less noticeable clockwise loop (Figs. 1b, S2). This feature remained unchanged under further repeated voltage sweeps (Fig. S2). It shows the dependence of *G* on the history of applied *V*, that is, a memory effect. The result is unexpected because normally a linear response without hysteresis is expected for such µm-size pores where bulk transport should dominate. The anomalous observation hints to a different mechanism at play and, presumably, involves surface transport (see below). The loop area $A_{loop}$ enclosed by the *I-V* curve serves as a measure for the observed memory effect. We find it depends on $r_s$, or the frequency *f* of *V*(*t*) (Fig. 1c). Increasing *f* from 0.4 mHz to 200 mHz results in a rapid decrease in $A_{loop}$ by about ten-folds (Fig. 1c). The pinched hysteresis loop (Fig. 1b) and its *f* dependence (Fig. 1c) are hallmarks of a memristor[3,4]. These two hallmarks were fully reproduced by fitting the experimental *I-V* curves with an equivalent circuit model constructed by a memristor in series with a direct-current (DC) power source (Figs. 1c, S3, "Equivalent circuit model" in SI). Specially, the fitting allows us to extract the memory timescale $\tau_m$ and the extracted $\tau_m$ at different *f* display an over three-orders-of-magnitude decrease and from ~200 s down to ~65 ms in the same range of *f* (inset of Fig. 1c). We note that to observe a pronounced hysteretic *I-V* loop and in particular, measure the *f* dependence with sufficient accuracy, our *I-V* measurements were performed at relatively low *f* (Figs. 1, S2). However, one of the key features of a memristor is that discrete voltage spikes can be used to tune its conductance states, which is advantageous over continuous voltage sweeps from a perspective of energy consumption. We will show later that our devices could reliably discern and process voltage spikes of physiologically relevant amplitudes and frequencies. Similar results were observed for NaCl solutions (inset of Fig. 1b). In stark contrast to KCl and NaCl, the *I-V* curves for $MgCl_2$ and $CaCl_2$ of the same *C* display inversely rectified and directional hysteresis loops (Fig. 1d). The results indicate that the observed memory effect is ion-selective.

## *Understanding the memory effect and its ion selectivity*

To quantify the ion selectivity, which can be characterized by the ratio of transport numbers for cations and anions, $t_+/t_-$, we note that the zero-current voltage $V_0$ observed in the hysteretic *I-V* curves (Figs. 1b, d) includes contributions from redox reactions at the electrodes and the membrane potential $V_m$ due to selective ion transport. Subtracting the electrode contribution allows us to extract $V_m$ for the tested electrolytes (inset of Fig. 1d). Accordingly, the ion selectivity $t_+/t_-$ can be calculated from $V_m$ using the Henderson equation[32]

$$\frac{t_+}{t_-} = -\frac{z_+}{z_-}\frac{\ln\left(\frac{C_{high}}{C_{low}}\right)+z_-\frac{eV_m}{k_BT}}{\ln\left(\frac{C_{high}}{C_{low}}\right)+z_+\frac{eV_m}{k_BT}} \quad (1)$$

where $z_+$ and $z_-$ are valences of cations and anions, respectively, $k_B$ is the Boltzmann constant, *e* is the elementary charge and $T$ = 297 ± 3 K. As per our setup (Fig. 1a) and eq. 1, the found $V_m$ based on statistics for several devices translate into $t_+/t_-$ of 1.5 – 2 for KCl and NaCl, and 0.5 – 0.9 for $MgCl_2$ and $CaCl_2$, respectively. The origin of the ion selectivity is discussed in SI ("Ion selectivity"). The key features found for different electrolytes including the directions of rectification and hysteresis loops, were fully reproduced using the described equivalent circuit (Fig. S3) and the fitting parameters (Table S1) effectively capture the ion selectivity ("Equivalent circuit model" in SI).

Although the found $t_+/t_-$ only slightly deviates from unity, its impact on the memory effect is significant. Our results (Figs. 1b, d) show that the directions of rectification and hysteresis loops in the *I-V* curves are practically determined by whether cations or anions in the tested electrolyte are the preferred charge carriers. This is further supported by our following control experiments where a thin layer of polyelectrolytes (diallyldimethylammonium chloride, PDDA) was grafted on graphene. In an aqueous environment, the adsorbed PDDA molecules would dissociate into adsorbed polycations and charge-balancing $Cl^-$ anions and hence, positively charge the graphene surface. Figure S2f compares representative *I-V* curves obtained from such a functionalized sample and a pristine graphene sample.

In KCl solutions, the functionalized device exhibited similar features as those found for the untreated samples in $MgCl_2$ and $CaCl_2$ solutions, along with $t_+/t_- < 1$ as evidenced by $V_m > 0$ (inset of Fig. S2f). A wider range of $t_+/t_-$ was yielded by adjusting solutions' pH and its influence on the memory effect is detailed in SI ("pH dependence", Fig. S4).

To gain further insight, we performed four additional experiments. First, we flipped the device while using the same setup as in Fig. 1a, and observed essentially the same *I-V* characteristics (Fig. S2). This shows that the characteristics were not caused by the possible asymmetry in the device structure or property. Second, we measured the same device but in KCl solutions of equal *C*. Both *C* = 1 mM and 1 M yielded linear and cross-zero *I-V* responses (Fig. 1b). Third, we deliberately removed the rippled graphene edges by dry etching (Fig. S1) and found a sublinear *I-V* response without hysteresis under the same conditions (Fig. 1b). Finally, in addition to graphene samples prepared by CVD, we also tested those made by mechanical exfoliation (Fig. S2). The latter are known to carry fewer defects and hence, surface charges in water. We found no noticeable difference in their *I-V* curves (Fig. S2), thereby ruling out defects as the primary reason for the memory effect. Instead, these qualitative experiments unambiguously corroborate that the memory originates from rippled graphene edges and concentration difference, which is supported by our theoretical analysis (see below, Figs. S5-S8).

The importance of rippled graphene edges implies that ion flows near the pore rim should be highly non-trivial. Trying to figure out the spatial distribution of ion current in our experimental setup, we performed finite element analysis (FEA) to solve the Poisson-Nernst-Planck (PNP) equations ("Spatial distribution of ion current" in SI). The key finding is that most of the ion flows are concentrated within a finite thickness around the pore rim rather than through the bulk space in the pore interior (Fig. S5). This conclusion is somewhat counterintuitive but overlooked before, presumably because for such µm-size pores, one may intuitively consider that the bulk current should dominate. Nonetheless, the conclusion is supported by our molecular dynamics (MD) simulations (Fig. S6, "Atomistic simulations" in SI), although a much smaller simulation model was used due to limited length and time spans and only qualitative agreement is reached. More importantly, statistics over all the ion trajectories in the latter simulations allows us to find out a characteristic transport pathway that contributes to over 90% of the current across the pore edge. In this pathway, ions first adsorb on the graphene surface and then diffuse to the pore edge (Fig. S6). The found pathway implies that the observed effect crucially involves surface transport. The latter's importance is in turn supported by our measurements for polyelectrolyte-modified devices and for pH dependences (Figs. S2 and S4), where charged ionic species are present on the devices' surfaces, and also, supported by our following theoretical analysis.

More quantitative analysis on the spatial distribution of ion flows as visualized in FEA (Fig. S5) reveals that more than half of the total current is through an annular region about 100 nm thick in a 2-µm-diameter pore. On the other hand, our electron micrographs (Figs. 1 and S1) show that the rippled graphene structure extends by about 30 – 50 nm from the sharp edges of a similar-size $SiN_x$ pore. These comparable length scales suggest that the transported ions could effectively contact and interact with the rippled graphene, and hence, get trapped inside the structure, as evidenced by the post-measurement electron micrograph shown in Fig. S1. In this context, the characteristic dimension $d^*$ for assessing the importance of surface transport should practically be the size of empty space formed by corrugated and folded graphene sheets (a few nm in height, Fig. S1) rather than the pore's µm-scale dimension. Indeed, our analysis by comparing the estimated Debye length[33,34] $\lambda_D$ with $d^*$, and Dukhin number *Du* with unity, suggests that surface effect, primarily the transport in and across rippled rim structure, is important ("Atomistic simulations" in SI). Taken together, these theoretical results allow us to conclude that nanoconfinement, typically imposed in experiments by restricting the pore size to nm-scale, can be accessed equivalently in µm-size pores by engineering the spatial distribution of ion flows and the nanostructure of pore rim.

With these theoretical results in hand, our further MD simulations by taking the rippled graphene edges into account show that the ion transport can be delayed by adsorption on the graphene surface and, crucially, the delay effect is enhanced by ion accumulation inside the rippled structure ("Atomistic simulations" in SI). The described slow ion dynamics is the primary origin of observed memory effect. Accordingly, the complete ion transport pathway should involve surface adsorption and diffusion, trapping and residing in the rippled edges, and escaping (Fig. S6), which is visualized by the spatial distribution of salt precipitates in the post-measurement electron micrograph (Fig. S1). The memory timescale is then expressed as $\tau_m \approx \tau_s\ \tau_e/\tau_t$, with $\tau_s$, $\tau_t$ and $\tau_e$ timescales for surface diffusion, ion trapping into and escaping from the rippled structure, respectively. Further analysis reveals $\tau_e >> \tau_t$ ("Atomistic simulations" in SI), and because $\tau_s \propto L_D^2$ with $L_D$ the length of diffusion pathway, the observed $\tau_e >> \tau_t$ translates into much longer distance of random motion within the rippled structure than that of the diffusion pathway itself, that is, the trapped ion can cross an individual ripple multiple times and bidirectionally before its escaping. This is fully consistent with our Kinetic Monte Carlo simulations (Fig. S7, "Kinetic Monte Carlo simulations based on the hopping diffusion model" in SI). The found $\tau_e >> \tau_t$ also translates into a finite difference in the free energy barrier $G$ between the escaping and trapping steps ("Atomistic simulations" in SI), which we attribute to the interactions of ions with and inside the strongly confined rippled graphene edges ("Ion selectivity" in SI). On the other hand, the analysis on the concentration effects suggests that different $C$ across the pore lead to ion accumulation and depletion under an alternative electric field (Fig. S8, "Concentration effects" in SI) and the time difference between the escaping and trapping steps ($\tau_e >> \tau_t$) makes the dynamic process lagged with respect to changes in the electric field and hence, yields a rectified memory effect, as manifested in all the shown *I-V* curves.

### *Constructing fluidic circuits and quantifying the memory effect*

The devices are easy to fabricate and the observed memory effect is highly reproducible. We find 89 out of 97 of our measured devices (yield of ~92%) showed quite similar hysteretic *I-V* curves under the same experimental conditions (detailed measurements with each device see Table S2). For example, Figs. 2a and S2 show the full sets of cyclic *I-V* measurements ($f \approx 0.4$ mHz) for three of our devices. The general evolution is that after a few cycles of cyclic voltage sweeps, hysteresis loops appeared and after 10 – 30 cycles, the hysteretic *I-V* curves became stable. Statistics over the stable *I-V* curves for the shown devices yields $A_{loop}$ of 0.25 ± 0.05 µW, 0.22 ± 0.02 µW and 0.20 ± 0.06 µW, respectively, and summarizing them all, a device-to-device variation not exceeding 10.4% is obtained, which is defined as standard deviation (SD) of $A_{loop}$ divided by its mean (that is, the coefficient of variation). The small variation agrees with that extracted from statistics of HRS/LRS ratios for 43 devices (~10.3%, Fig. S2, "Additional *I-V* curves" in SI). Furthermore, similar statistics over all the zero-current voltage $V_0$ extracted from stable *I-V* curves yields $V_0$ = -0.205 ± 0.011 V, -0.194 ± 0.018 V and -0.219 ± 0.007 V, respectively. The results translate into a variation of only a few percent in $V_0$ under cyclic *I-V* measurements on a single device lasting for days and across multiple devices, which support the stability of our electrodes. Once the memory effect emerged, it could be stable for at least a few days and up to one month under cyclic voltage sweeps (Fig. 2b). Longer-term measurements led to either a notable smaller *I* (presumably due to pore clogging), or the complete disappearance of hysteresis loops. We believe the latter was caused by the blockage of ion diffusion pathway due to the accumulation of salt precipitates in the rippled structure (Fig. S1). This is supported by the observation that after dissolving the precipitates in water at an elevated temperature *T* ("Additional *I-V* curves" in SI), 32 out of 34 treated devices recovered their stable memory. The testing-cleaning procedures could be repeated many times, and devices with a longer working duration typically involved multiple cleaning steps (Fig. 2c). The longest working duration recorded for our tested devices was about three months whereby the devices were subjected to over 20 of the described water-cleaning steps (Fig. 2c). Also, the influence of cleaning on the observed memory was negligibly small, as evidenced from the compilation of stable *I-V* curves after each cleaning step for one of our devices (inset of Fig. 2c). The comparison yields a variation coefficient of ~7% in their $A_{loop}$, which is again consistent with the observed small device-to-device variations. The described features including high yield, long-term stability and recyclability are long sought-after and underpin our following experiments.

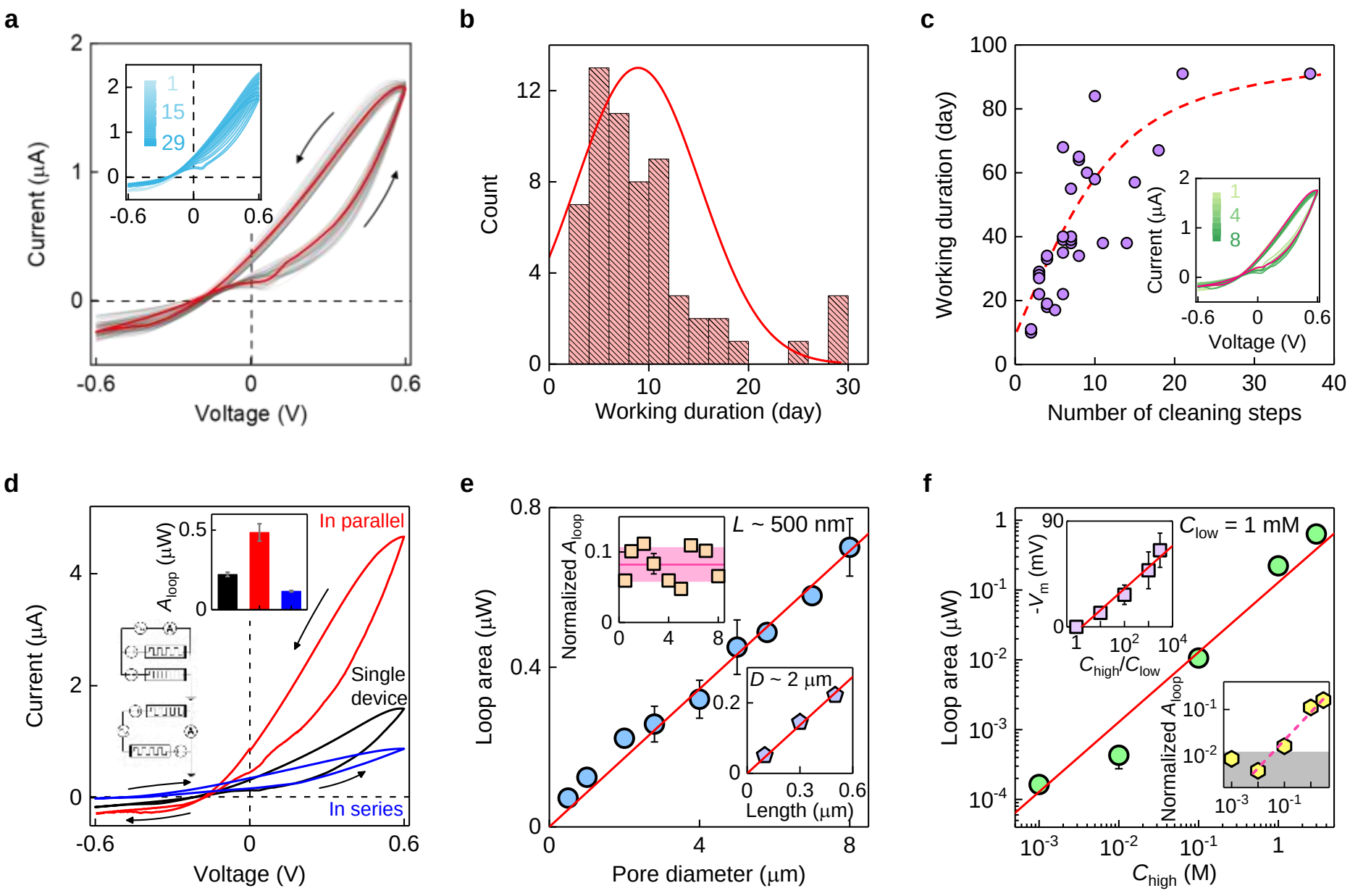

**Fig. 2: Robustness, integration, and quantification of the memory effect. a** Full set of cyclic *I*-*V* measurements ($f \approx 0.4$ mHz) for a representative device over a few days. Main figure, *I*-*V* curves after reaching stabilization. The compilation includes 142 curves in total with the red one highlighting the characteristic memristive behavior. Inset, before stabilization (color coded). **b** Histogram of the working duration after reaching stabilization for 60 devices that subjected to cyclic *I*-*V* measurements such as in (a). **c** Working duration as a function of the number of cleaning steps. Each symbol represents a different device and in total, 32 devices were shown, which were subjected to cleaning in water at $T = 60$ °C. Dashed red curve, guide to the eyes. Inset, stable *I*-*V* curves after each cleaning step (color coded) for one of our devices with the red one denoting that before cleaning. **d** Stable *I*-*V* curves for parallel and series circuits constructed by two devices working in the same concentration direction ($C_{high}$ = 1 M, $C_{low}$ = 1 mM) and compared with that for a single device (color coded). Lower insets, schematics for the integrated circuits. Upper inset, comparison of $A_{loop}$ (same color coding). **e** Main figure, $D$ dependence of $A_{loop}$. Length, $L \approx 500$ nm. Top inset, same dependence but for normalized $A_{loop}$. Same *x*-axis as in the main figure. Solid pink line, average over all data with the shaded area indicating SD. Bottom inset, $L$ dependence, $D \approx 2$ μm. Same *y*-axis as the main. Solid red lines in the main and bottom inset, best linear fits. For all measurements in (e), $C_{high}$ = 1 M, $C_{low}$ = 1 mM. **f** Main, $A_{loop}$ at fixed $C_{low}$ = 1 mM but varied $C_{high}$ with the upper inset plotting $V_m$ as a function of $C_{high}/C_{low}$. Solid red lines in the main and top inset are linear fits. Note the slope in the main figure is fixed at 1. Bottom inset, same $C_{high}$ dependence as in the main figure but for normalized $A_{loop}$. Pink dashed line, guide to the eyes. Grey shaded area marks the normalized $A_{loop}$ smaller than that at $C_{high}$ = $C_{low}$ = 1 mM. KCl solutions were used for all the measurements. The shown *I*-*V* curves were acquired under $V$ from -0.6 V to +0.6 V and at $r_s$ = 1 mV/s; the arrows indicate the same loop directions; black dashed lines mark the positions of zero $I$ and $V$. All the $A_{loop}$ data in (d-f) are averaged over 10 independent *I*-*V* measurements with error bars indicating SD. In (d, f), $D \approx 2$ μm, $L \approx 500$ nm.

To prove the feasibility of device integration on a circuit level, we used two similar devices to construct fluidic circuits (lower insets of Fig. 2d). Representative *I*-*V* curves for the series and parallel circuits with two devices in the same concentration directions are shown in Fig. 2d and compared with that for a single device. As expected, $A_{loop}$ for the parallel and series circuits are about twice and half that for a single device, respectively (upper inset of Fig. 2d). In contrast, if the concentration directions in the two integrated devices are opposite and the resulting $V_0$ essentially cancel out, the *I*-*V* curves for both the series and parallel circuits are sublinear and cross through the origin (not shown).

To quantify how the memory effect evolves with its two determinants, namely, rippled graphene wrapping the pore rim and different concentrations across the pore, we carried out the following measurements using KCl solutions. First, we extracted $A_{loop}$ for pores having different diameters $D$ (0.5 – 8 μm) but the same $L \approx 500$ nm. Results show that the absolute values of $A_{loop}$ increase linearly with $D$. Because the conductance $G$ in such large pores is also expected to scale linearly with $D$[35,36] (Fig. S9), to subtract the effect of $G(D)$, we followed the previous methods[16,23] and normalized the absolute value of $A_{loop}$ with a characteristic conductance $G^*$, as defined by $G^* = I(V = V_{max})/(V_{max} - V_0)$ (details see "Additional *I-V* curves" in SI). The normalized $A_{loop}$ is independent of $D$ within a scatter of ~30% (upper inset of Fig. 2e), indicating that the observed linear-in-$D$ dependence of absolute $A_{loop}$ is simply caused by the linear $G(D)$ (Fig. S9), whereas the memory effect is practically independent of pore size. The linear $G(D)$ translates into constant ion flow per unit perimeter and our theoretical analysis reveals that the ion flow across the pore rim dominates over that through the pore interior (Figs. S5, S6). Accordingly, these observations suggest that the memory effect scales linearly with the ion flow per unit perimeter. Furthermore, we note that the experimental scatter of ~30% (upper inset of Fig. 2e) cannot be explained by the observed variations among different devices (about 10%). However, it agrees well with the deviation of $G^*(D)$ with respect to the best linear fit ("Additional *I-V* curves" in SI, Fig. S9). Second, for a fixed $D \approx 2$ μm, both the absolute and normalized values of $A_{loop}$ display linear-in-$L$ dependences (lower inset of Fig. 2e, Fig. S9), which unambiguously corroborate that the memory effect further scales linearly with the length of ion transport pathway. A longer pathway allows the transported ion to be trapped inside the longer rippled structure for a prolonged duration before its escaping, resulting in a larger memory effect (Fig. S6). The described longer residence is further supported by the smaller $G^*$ observed in the corresponding *I-V* curves (Fig. S9).

Next, we fixed $C_{low} = 1$ mM and varied $C_{high}$ and found that absolute values of $A_{loop}$ scale linearly with $C_{high}$ spanning over 4 orders of magnitude (Fig. 2f). Same as in Fig. 2e, we normalized $A_{loop}$ and found that the data still follow an increasing trend within our experimental accuracy (lower inset of Fig. 2f). The observation indicates that the memory effect is determined by either $C_{high}$ or the concentration gradient $C_{high}/C_{low}$. To disentangle the two possible contributions, we fixed $C_{high}/C_{low} = 10^3$ and varied the absolute values of both $C_{high}$ and $C_{low}$. A second linear increase of absolute $A_{loop}$ with $C_{high}$ is seen (Fig. S9) and interestingly, the same $A_{loop}(C_{high})$ data but after normalization by $G^*(C_{high})$ display a decreasing trend. The latter trend indicates that the memory effect is weakened by increasing $C$ in both reservoirs, which is in line with the expected concentration evolutions of both $\lambda_D$ and $Du$ and therefore, suggests the importance of surface effect. This again echoes both our experimental measurements (Figs. S2, S4) and theoretical analysis (Figs. S6). Note the membrane potential $V_m$ increases linearly with $\ln(C_{high}/C_{low})$ (upper inset of Fig. 2f). According to eq. 1, this indicates the ion selectivity $t_+/t_-$ is insensitive to concentration changes and remains essentially as a constant.

### *Emulating synaptic plasticity*

The observed memory effect enables emulating many functionalities found in biological synapses[1,2]. Specially, one of the mechanisms underlying learning and memory is the reversible modification of synaptic strength by the incoming voltage spikes (synaptic plasticity) and its retention over time[1,2]. To mimic this plasticity, we used a series of discrete write ($V_{write}$) or erase ($V_{erase}$) spikes with programmable amplitude $A$, duration $t_d$ and time interval $\Delta t$, instead of continuous voltage sweeps. The resulting evolutions in $I$ were directly recorded or translated into evolutions in $G$ by the following read spikes ($V_{read}$), which were negligibly small and had negligible perturbations on $G$ compared to $V_{write}$ or $V_{erase}$. To assess short-term plasticity, we applied a pair of voltage spikes of $A = 0.6$ V, $t_d = 5$ s and varying $\Delta t$ (Figs. 3a, b), and then recorded the ion current ratio after the 1st and 2nd spikes, $I_2/I_1$. For KCl solutions, Fig. 3a shows that two positive (negative) spikes increase (decrease) the current, resulting in so-called paired-pulse facilitation (PPF) and depression (PPD), respectively. By contrast, an opposite trend is seen for $MgCl_2$ solutions. These results align with the inverse hysteretic loops in the *I-V* curves detected for $MgCl_2$ and KCl solutions (Figs. 1b, d). More quantitatively, we find that $I_2/I_1$ decays exponentially with increasing $\Delta t$. Fitting the data with $I_2/I_1 \propto \exp(-t/\tau)$ yields a characteristic time constant $\tau$ of a few seconds for both $MgCl_2$ and KCl solutions.

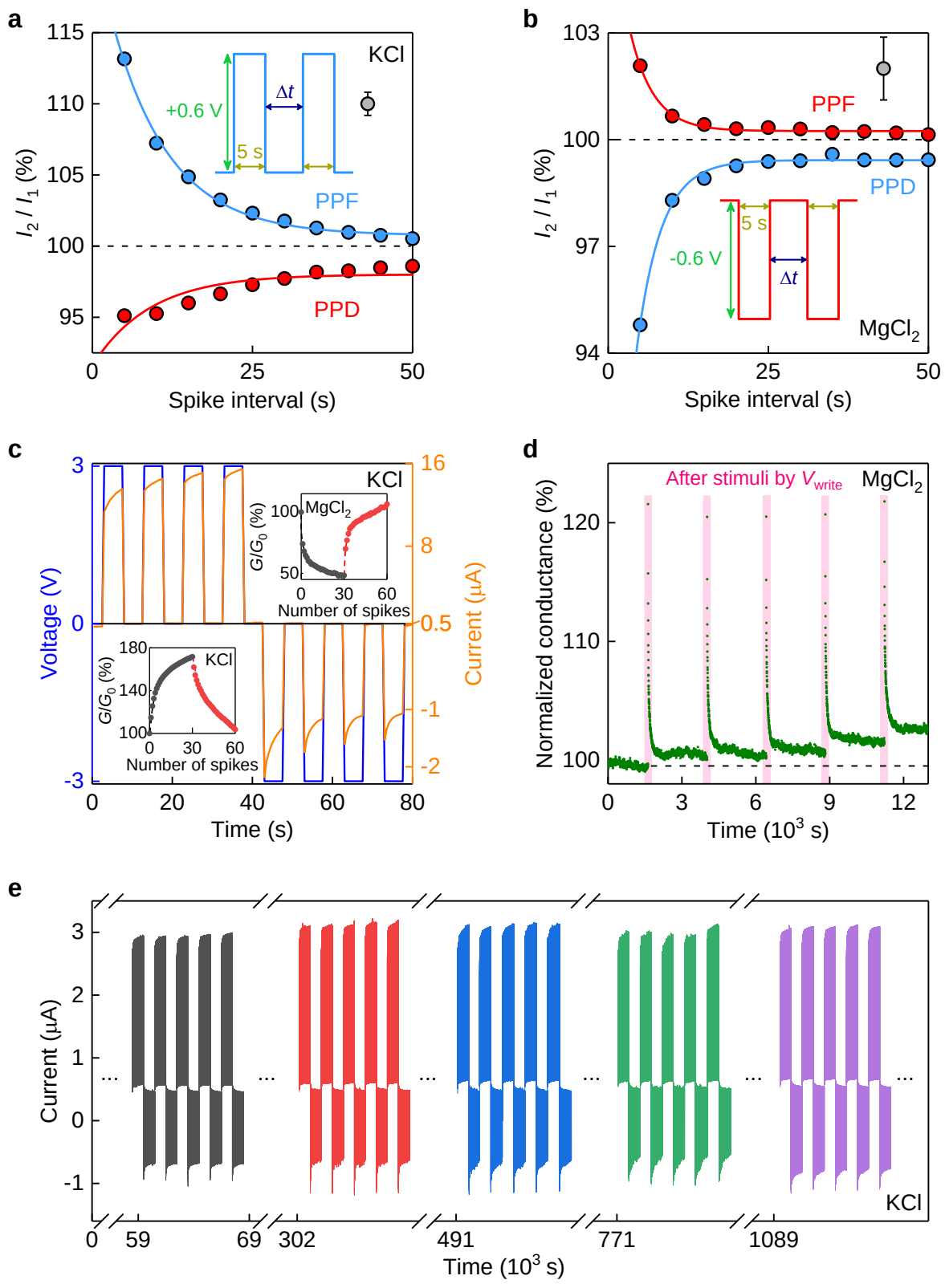

**Fig. 3: Modulating the conductance states using voltage spikes.** Current ratio $I_2/I_1$ in response to a pair of positive (blue) or negative (red) spikes versus the time interval $\Delta t$ for **a** KCl and **b** $MgCl_2$ solutions, respectively. Symbols, measurement data. Error bars on the grey symbols in (a, b) indicate SD obtained from repeated measurements performed on the same device. Solid curves, best exponential fits. Black dashed lines mark the position of $I_1/I_2$ = 1. Insets in (a, b), details of the applied spikes (same color coding as in the main figure). **c** Current evolution (orange, right-*y*) under successive positive and negative spikes (blue, left-*y*). Insets, reversible modification of *G* (normalized by that before the first spike, $G_0$) under successive positive (black) and negative (red) spikes for KCl (bottom) and $MgCl_2$ (top) solutions, respectively. Spikes ($V_{write}$ or $V_{erase}$) of $A$ = 3 V and $t_d$ = 5 s were applied to modify *G*, which were read by the following $V_{read}$ of $A$ = 0.02 V and the same $t_d$. Time interval, $\Delta t$ = 5 s. Dashed curves in the insets, guides to the eyes. **d** Changes in *G* under repeated stimuli by a negative $V_{write}$ ($A$ = 3 V, $t_d$ = 1 s) in $MgCl_2$ solutions. Symbols are *G* recorded by $V_{read}$ ($A$ = 0.02 V, $t_d = \Delta t$ = 1 s) and normalized by $G_0$. The abrupt changes in *G* after stimuli by $V_{write}$ are shaded in pink, highlighting short-term plasticity. Black dashed line marks $G/G_0$ = 1. **e** Successive current responses of one of our tested devices to alternative positive and negative spikes ($A$ = 0.6 V, $t_d = \Delta t$ = 0.5 s) in KCl solutions. In a single cycle, 1,000 positive and negative spikes were sequentially applied, respectively. The cycle was repeated over $10^5$ s until the current responses became unstable. Then we cleaned the device in water at $T$ = 60 °C and continued the test again (color coded). In total, the device reliably responded to more than $10^6$ spikes and the process lasted over $10^6$ s. Shown are randomly chosen segments along the entire time span, highlighting the steady increase and decrease in *I* under successive positive and negative spikes, respectively. The rest are folded into breaks along the x-axis for clarity. Devices in all our measurements, $D \approx 2$ μm, $L \approx 500$ nm; electrolytes, $C_{high}$ = 1 M, $C_{low}$ = 1 mM.

To assess long-term plasticity, we show in Fig. 3c that in KCl solutions, successive stimuli of positive (negative) spikes lead to progressive increase (decrease) in *I*, and hence, reversible modification of *G* (lower inset of Fig. 3c). The absolute values of *I* under positive spikes are notably higher than those under negative ones. This correlates well with the rectified *I-V* characteristics under continuous voltage sweeps (Figs. 1b, S2). In contrast, same voltage spikes but in $MgCl_2$ solutions yield an opposite evolution in $G$ (upper inset of Fig. 3c), in agreement with that observed for short-term plasticity (Figs. 3a, b). Specially, under exposure to $V_{write}$ of -3 V, the recorded $G$ in $MgCl_2$ solutions abruptly increase and then relax in a few minutes to a long-term value above its initial state $G_0$. This shows the presence of mixed short- and long-term plasticities. Repeated $V_{write}$ lead to multiple long-term $G$ that are clearly distinguishable and gradually step up during the measurement for over $10^4$ s (Fig. 3d). Furthermore, we would like to note a few experimental facts. First, from the evolutions of either $I$ or $G$ under the stimuli of (at least a limited number of) voltage spikes up to 3 V, we believe no irreversible or semi-reversible electrochemical processes related to electrodes occurred, as evidenced by the absence of either irreversible drift of the $I/G$ baselines or the loss of successive potentiation/depression responses. Second, we tested 39 devices in total, using alternative series of positive and negative spikes, respectively, and found that all of them showed stable current evolutions in response to at least $10^5$ successive spikes lasting more than $10^5$ s. Longer-term stimuli by a larger number of spikes eventually led to the loss of stable increase and decrease in $I$. However, we believe the observed instability was irrelevant with any possible electrode processes, but rather could be explained by the accumulation of salt precipitates in the rippled structure (Fig. S1b) that blocked the ion diffusion pathways (Fig. S6, "Atomistic simulations" in SI). The latter is supported by the fact that after simply cleaning the used devices in water as described above, 30 out of 32 recovered the long-term plasticity, which ensured stable response to over $10^6$ spikes and lasting over 10 days (Fig. 3e). This is perhaps the best achievable endurance so far for fluidic memristors in response to voltage spikes, which is highly sought-after for neuromorphic applications. The long-term plasticity could be finely tuned via the parameters of input spikes (Fig. S10). Stronger stimuli using spikes having higher $A$, longer $t_d$ and shorter $\Delta t$ (or higher frequency $f$) generally lead to greater changes in $I$ (Fig. S10). In those measurements, we used spikes up to a few volts and lasting for seconds so that pronounced evolutions in $I$ could be observed. Nonetheless, we found that using biologically relevant spikes[9,10] down to a few tens of mV and lasting about 10 ms could still lead to successive increase or decrease in $I$, with $\Delta I$ between adjacent spikes slightly above our lower detection limit (~10 pA) (Fig. S11, "Influence of spike parameters on the current modulation" in SI).

### *Neuromorphic applications*

The mixed short-term and long-term memories and plasticities (Fig. 3) allow us to implement many neuromorphic applications by storing, transmitting and processing event-based information encoded in the temporal domain as evolutions of $G$. For example, we designed image identification experiments (Figs. S12-S13, details see "Image identification" in SI) utilizing the devices' long-term memory and plasticity, and taking advantage of their long-term endurance (Fig. 3). The basic idea is to encode the intensities of tested images in two selected datasets[37,38], MNIST and CIFAR-10, into long spike trains, then transmit them through the constructed fluidic circuit and convert them into $G$ evolutions, and finally, implement image identification using the latter as features and with the assistance of our selected neural network models. The key observation is that the accuracies after processing by our fluidic circuit were close to that directly using the original data (Figs. S12-S13, Table S3), which proves the high fidelity and reliability of our devices for storing and transmitting largescale and event-based information over a long time.

The short-term memory and plasticity also enable our devices to implement reservoir computing for analyzing the information encoded in the temporal domain of transmitted neural spike trains, for example, recognizing different neural firing patterns and identifying synchronization states among different neurons. This ability is important for understating biological processes such as neural communications, memory and consciousness[39,40]. Conventional analysis approaches mostly involve transmitting and storing large amount of neural data[41] and then post-processing them separately by, for example, neural network models[42]. Real-time analysis was recently demonstrated using a solid-state memristor[43]. In that example, the developed neural network model was first trained on a simulation dataset and then tested on the real experimental data. This design inherently suffers from the so-called 'sim-to-real gap'. The neural network models optimized from idealized mathematical approximations cannot fully capture the non-ideality of experimental devices and the complexity of operation environment, which typically results in unsatisfactory accuracies. To bridge this gap, one needs to train and test the neural network models using real experimental data, which ultimately requires the devices to have a good long-term stability so that sufficient experimental data can be reliably acquired. In this context, we expect our fluidic memristors can achieve a better performance.

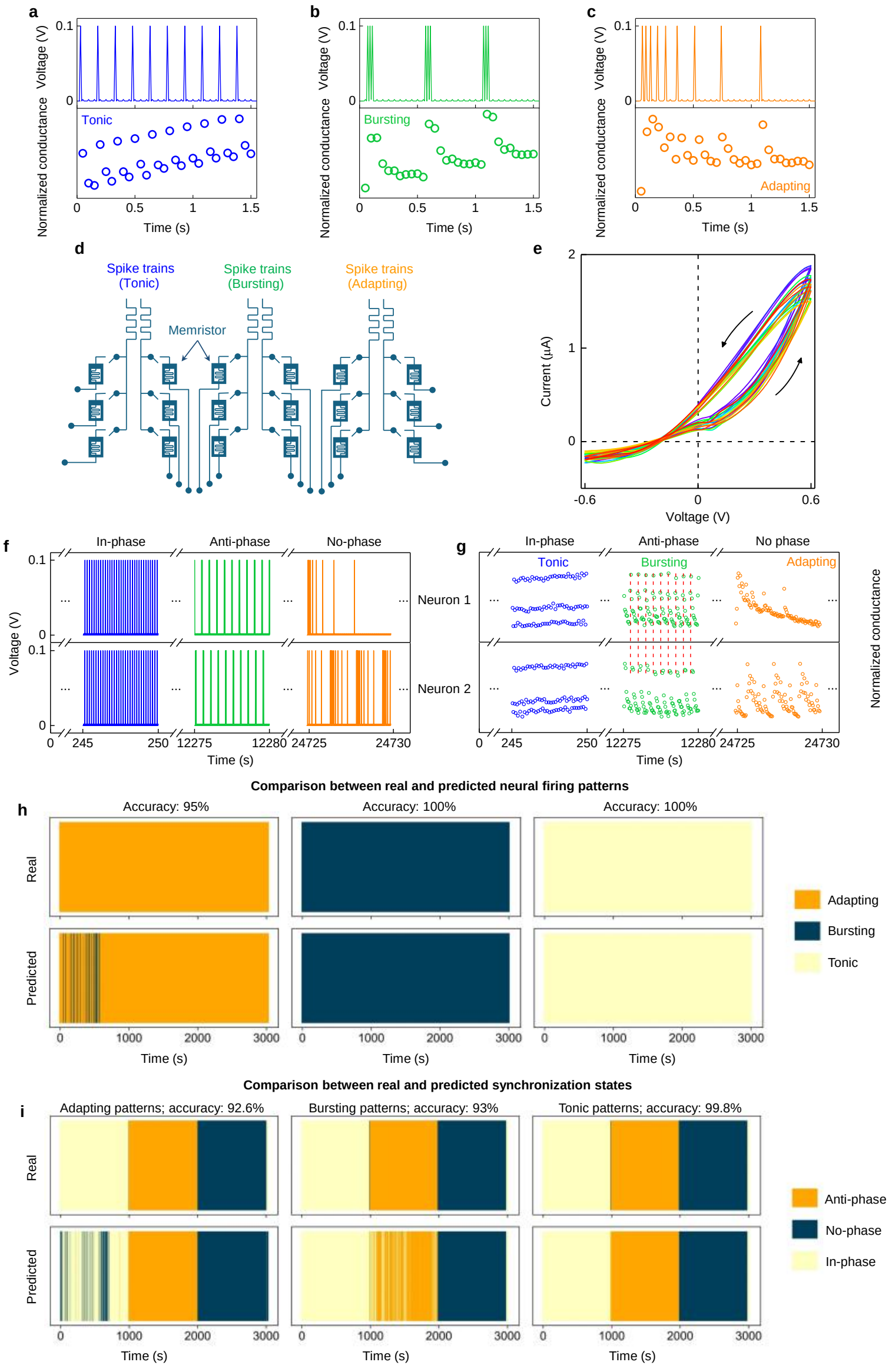

**Fig. 4: Real-time analysis of neural activities. a-c** Top panels, programmed short spike trains featuring tonic, bursting and adapting neural firing patterns (color coded), respectively. Bottom, corresponding time evolutions of $G$ (normalized by $G_0$). **d** Schematic of the integrated circuit constructed by 18 fluidic memristors. **e** Corresponding stable *I-V* curves for all devices in (d). Each curve represents a different device (color coded). Arrows indicate the same direction of hysteresis loops. Black dashed lines mark zero *I* and *V*. **f** Programmed 7.5 h-long spike train pair for generating the training dataset. They were used to emulate signals transmitted from two neurons. **g** Corresponding time evolutions of $G/G_0$. Shown in (f, g) are randomly selected segments featuring three neural firing patterns (tonic, bursting and adapting, color coded) and three synchronization states (in-phase, anti-phase and no-phase, from left to right) along the entire time span. Red dashed lines in (g) mark the observed time shift of ~100 ms in the anti-phase and bursting pair. **h & i** Simultaneous and real-time predictions of neural firing patterns and synchronization states, respectively, and their comparisons with the real data.

Because the devices can reliably work with spikes of physiologically relevant amplitudes and frequencies ($A < 130$ mV and $f = 10 - 100$ Hz[9,10]) (Fig. S11), for simplicity while retaining the key features of neural spikes, we emulated the real neural signals using $V_{write}$ of $A = 100$ mV and $t_d = 10$ ms. By varying the time intervals between individual $V_{write}$ or units composed of multiple $V_{write}$, spike trains featuring three characteristic firing patterns, namely, tonic, bursting and adapting, were constructed (Figs. 4a-c, top panels. Details see section "Real-time analysis of emulated neural signals" in SI). The resulting time evolutions of $G$ were then recorded by $V_{read}$ of $A = 2$ mV and $t_d = 10$ ms that were incorporated uniformly at a time interval $\Delta t = 40$ ms throughout the entire time line. Figures 4a-c (bottom) exemplify the evolutions of $G$ (normalized by $G_0$) in response to the three neural firing patterns (top). Clear differences in the recorded $G$ responses are shown, which in turn can be used as features for recognizing different neural firing patterns. To fully exploit the long-term stability of our devices, we further programmed spike trains of a few hours long by repeating these few-seconds long neural firing patterns along the same time line so that sufficiently large datasets of $G$ evolutions could be collected. In a single long spike train, each firing pattern spanned the same time length. Using a pair of such spike trains but with a relative time shift in their tonic, bursting and adapting segments, respectively, three synchronization states, namely, in-phase, anti-phase and no-phase could be constructed (Figs. 4f, g, "Real-time analysis of emulated neural signals" in SI). Different synchronization states in each firing pattern also occupied the same length. This structure ensured the developed neural network models at a later stage could learn to distinguish temporal dynamics without a class-based bias. In principle, a minimum number of two devices are required for simultaneously identifying the synchronization states between two neurons. To increase the efficiency, we customized a fluidic circuit using 18 properly connected devices (Figs. 4d, S14). Their representative *I-V* curves after reaching stabilization are shown in Fig. 4e. Statistics over all these curves yields $A_{loop}$ of $0.23 \pm 0.02$ µW, and hence, a variation of ~8.7%, in agreement with the observed device-to-device variations. This circuit enabled identifying the neural firing patterns and synchronization states between two neurons in an in-parallel and simultaneous manner. To be specific, two long spike trains featuring three firing patterns and three synchronization states were first divided into three pairs of shorter segments, with each pair in the same pattern (tonic, bursting or adapting). Then they were fed into the circuit's six inputs (Fig. 4d), respectively. Three devices were connected in parallel to the same input (Fig. 4d). They were used to successively process the three synchronization states within each pair of the shorter segments. For generating the training dataset, a pair of 7.5 h-long spike trains were employed for the experiment, whereas for the testing dataset, shorter spike trains having the same parameters but lasting about 2.5 h were used. Independent processing of these two pairs resulted in a total number of about $1.5 \times 10^6$ conductance samples ($1.1 \times 10^6$ for training and $0.4 \times 10^6$ for testing) constituting the entire experimental dataset, which is sufficiently large for both training and testing the developed neural network models.

Figures 4f and g compare the programmed 7.5 h-long spike train pair and the resulting time evolutions of $G$. From the shown data segments, it is evident that the constructed three firing patterns and three synchronization states can be clearly discerned from the observed $G$ evolutions. Nonetheless, to assess the quality of the entire training and testing datasets, we used the $G$ responses to different patterns as features and classified them using a simple CNN model (Fig. S14). The model contained a one-dimensional (1D) convolution layer with a Gaussian noise layer for data augmentation, and a fully connected layer (Fig. S14). It was trained on our prepared training and validation sets, which were obtained by splitting the two time series of $G$ responses into a length ratio of 4:1. Then online predictions were made on the testing data, yielding an overall high accuracy of 96.8%, which is notably higher than that reported for the aforementioned solid-state memristor (~87%)[43]. Details about the predictions are specified in the confusion matrix shown in Fig. S14. The bursting patterns were classified with 100% accuracy, whereas the adapting ones exhibited the highest misclassification rate with ~9.4% of all samples misclassified as bursting patterns. Furthermore, near-100% accuracy was obtained for the recognition of tonic patterns and few samples (<0.2%) were misclassified as adapting or bursting ones.

To implement simultaneous identification of neural firing patterns and synchronization states using shared features from the *G* responses to a pair of spike trains, we developed a multi-scale CNN model architecture having three parallel 1D convolutional branches with kernel sizes of 5, 20, and 50, respectively (Fig. S14, details see "Real-time analysis of emulated neural signals" in SI). This design allows extracting features of different timescales (that is, the short-term and sudden spiking, medium-term periodicities, and long-term general trends) concurrently from the evolutions of *G*. To train the multi-scale CNN model, we employed the same training and validation sets as prepared above but in a pairwise manner so that features corresponding to the three different synchronization states between two neurons could be extracted from the correlated time series pairs. Similarly, the testing dataset was also used in pairs for simultaneous online predictions of the neural firing patterns and synchronization states. Results and their comparisons with the real data are shown in Figs. 4h and i. Our model achieves high overall accuracies of ~98% for neural firing patterns and ~95% for synchronization states. Specifically, the bursting and tonic patterns were identified with 100% accuracy, whereas a slightly lower accuracy of 95% was seen for predicting the adapting ones. This result is in good agreement with the prediction for neural firing patterns alone (Fig. S14). On the other hand, identification of the three different synchronization states gave an accuracy of higher than 92% across all firing patterns and in particular, a near-perfect accuracy of 99.8% was achieved between two spike trains featuring tonic patterns. For the adapting and bursting ones, predictions for the no-phase segments remained accurate but misclassification primarily occurred in the in-phase and anti-phase segments.

To conclude, micrometer size pores with sharp and smooth walls yield a sublinear *I-V* response under a concentration difference. However, the presence of nanoporous and jagged edges such as those constructed by corrugated graphene can effectively confine the permeating ions in the nanostructures and slow down the cross-pore transport, resulting in a memory effect. This highlights engineering the morphology of pore walls as additional degree of freedom beyond the well-documented pore's charge and size for new insights and applications. Fundamentally, the underlying mechanism is instructive. It shows that the ion flows across a micropore are concentrated in the vicinity of pore edges rather than in the bulk space of pore interior. Thanks to this rather nonuniform spatial distribution, the strong nanoconfinement important for an ionic memory effect can be effectively grafted to the pore edges. This allows lifting up the restrictions of using nanopores or nanochannels for fluidic memristors and circumventing many issues intrinsic to nanofluidics leading to poor reproducibility. More generally, the described mechanism can be extended to other two-dimensional materials, which can in principle form similar rippled structures around the pore rim, and, also, to a wider range of nanoporous structures decorating the pore walls beyond those assembled by two-dimensional materials in general. From an engineering standpoint, our devices are scalable and integrable. Rippled graphene pore arrays could be fabricated on a wafer scale thanks to the well-established synthesis and transfer of graphene films, and microfabrication techniques. Then fluidic crossbar arrays could potentially be assembled by introducing parallel microchannels to the pore arrays but perpendicularly aligned on the opposite sides of the wafer, mimicking their solid-state counterparts. The described microchannels could be well-defined in PDMS layers by standard microfabrication techniques. On the other hand, the memory and plasticity can be tailored using different ions as charge carriers. These characteristics promise the use of fluidic circuits and multiple ions for in-parallel information processing. More importantly, the exceptional stability and endurance of our devices enable reliably transmitting, storing and processing event-based information encoded by large amount of voltage spikes over a long time. Combined with the advancement of artificial intelligence models, we expect the devices and their integrated circuits may be useful in more complex neuromorphic applications.

1. Voglis G, Tavernarakis N. The role of synaptic ion channels in synaptic plasticity. *EMBO Rep* 2006;**7**:1104–10. https://doi.org/10.1038/sj.embor.7400830.
2. Gerstner W, Kistler WM, Naud R *et al. Neuronal Dynamics: From Single Neurons to Networks and Models of Cognition*. Cambridge: Cambridge University Press, 2014.

3. Chua L. Memristor-The missing circuit element. *IEEE Trans Circuit Theory* 1971;**18**:507–19. https://doi.org/10.1109/TCT.1971.1083337.
4. Chua LO, Sirakoulis GC, Adamatzky A (eds). *Handbook of Memristor Networks*. Cham: Springer, 2019.
5. Wang Z, Wu H, Burr GW *et al.* Resistive switching materials for information processing. *Nat Rev Mater* 2020;**5**:173–95. https://doi.org/10.1038/s41578-019-0159-3.
6. Sebastian A, Le Gallo M, Khaddam-Aljameh R *et al.* Memory devices and applications for in-memory computing. *Nat Nanotechnol* 2020;**15**:529–44. https://doi.org/10.1038/s41565-020-0655-z.
7. Lanza M, Sebastian A, Lu WD *et al.* Memristive technologies for data storage, computation, encryption, and radio-frequency communication. *Science* 2022;**376**:eabj9979. https://doi.org/10.1126/science.abj9979.
8. Mayer SF, Mitsioni MF, Robin P *et al.* Lumen charge governs gated ion transport in β-barrel nanopores. *Nat Nanotechnol* 2026;**21**:116–24. https://doi.org/10.1038/s41565-025-02052-6.
9. Laughlin SB, De Ruyter Van Steveninck RR, Anderson JC. The metabolic cost of neural information. *Nat Neurosci* 1998;**1**:36–41. https://doi.org/10.1038/236.
10. Fu S, Gao H, Wang S *et al.* Constructing artificial neurons with functional parameters comprehensively matching biological values. *Nat Commun* 2025;**16**:8599. https://doi.org/10.1038/s41467-025-63640-7.
11. Emmerich T, Ronceray N, Agrawal KV *et al.* Nanofluidics. *Nat Rev Methods Primers* 2024;**4**:69. https://doi.org/10.1038/s43586-024-00344-0.
12. Sheng Q, Xie Y, Li J *et al.* Transporting an ionic-liquid/water mixture in a conical nanochannel: a nanofluidic memristor. *Chem Commun* 2017;**53**:6125–7. https://doi.org/10.1039/C7CC01047H.
13. Bu Y, Ahmed Z, Yobas L. A nanofluidic memristor based on ion concentration polarization. *Analyst* 2019;**144**:7168–72. https://doi.org/10.1039/C9AN01561B.
14. Zhang P, Xia M, Zhuge F *et al.* Nanochannel-Based Transport in an Interfacial Memristor Can Emulate the Analog Weight Modulation of Synapses. *Nano Lett* 2019;**19**:4279–86. https://doi.org/10.1021/acs.nanolett.9b00525.
15. Robin P, Kavokine N, Bocquet L. Modeling of emergent memory and voltage spiking in ionic transport through angstrom-scale slits. *Science* 2021;**373**:687–91. https://doi.org/10.1126/science.abf7923.
16. Robin P, Emmerich T, Ismail A *et al.* Long-term memory and synapse-like dynamics in two-dimensional nanofluidic channels. *Science* 2023;**379**:161–7. https://doi.org/10.1126/science.adc9931.
17. Xiong T, Li C, He X *et al.* Neuromorphic functions with a polyelectrolyte-confined fluidic memristor. *Science* 2023;**379**:156–61. https://doi.org/10.1126/science.adc9150.
18. Emmerich T, Teng Y, Ronceray N *et al.* Nanofluidic logic with mechano–ionic memristive switches. *Nat Electron* 2024;**7**:271–8. https://doi.org/10.1038/s41928-024-01137-9.
19. Kamsma TM, Kim J, Kim K *et al.* Brain-inspired computing with fluidic iontronic nanochannels. *Proc Natl Acad Sci USA* 2024;**121**:e2320242121. https://doi.org/10.1073/pnas.2320242121.
20. Ling Y, Yu L, Guo Z *et al.* Single-Pore Nanofluidic Logic Memristor with Reconfigurable Synaptic Functions and Designable Combinations. *J Am Chem Soc* 2024;**146**:14558–65. https://doi.org/10.1021/jacs.4c01218.
21. Yu SY, Hu J, Li Z *et al.* Metal–Organic Framework Nanofluidic Synapse. *J Am Chem Soc* 2024;**146**:27022–9. https://doi.org/10.1021/jacs.4c08833.
22. Song R, Wang P, Zeng H *et al.* Nanofluidic Memristive Transition and Synaptic Emulation in Atomically Thin Pores. *Nano Lett* 2025;**25**:5646–55. https://doi.org/10.1021/acs.nanolett.4c06297.
23. Ismail A, Nam GH, Lokhandwala A *et al.* Programmable memristors with two-dimensional nanofluidic channels. *Nat Commun* 2025;**16**:7008. https://doi.org/10.1038/s41467-025-61649-6.

24. Cohen LD, Ziv NE. Neuronal and synaptic protein lifetimes. *Curr Opin Neurobiol* 2019;**57**:9–16. https://doi.org/10.1016/j.conb.2018.12.007.
25. Mohar B, Michel G, Wang YZ *et al.* DELTA: a method for brain-wide measurement of synaptic protein turnover reveals localized plasticity during learning. *Nat Neurosci* 2025;**28**:1089–98. https://doi.org/10.1038/s41593-025-01923-4.
26. Wahab OJ, Daviddi E, Xin B *et al.* Proton transport through nanoscale corrugations in two-dimensional crystals. *Nature* 2023;**620**:782–6. https://doi.org/10.1038/s41586-023-06247-6.
27. Wu ZF, Sun PZ, Wahab OJ *et al.* Proton and molecular permeation through the basal plane of monolayer graphene oxide. *Nat Commun* 2023;**14**:7756. https://doi.org/10.1038/s41467-023-43637-w.
28. Ji Y, Hao GP, Tan YT *et al.* High proton conductivity through angstrom-porous titania. *Nat Commun* 2024;**15**:10546. https://doi.org/10.1038/s41467-024-54544-z.
29. Schnurr B, Gittes F, MacKintosh FC. Metastable intermediates in the condensation of semiflexible polymers. *Phys Rev E* 2002;**65**:061904. https://doi.org/10.1103/PhysRevE.65.061904.
30. Xu Z, Buehler MJ. Geometry Controls Conformation of Graphene Sheets: Membranes, Ribbons, and Scrolls. *ACS Nano* 2010;**4**:3869–76. https://doi.org/10.1021/nn100575k.
31. Zhao Y, Liao J, Bu S *et al.* Automated processing and transfer of two-dimensional materials with robotics. *Nat Chem Eng* 2025;**2**:296–308. https://doi.org/10.1038/s44286-025-00227-5.
32. Perram JW, Stiles PJ. On the nature of liquid junction and membrane potentials. *Phys Chem Chem Phys* 2006;**8**:4200. https://doi.org/10.1039/b601668e.
33. Zhai L, Xu Z. Information processing by digitalizing confined ion transport. *Iontronics* 2026;**2**:11. https://doi.org/10.20517/iontronics.2026.11.
34. Li X, Wang ZL, Wei D. Iontronic logic control driven by dynamic electrical double layer regulation. *Iontronics* 2025;**1**:2. https://doi.org/10.20517/iontronics.2025.02.
35. Hall JE. Access resistance of a small circular pore. *J Gen Physiol* 1975;**66**:531–2. https://doi.org/10.1085/jgp.66.4.531.
36. Zhang W, Zhang A, Zhou W *et al.* Revisiting ion transport through micropores: significant and non-negligible surface transport. *Nanoscale Horiz* 2026;**11**:795–802. https://doi.org/10.1039/D5NH00582E.
37. LeCun Y, Cortes C, Burges C. The MNIST database of handwritten digits. 1998. http://yann.lecun.com/exdb/mnist/.
38. Krizhevsky A, Hinton G. *Learning Multiple Layers of Features from Tiny Images*. University of Toronto Technical Report, 2009.
39. Bean BP. The action potential in mammalian central neurons. *Nat Rev Neurosci* 2007;**8**:451–65. https://doi.org/10.1038/nrn2148.
40. Brown EN, Kass RE, Mitra PP. Multiple neural spike train data analysis: state-of-the-art and future challenges. *Nat Neurosci* 2004;**7**:456–61. https://doi.org/10.1038/nn1228.
41. Kruskal PB, Jiang Z, Gao T *et al.* Beyond the Patch Clamp: Nanotechnologies for Intracellular Recording. *Neuron* 2015;**86**:21–4. https://doi.org/10.1016/j.neuron.2015.01.004.
42. Anumanchipalli GK, Chartier J, Chang EF. Speech synthesis from neural decoding of spoken sentences. *Nature* 2019;**568**:493–8. https://doi.org/10.1038/s41586-019-1119-1.
43. Zhu X, Wang Q, Lu WD. Memristor networks for real-time neural activity analysis. *Nat Commun* 2020;**11**:2439. https://doi.org/10.1038/s41467-020-16261-1.

## Supplementary information

### Synthesis of graphene

The graphene films were synthesized using commercially supplied Cu foils (purity >99.96%; thickness, 25 µm; from *Dongguan Lidong Metal Foil Co., Ltd.*) as substrates and in a well-established low-pressure chemical vapor deposition (CVD) system. Prior to growth, we first performed electrochemical polishing for the Cu foil (A3 or A4 size), followed by washing it in deionized water and blow-drying with nitrogen gas. Then we scrolled the foil into a cylinder shape, placed it into the CVD system and heated from room temperature (*T*) to 1000 °C in about 45 min under the protection of a $H_2$ atmosphere (20 sccm, ~100 Pa). We maintained *T* at 1000 °C for 30 min to remove contaminations on the Cu surface. Then $CH_4$ was introduced at a flow rate of 20 sccm to initiate graphene growth. This process lasted for 20 min and the resulting graphene was in high quality, polycrystalline and continuous. Once the growth process was completed, the system underwent rapid cooling from 1000 °C down to room *T* by quickly pulling the quartz tube out of the heating zone and exposing it to the air.

### Device fabrication

The synthesized graphene film was transferred onto a micrometer size aperture perforated in a freestanding silicon nitride ($SiN_x$) membrane. Fabrication of the latter structure was well-documented before[1-3]. In brief, a piece of silicon wafer (~500 µm thick) with its double sides coated by $SiN_x$ was employed for the fabrication. The thickness of the $SiN_x$ layer was either 100 nm, 300 nm or 500 nm, which determined the length *L* of the final aperture. An array of 800 × 800 µm² squares were defined by photolithography on one side, followed by dry etching to completely remove $SiN_x$ in the squares. Then the exposed silicon was etched through by a concentrated KOH solution (30 wt% in water) at 80 $^{o}$C for 8 h, resulting in an array of freestanding $SiN_x$ squares on the other side of the wafer, each about 100 µm in size. To make apertures in the freestanding $SiN_x$ membranes, a second round of photolithography and dry etching was employed for diameters $D > 1$ µm. For $D \leq 1$ µm, focused ion beam (from *ZEISS*) was used instead. Its acceleration voltage was 30 kV and the ion beam current was 0.1 – 3 nA. A sufficiently large dose of 600 mC/cm² was imposed to ensure completely etching through the $SiN_x$ membrane. After the perforation, the resulting aperture was examined under an electron microscope (from *ZEISS*) to determine its real *D*.

To transfer the graphene film from the underlying Cu foil to the $SiN_x$ aperture, we spin-coated a layer of poly(methyl methacrylate) (PMMA, 4 wt% in anisole, from *Microchem*) on top of the graphene-on-Cu structure, followed by baking it at 130 °C for 5 minutes. Because graphene grew on both surfaces of the Cu foil, the uncoated graphene on the bottom side was wiped off. Then the Cu foil was etched away in an ammonium persulfate solution (0.2 M, from *Macklin*) for a couple of hours. The resulting PMMA-coated graphene film was thoroughly washed by deionized water to remove etchant residues and transferred over the $SiN_x$ aperture following procedures standard for van der Waals assembly[4,5]. After the transfer, the PMMA layer was dissolved by acetone and then the whole assembly was rinsed with isopropanol and blow-dried by nitrogen. Because of surface tension and the rapid evaporation of organic solvent, the freestanding graphene membrane spanning over the aperture easily broke and the resulting segments tended to crimple around the rim (Fig. 1a). Finally, the device was annealed at 400 $^{o}$C in a $H_2$/Ar mixture ($H_2$, 10 sccm; Ar, 100 sccm) for 3 h to remove polymer residues.

### Transmission electron microscopy

The rippled edge structure such as in Fig. 1a was further characterized in more detail under a transmission electron microscope (from *Tecnai*). Figure S1a shows one of such examples. Although the formation of this rippled structure was rather random, we could still get some useful information. For example, within the rippled region extending outside the $SiN_x$ pore edge by a few tens of nm, many domains composed of parallel fringes and approximately oriented along the pore periphery were observed (Fig. S1a). These fringes were imaged due to the parallel alignment of graphene basal planes with respect to the incident electron beam and therefore, indicate the positions of highly curved structure, that is, the ripple. Statistical analysis on several of them yields the interlayer spacing of 0.36 nm ± 0.02 nm, which is close to that of graphite (~0.335 nm). This suggests that the rippled structure around the pore rim was formed by tightly stacking the corrugated graphene segments.

**Electrical measurements and additional *I-V* curves**

Prior to measurements using the setup in Fig. 1a, we first filled both reservoirs with isopropanol to properly wet the device's two surfaces. Then the isopropanol was gradually replaced by deionized water. To measure ion transport through the device, deionized water was further replaced by an electrolyte and a pair of Ag/AgCl electrodes were inserted. The tested electrolyte solutions included KCl, NaCl, $MgCl_2$ and $CaCl_2$, with concentrations $C$ of 0.1 mM – 3 M. Normally, the solutions' pH was not modified, but for pH dependence measurements (in KCl), it was adjusted by titrating HCl or KOH solutions of $C = 10^{-2} – 10^{-3}$ M and monitored by a pH meter. The time responses of ion current $I(t)$ to both continuous voltage sweeps $V(t)$ and sequentially applied voltage spikes were measured using a source meter (*Keithley 2636B*). Detailed parameters of the applied $V$ were programmed using the software LabVIEW. The lower current detection limit was ~10 pA, as determined using a control sample without an aperture.

To observe an ionic memory effect, a concentration difference across the rippled graphene pore was required. To this end, one of the reservoirs was filled with a tested electrolyte of lower concentration, $C_{low}$, and the other was of higher concentration, $C_{high}$. The positive electrode was always placed in the $C_{high}$ reservoir (Fig. 1a). Unless otherwise specified, $C_{high}$ = 1 M and $C_{low}$ = 1 mM were used except in the $C$ dependence measurements (Figs. 2f, S9c, d). Figure S1b shows a representative electron micrograph for one of our devices after long-term measurements. Prior to characterizations, the device was thoroughly rinsed with deionized water. Nonetheless, many nanoscale particles that were either wrapped inside the rippled graphene edges or adsorbed on the surface of graphene (Fig. S1b) were still observed. These particles were adsorbed salt precipitates that could be effectively removed by further heating the devices in water at 60 – 80 $^{o}$C for 1 – 2 h. This water treatment process was adopted for cleaning and recycling the used devices.

Taking KCl solutions as an example, we show in Figs. 2a and S2a, b, the full sets of cyclic *I-V* curves for three representative devices. In those measurements, we applied a periodic triangular voltage wave with amplitude $A$ = 0.6 V and sweep rate $r_s$ = 1 mV/s. The continuous measurements typically lasted for a few days. All devices displayed similar *I-V* evolutions, that is, pronounced hysteresis loops appeared after a few cycles, which became relatively stable after 10 – 30 cycles. Statistics over the stable *I-V* curves for the shown devices yields 1) loop area $A_{loop}$ of 0.25 ± 0.05 µW (Fig. 2a), 0.22 ± 0.02 µW (Fig. S2a) and 0.20 ± 0.06 µW (Fig. S2b), respectively; 2) rectification ratio of 7.53 ± 2.07, 8.24 ± 0.88 and 7.49 ± 1.24, respectively; 3) zero-current voltage $V_0$ of -194 ± 18 mV, -219 ± 7 mV and -205 ± 11 mV, respectively. We believe such statistics is important, which allows us to quantitatively assess the impact of structural randomness on the studied ion transport. Based on these three electrical metrics, a maximum device-to-device variation of ~10.4% is obtained, which is calculated as the standard deviation (SD) divided by the mean value of a specific metric among different devices (that is, the coefficient of variation). Also, a variation not exceeding 30% is obtained from repeated *I-V* scans on a single device, which serves as a reference for the dynamic variations of our fabricated devices in continuous electrical measurements.

The observed device-to-device variation is further consistent with our statistics over the ratio between high- and low-resistance states (HRS/LRS) (Fig. S2c). To find out HRS/LRS, we transformed the recorded *I-V* curves into *G-V* curves by calculating the differential conductance $G(V) = dI/dV$. Because our devices displayed analog-type memory effects with continuum changes in resistance states, the extracted ratios of HRS/LRS also varied as a function of applied $V$. We found the observed maximum

HRS/LRS ratios were mostly in the range of 0.03 – 0.06 V and hence, we used the measured HRS/LRS at 0.045 V for statistics (Fig. S2c). Results show that the extracted HRS/LRS ratios from the stable *I-V* curves at different frequencies *f* fall into a decreasing trend with increasing *f*, which is gradually convergent to unity. The decreasing trend is in agreement with that observed for the *f* evolution of $A_{loop}$ and both of them lead to the conclusion that the memory effect is weakened with increasing *f*. Specially, the inset in Fig. S2c shows that the extracted HRS/LRS ratios at $f$ = 0.4 mHz are mostly located in the range of 2 – 4 and statistics for 43 of our tested devices yields an average HRS/LRS ratio of 2.9 with an SD of 0.3. This translates into a variation coefficient of ~10.3%, in agreement with that observed in the cyclic *I-V* measurements for multiple devices (~10.4%).

The observed rectification in all the shown *I-V* curves could, in principle, be traced to the asymmetric pore structure and surface property, or the concentration difference across the pore[6,7]. To find out which of the two was the origin, we flipped the device and kept all the other conditions identical. Figure S2d shows that no noticeable differences could be found in the recorded *I-V* curves, indicating that the rectification was mainly caused by the concentration difference. In the absence of a concentration difference, linear *I-V* responses were detected (Fig. 1b).

To show the importance of rippled graphene edges for the observed memory effect, we fabricated control samples with no such structure. To this end, the standard samples were placed upside down and exposed to oxygen plasma for 2 hours so that the graphene segments at the pore rim were completely etched away using the $SiN_x$ aperture as a mask (Fig. S1c). The control samples yielded sublinear *I-V* curves without hysteresis (Fig. 1b). Similar results were also observed for bare $SiN_x$ pores (without graphene). Qualitatively, the conductance of our rippled graphene pore is lower than that of a bare $SiN_x$ pore having the same $D \approx 2$ µm, $L \approx 500$ nm and under the same $C_{high} = 1$ M, $C_{low} = 1$ mM (Fig. 1b). To quantify this difference, we define a characteristic conductance as $G^* = I(V_{max})/(V_{max} - V_0)$, where $I(V_{max})$ is the current at the maximum voltage $V_{max}$ and $V_0$ is the voltage at zero $I$. Figure S9a compares $G^*$ for bare $SiN_x$ pores and rippled graphene pores at different $D$ and shows that both the $D$ dependences follow approximately a linear trend. The deviation of $G^*(D)$ for rippled graphene pores with respect to the best linear fit is estimated to be ~30% (Fig. S9a), which is notably larger than that of bare $SiN_x$ pores. Furthermore, the extracted slope from the best linear fit of $G^*(D)$ for rippled graphene pores is ~60% that for bare $SiN_x$ pores. The observed lower $G^*$ in the former case indicates a slower ion dynamics, presumably due to ions being trapped in the rippled graphene edges (Fig. S1b).

The described *I-V* measurements for pores having different $D$ and $L$, and in solutions of different $C_{high}$ and $C_{low}$ typically involved evolutions of pore conductance that contributed to the absolute values of $A_{loop}$. To subtract the influence of conductance changes, we followed the previous studies[8,9] and normalized the absolute value of $A_{loop}$ by $G^*$. The normalized data are plotted in Figs. 2 and S9, which allow more insights into the underlying mechanism of observed memory effect, as detailed in the main text.

**Ion selectivity**

In our setup (Fig. 1a), the extracted $V_m$ (inset of Fig. 1d) and calculated $t_+/t_-$ indicate cation and anion selectivities, respectively, for chloride solutions containing monovalent (NaCl, KCl) and divalent cations ($MgCl_2$, $CaCl_2$). The cation selectivity observed for NaCl and KCl is consistent with previous measurements[10] for graphene pores notably larger than the Debye length $\lambda_D$. It was attributed to negatively charged groups on graphene due to its structural defects or surface adsorbed hydrocarbon contaminations[11] that are unavoidable in air, thereby leading to stronger interaction of $Na^+$ and $K^+$ than that of $Cl^-$, and a locally higher cation concentration around the pore. The observed cation selectivity is also consistent with the reported stronger interaction of $K^+$ and $Na^+$ with the six-membered carbon rings in graphene (the well-known cation-π interaction)[12,13] than that of $Cl^-$. Following the same line of thought, the anion selectivity observed for $MgCl_2$ and $CaCl_2$ could be attributed to much stronger interactions of these divalent cations with graphene than those of monovalent ones, resulting in immobile and positive surface charges (so-called charge inversion)[14,15]. However, we also cannot rule out the possible origin of notably higher hydration energies[16] of $Mg^{2+}$ and $Ca^{2+}$ than that of $Cl^-$. This could result in higher barriers for these divalent cations accessing the strongly confined and rippled graphene edges (when dehydration is involved), or alternatively, stronger screenings by their hydration shells and hence, reduced interactions with graphene (when dehydration is insignificant).

The possible origins for the observed ion selectivity are mostly discussed based on the available literature. Nonetheless, our experiments could at least distinguish whether the structural defects of graphene were important or not. A small number of defects are practically unavoidable in the graphene samples prepared by CVD. They could potentially transform into surface charges in an aqueous environment. On the other hand, the samples made by mechanical exfoliation are known to have fewer defects than those by CVD. Having these in mind, we instead fabricated devices using mechanically exfoliated graphene. Measurements under the same conditions showed that the two samples exhibited no noticeable differences in their *I-V* curves after reaching stabilization (Fig. S2e), thereby ruling out the importance of structural defects for the observed memory effect, and, also, the ion selectivity.

The selectivity between cations and anions, $t_+/t_-$, could be altered using different electrolytes, as evidenced in Figs. 1b and d. Alternatively, this could also be achieved by grafting charged molecular species on the surface of graphene. To this end, we chose a polyelectrolyte molecule, Poly(diallyldimethylammonium chloride) or PDDA, to post-functionalize the fabricated devices. They were immersed in a PDDA aqueous solution (20 g/L, pH ≈ 9) for 1 hour and then rinsed with abundant deionized water to remove excessive PDDA molecules and blow-dried with nitrogen gas. Figure S2f shows a representative *I-V* curve obtained using such a functionalized sample in KCl solutions. In stark contrast to those made from pristine graphene, the functionalized sample displayed inversely rectified and directional hysteresis loops, which correlate well with the observed anion and cation selectivities, as evidenced by $V_m > 0$ and $< 0$ for functionalized and pristine graphene samples, respectively, as shown in the inset of Fig. S2f.

**PH dependence**

The described comparison between measurement results for polyelectrolyte-modified and pristine graphene samples (Fig. S2f) indicates that surface transport is important. This in turn suggests adjusting the solutions' pH could yield a wider range of $t_+/t_-$ because excessive protons or hydroxide ions would adsorb on graphene and tune the relative concentrations of surface adsorbed cations and anions. Taking KCl solutions as an example, Fig. S4a exemplifies stable *I-V* curves recorded at different pH, from which $t_K/t_{Cl}$ and $A_{loop}$ could be extracted (Figs. S4b, c). A cation selectivity of $t_K/t_{Cl} \approx 2$ is seen at neural pH, whereas $t_K/t_{Cl}$ increases monotonically from 1.2 to 4.3 with increasing pH from 4 to 10 (Fig. S4b). The smaller $t_K/t_{Cl}$ observed at acidic pH can, in principle, be traced to either a reduced $t_K$, due to weakened surface attraction and/or enhanced repulsion of $K^+$ ions in the presence of excessive protons, or an increased $t_{Cl}$, due to stronger attraction of $Cl^-$ ions. Similarly, the larger $t_K/t_{Cl}$ at basic pH can be explained by stronger surface attraction of $K^+$ ions and/or repulsion of $Cl^-$ ions by the present hydroxide ions. In contrast, the extracted $A_{loop}$ follows a different but non-monotonic trend as a function of pH (Fig. S4c). A minimum $A_{loop}$ is recorded at pH = 7 – 8, and referring to this minimum, $A_{loop}$ increases by a few folds under both acidic and basic conditions (Fig. S4c). To further subtract the effect of conductance changes caused by adjusting pH, we normalized the $A_{loop}$(pH) data by the characteristic conductance $G^*$ extracted from corresponding *I-V* curves (Fig. S4a). A similar trend is also found for the normalized $A_{loop}$(pH) (Fig. S4d). The described rather different trends of (normalized) $A_{loop}$ and $t_K/t_{Cl}$ suggest that changes in ion selectivity at different pH are not the primary cause for the observed changes in memory effect. Trying to understand this pH effect, we note that the memory is rooted at slow ion dynamics in the rippled graphene edges, that is, slower desorption of the transported ions than their adsorption (details see below section "Atomistic simulations"). Accordingly, the observed $A_{\text{loop}}$(pH) suggests that the ion dynamics in the presence of excessive protons or hydroxide ions is slower than that under neutral conditions. We attribute the slower dynamics to a stronger interaction between the transported ions ($K^+$ or $Cl^-$) and the graphene surface, that is, stronger attraction of $Cl^-$ and $K^+$, respectively, at acidic and basic pH conditions. This is consistent with the observed inverse trend of $G^*$(pH), showing that $G^*$ at pH = 7 – 8 is higher than that at both acidic and basic pH (Fig. S4a). The smaller $G^*$ in the latter two regimes serve as a direct proof for the presence of fewer freely moving ions that contribute to conduction and hence, the stronger ion-graphene interactions.

**Equivalent circuit model**

We developed an equivalent circuit model (ECM) to fit the experimental *I-V* curves (Fig. S3). The ECM was constructed by a memristor in series with a direct-current (DC) power source $V_{\text{DC}}$ (inset of Fig. S3a) and powered by a triangular voltage wave with the same amplitude $A$ = 0.6 V and frequency $f$ = 0.4 mHz – 200 mHz as those used in experiments. The fitting parameters are expected to provide more information about the ion transport dynamics and the memory effect. ECMs have been well-documented for analyzing devices such as thermistors[17] and nanofluidic memristors[18]. Transient simulation of ECM was performed using Simulation Program with Integrated Circuit Emphasis (SPICE)[19].

To describe how the memristor's conductance $G$ depends on the history of voltage $V(t)$, we followed the previous work[8] and formulated $G$ as

$$G(t) = \int_0^\infty G_{\text{qs}}[V(t-s)]\,\frac{e^{-s/\tau_{\text{m}}}}{\tau_{\text{m}}}\,\mathrm{d}s \tag{S1}$$

For simplicity, the quasi-static conductance $G_{\text{qs}}(V)$ is approximated as a linear function of $V$, which is bounded by ON and OFF conductance states, $G_{\text{on}}$ and $G_{\text{off}}$, respectively. That is,

$$G_{\text{qs}}(V) = kV + b, G_{\text{off}} \le G_{\text{qs}}(V) \le G_{\text{on}} \tag{S2}$$

Time differential of eq. S1 yields

$$\frac{\mathrm{d}G}{\mathrm{d}t} = g(G,V) = \frac{G_{\text{qs}}(V) - G}{\tau_{\text{m}}} \tag{S3}$$

Equation S3 conforms to the definition of a generic memristor[20], which is defined by $I = G(x)V$, $\frac{\mathrm{d}x}{\mathrm{d}t} = g(x,V)$, with $x$ (= $G$) being a state variable.

Taking KCl and $MgCl_2$ solutions of $C_{\text{high}}$ = 1 M and $C_{\text{low}}$ = 1 mM as an example (Figs. 1b, d), we show in Fig. S3 that the experimental *I-V* curves acquired at $f \approx 0.4$ mHz (corresponding to $r_{\text{s}}$ = 1 mV/s) can be well fitted using the described ECM (inset of Fig. S3a) and eqs. S1-S3. The fitting effectively captures key features such as rectification, hysteresis loops and their directions. The fitting parameters are summarized in Table S1 and comparisons for KCl and $MgCl_2$ can provide more information about the electrolyte-specific characteristics. For example, the slope $k$ in eq. S2 correlates with the rectification characteristics of the *I-V* curves and therefore, the ion selectivity, as detailed in below section "Concentration effects". A positive $k$ translates into a higher $G_{\text{qs}}$ at $V > 0$ than that at $V < 0$, which as per our setup (Fig. 1a), indicates the preferential transport of cations over anions. Similarly, a negative $k$ supports anions as the preferred charge carriers. The fitting yields $k > 0$ for KCl and $k < 0$ for $MgCl_2$, which are fully consistent with their cation and anion selectivities (inset of Fig. 1d), respectively. On the other hand, the memory timescale $\tau_{\text{m}}$ fitted for KCl is longer than that for $MgCl_2$. The memory timescale $\tau_{\text{m}}$ correlates with the difference in the free energy barriers $G$ between the escaping and trapping of transported ions in the rippled edge structure, as detailed in below section "Atomistic simulations", and follows $\tau_{\text{m}} \sim \exp(G/k_{\text{B}}T)$, whereas $G$ characterizes the overall interaction strength for the tested electrolyte with the rippled graphene. Accordingly, the fitting result suggests that the overall interaction in the KCl – graphene system, where the transport of $K^+$ is preferred over $Cl^-$, is stronger than that in the $MgCl_2$ – graphene system, with $Cl^-$ the preferred charge carrier.

Similar fittings but to the *I-V* curves acquired at different $f$ can also reproduce the frequency dependence observed in experiments (Fig. 1c). In addition to changes in $A_{\mathrm{loop}}$, the fittings allow us to extract the evolution of $\tau_{\mathrm{m}}$ as a function of $f$ (inset of Fig. 1c), showing that $\tau_{\mathrm{m}}$ decreases by over 3000 folds and from ~200 s to ~65 ms with increasing $f$ from 0.4 mHz to 200 mHz.

**Spatial distribution of ion current**

The observed memory effect suggests an underlying fast-slow ('stop-and-go') kinetics along the ion transport pathway[21], which is unexpected from a bulk flow through a µm-size pore. To provide theoretical insights into the ion transport pathway, we first analyzed the spatial distribution of ion current $I$ in our experimental setup. To this end, we employed finite element analysis (FEA) to solve the Poisson-Nernst-Planck (PNP) equations using the FEniCS software[22,23].

Taking KCl solutions as an example, the PNP equations

$$\nabla \cdot (-\epsilon \nabla V) = F(c^+ - c^-) \tag{S4}$$

$$\nabla \cdot \left(-\nabla c^{\pm} \mp \frac{e}{k_{\mathrm{B}}T} c^{\pm} \nabla V\right) = 0 \tag{S5}$$

were solved using the first-order Lagrange element and non-uniform meshes, where $\epsilon$ is the dielectric constant of water, $F$, $e$, and $k_{\mathrm{B}}$ are the Faraday constant, elementary charge, and Boltzmann constant, respectively, $V$ is the electric potential and $c^+$, $c^-$ are concentrations of $K^+$ and $Cl^-$, respectively. The size of the meshes was set as 0.1 nm, which is smaller than the Debye length $\lambda_{\mathrm{D}}$ of ~0.3 nm corresponding to $C_{\mathrm{high}}$ = 1 M. Figure S5a (left) illustrates the solution domain, which considers a cylindrical region with radius AB (GH) = 5 µm and height BC (FG) = 5 µm on each side of the device. AH is the symmetric boundary. AB and GH are boundaries for the bulk KCl solutions and an electrical potential difference $\Delta V$ was applied between them. Specifically, Dirichlet boundary conditions were given on AB and GH. The electric potentials applied to AB and GH were set to $V$ = +0.3 V and -0.3 V, respectively, yielding an overall cross-pore bias of 0.6 V consistent with that used for *I-V* measurements such as those in Figs. 1 and S2. Also, same as $C_{\mathrm{high}}$ and $C_{\mathrm{low}}$ used in those measurements, the ion concentrations in reservoirs ABCD and EFGH were set as $c^+ = c^-$ = 1 M and $c^+ = c^-$ = 1 mM, respectively. On the device surface CDEF, we set the charge density and normal ion flux to zero so that

$$\nabla V \cdot \boldsymbol{n} = 0,$$

$$\left(-\nabla c^{\pm} \mp \frac{e}{k_{\mathrm{B}}T} c^{\pm} \nabla V\right) \cdot \boldsymbol{n} = 0$$

Here we neglected the surface charge effect because $\lambda_{\mathrm{D}}$ < 10 nm is much smaller than the pore radius $r$ (DM). To find out the influence of pore size on the spatial distribution of $I$, we varied $r$ = 0.1 – 4 µm and $L$ = 2 – 500 nm, respectively.

Figure S5a (right) exemplifies the spatial distribution of $I$ across the pore with $r$ = 1 µm. Unexpectedly, most of the streamlines are concentrated in a finite thickness around the pore rim and the effect becomes more pronounced in smaller pores. This shows that the ion flux across the pore edges dominates over that through the pore interior even in µm-size pores.

To further quantify the relative contribution of pore edge and its interior, we note that the total current through the pore DM is sum of that directly traversing the cylindrical region ANDM and crossing its edge ND (Fig. S5a, left). Accordingly, we define the latter contribution as edge current $I_{\mathrm{edge}}$ and compare it with the total current $I_{\mathrm{total}}$. Figure S5b shows one of our examples for such comparison at different $r$ but fixed $L$. We find that $I_{\mathrm{edge}}$ is quite close to $I_{\mathrm{total}}$ for $r \leq 1$ µm and the maximum deviation is only about 5% at the largest $r$ = 1 µm. Same as predictions from the theory[24], both currents increase linearly with $r$. Similar dependences also hold for other $L$ up to 500 nm and the maximum deviation between the edge and total currents never exceeds a few percent. Our measurements and below theoretical analysis ("Atomistic simulations") suggest that the memory effect originates from the slow dynamics, namely, the slower desorption of transported ions than their adsorption within the rippled graphene edges. Therefore, as a characteristic measure for the memory effect, the normalized hysteresis area $A_{loop}$ should increase with the ratio $I_{edge}/I_{total}$. In the tested diameter $D$ range of 0.5 – 8 µm, we find the extracted normalized $A_{loop}$ is independent of $D$ within the experiment scatter (Fig. 2e), suggesting that $I_{edge}/I_{total}$ exhibits negligible changes.

On the other hand, we find whether or not $I_{edge}$ dominates over $I_{total}$ depends on the ratio between the size of the pore ($r$) and that of the bulk region ($R$). This is evident from comparisons between $I_{edge}$ and $I_{total}$ at a wider range of $r$ (1 – 4 µm) in Figs. S5c, d. For a fixed half bulk size $R$ = 5 µm, Fig. S5c shows that the ion flow through the pore interior in turn dominates over that across the pore edge only when the sizes of the reservoir and the pore are comparable. To be more quantitative, Fig. S5d shows that $I_{edge}/I_{total}$ decreases with increasing $r/R$ and specially, $I_{edge}/I_{total}$ = 0.5 at $r/R \approx 0.7$. Accordingly, a weakened memory effect is expected with increasing $r/R$ and for comparable sizes of $r$ and $R$, the dominant bulk ion transport is expected to mask the memory effect. Nonetheless, considering the fact that the pore sizes of all our measured devices were far smaller than the reservoirs, we therefore conclude that the current across the pore edges should always dominate over that through the pore interior in our experimental setup. The above calculations were performed at zero surface charge density ($\sigma$). Nonetheless, our FEA by considering realistic $\sigma$ of up to 200 mC/m$^2$ shows negligible impact to $I_{edge}$ and $I_{total}$.

**Atomistic simulations**

Having the described FEA results in hand, we then carried out a molecular dynamics (MD) simulation using the Large-scale Atomic/Molecular Massively Parallel Simulator (LAMMPS)[25] to show the distribution of ion current across a pore's open space under an external electric field. The inset of Fig. S6a illustrates the simulation system. In this specific example, a graphene sheet having a pore of $D$ = 4 nm in the center separated a periodic box (10 nm × 10 nm × 20 nm) into halves. The box was filled with KCl solutions (for example, at $C$ = 0.4 M). Water molecules were described by the extended simple point charge (SPC/E) force fields[26]. To constrain the shape of water molecules, we applied the SHAKE algorithm[27] with a precision of 0.0001. Graphene and ions were modeled by the All-Atom Optimized Potential for Liquid Simulations (OPLS-AA) force fields[28]. The interactions between ions and graphene were described by the scaled Lennard-Jones potentials, which were amplified by a factor of 10 to include the cation-π effect[12,13,29]. To compute the long-range Coulomb interactions, we used the PPPM (Particle-Particle Particle-Mesh) method[30] with a precision of 0.0001. The simulation time step was 1 fs. The system was first allowed to equilibrate for 0.5 ns in the isothermal–isobaric (NPT) ensemble. Then an electric field of 0.5 V/nm was applied to the system and in total, the simulation was run for 3 ns in the canonical (NVT, namely, fixed number of particles $N$, constant volume $V$ and temperature $T$) ensemble. The trajectories of ions were recorded and the resulting ion current in every small area d$A$ equivalent to 1/20 the open pore area was translated into current density $j$, so that the latter's spatial distribution in the pore could be mapped. Figure S6a shows that the current density in the vicinity of the pore edge $j_{edge}$ is notably higher than that in the pore interior, in qualitative agreement with our FEA results (Fig. S5). More quantitative analysis reveals that about 75% of the total current passes through an annular region about 0.6 nm thick extending from the pore edge. Working in this edge region, the MD simulation allows us to perform statistics over the contributed ion trajectories. Results show that over 90% of this current is contributed by a characteristic pathway where ions first adsorb on the graphene sheet and then diffuse to the pore edge. Figure S6b exemplifies a snapshot for mapping the distribution of adsorption sites. It shows that the ions adsorb over the entire surface of graphene. In particular, the adsorbed ions are relatively concentrated near the pore edge and the density becomes slightly decayed with increasing the distance from the pore (Fig. S6c). This spatial distribution suggests a possible current origin of ions first being absorbed far away and then diffusing along the surface to the pore. The found ion pathway also suggests that the observed effect crucially involves surface transport, which in turn agrees with measurements for the polyelectrolyte-functionalized devices in Fig. S2f and for pH dependences in Fig. S4.

On the other hand, both the transmission electron and scanning electron micrographs (Figs. 1a and S1a) show that the rippled graphene edges extend by about 30 – 50 nm from the $SiN_x$ pore rim, whereas further quantitative analysis of our FEA results indicates that over half of the ion current is concentrated within a thickness of about 100 nm around the rim of a similar µm-size pore. These comparable length scales suggest that the transported ions could effectively contact and interact with the highly corrugated and folded graphene edges. The latter structure was formed by graphene segments that cleaved from the pore's top and crimpled toward the pore rim (Figs. 1a, S1a). Assuming a tubular space formed between stacked graphene ripples, its characteristic size $d^*$ could be roughly estimated as a few nm (Fig. S1a), in agreement with the experimentally observed pore sizes (nm-scale) formed by corrugated graphene sheets[31] and theoretical predictions[32]. In this context, to assess the contribution from surface transport, one should compare $d^*$ with the Debye length[33-35] $\lambda_D$, rather than the pore radius $r$, or use $d^*$ to estimate the Dukhin number $Du$ and compare it with unity.The latter measures the relative importance of surface to bulk transport. To this end, typical concentrations $C$ of 1 mM – 1 M for a tested solution (e.g., KCl) translate into $\lambda_D$ of 9.6 nm – 0.3 nm. For typical surface charge densities $\sigma$ of 10 – 100 mC/m$^2$ and $C$ of 1 mM – 1 M, the Dukhin number ($Du \sim |\sigma|/eCd^*$, where $e$ is the elementary charge) is estimated as 0.1 – 1,000. Hence, the obtained ranges of $\lambda_D$ and $Du$ cover $d^*$ and unity, respectively, emphasizing the importance of surface effect.

To further elucidate the ion transport pathway by taking the rippled edge structure into account, we carried out additional molecular dynamics (MD) simulations. A simulation domain of 30 nm × 10 nm × 20 nm was established with periodic boundary conditions (PBCs) enforced in all directions. Practically, the real rippled graphene structure formed at the pore rim is unrealistic to model in MD simulations. However, because we are only interested in the ion dynamics confined in the rippled structure, we instead created a simplified model (Figs. S6d, e) aiming to capture the key physics. We started from a monolayer graphene sheet with length of 30 nm and width of about 10 nm (step 1 in Fig. S6d). It was cut off along the width direction, resulting in two pieces of graphene segments (step 2). Then they were scrolled and folded away from each other and along the length direction, yielding two graphene nanoripples separated by a 6 nm-wide trench (step 3). The curvature of the resulting ripples was about 1 nm$^{-1}$ and hence, the effective inner space for accommodating ions was about 2 nm in size. We further cut the top edges of the ripples into a wedge shape (Fig. S6e). This structure facilitated the access of ions into the ripples. The transiently trapped ions inside the ripples would eventually escape into the trench. A Nosé-Hoover thermostat was used to maintain $T$ = 300 K. To extract the ion trajectory and associated energy profile, we performed a 1-ns run with time step of 1 fs under an applied electric field of 1 V/nm. Based on 10 such trajectories, we calculated the mean residence time $\tau$ for ion migration on the graphene surface and through the ripples, respectively, following the previous procedures[36].

Our simulations capture the characteristic ion transport pathway including the following steps: adsorption and diffusion on graphene's surface, trapping by the ripple, residing inside the ripple and finally, escaping into the trench (number labelled in Fig. S6e). In particular, under an electric field of 1 V/nm, the ion residence time $\tau$ on graphene and inside the ripple is ~70 ps and ~140 ps, respectively. The prolonged residence inside the ripple suggests that a difference in the free energy barrier $G$ is involved in the escaping and trapping steps and the characteristic timescale follows $\tau \sim \exp(G/k_BT)$. The simulations allow us to construct a qualitative free energy profile along the transport pathway (Fig. S6f), which effectively captures the key features of ion dynamics such as fast trapping and slow escaping. This ion dynamics underlies the observed memory effect. Following such a pathway, the memory timescale can be estimated as $\tau_m \approx \tau_s \tau_e/\tau_t$, with $\tau_s$, $\tau_t$ and $\tau_e$ the characteristic timescales for surface adsorption and diffusion, trapping by the ripple, and residing inside the ripple until escaping, respectively. Specially, $\tau_m$ can be obtained from fittings to the experimental *I-V* curves ("Equivalent circuit model" in Methods) and the described relation $\tau_m \approx \tau_s \tau_e/\tau_t$ in turn allows us to gain an estimate over the orders of magnitude for $\tau_e/\tau_t$ and *G* (see below).

The diffusion timescale can be expressed as $\tau_s = L_d^2/(2D_{ion})$, where $L_d$ is the path length and $D_{ion}$ is the diffusion coefficient of ions on graphene. Because the rippled graphene edges were formed by the broken freestanding graphene membrane over the µm-size pore that contracted and folded at the pore rim, $L_d$ is estimated to be on a µm scale. For the tested ions (e.g., $K^+$ or $Cl^-$), $D_{ion}$ is of the order of $10^{-9}$ m$^2$/s and can be smaller if an electric field[37] and/or a strong confinement[38] is imposed, as in our experiments. Accordingly, $\tau_s$ can be estimated to be on the ms-scale. Trying to estimate $\tau_e/\tau_t$, we note that the transported ions driven by voltage sweeps of high *f* do not have time to be trapped into the rippled graphene edges, which could result in a severely shortened $L_d$ and therefore, a practically underestimated $\tau_s$ than the ms scale. To get a better estimate of $\tau_e/\tau_t$ and the associated *G*, we use $\tau_m \approx 200$ s corresponding to the lowest $f \approx 0.4$ mHz, which translates into $\tau_e/\tau_t$ of the order of $10^5$, and hence, *G* of the order of $10k_BT$. Alternatively, a large equivalent diffusion length $L_d^*$ of a few hundreds of µm can be obtained according to $\tau_m \approx \tau_s = L_d^{*2}/(2D_{ion})$, which is over 2 orders of magnitude longer than $L_d$. These estimates indicate that the transported ions reside in the rippled graphene structure for a prolonged duration, and presumably, could cross an individual ripple multiple times during their random motion.

**Kinetic Monte Carlo simulations based on the hopping diffusion model**

The above conclusion suggests that the studied ion transport follows a hopping diffusion process, which can be analyzed using a kinetic Monte Carlo (kMC) simulation. The kMC simulation further allows us to have a better understanding on the observed wide range of $\tau_m$ in our experiments (inset of Fig. 1c and Fig. 3). In brief, the entire rippled edge structure was modelled as a one-dimensional (1D) transport pathway terminated by a source (S) and a drain (D) that were connected by a series of free sites(F) and trapping (T) sites (Fig. S7a). The latter two sites appeared in pairs and represented the adsorption positions outside and inside an individual ripple, respectively. Accordingly, the entire ion transport process across the rippled edge structure was simulated as a hopping process starting from the source and ending at the drain. In between the source and drain, the transported ion hopped over a series of free sites, and at every free site, it had a chance to get into the paired trapping site. The transport rate along such a pathway is thus determined by the energy barriers *G* associated with hopping between different sites. Specifically, we define the energy barrier for diffusion between neighboring free sites as $G_s$ and the applied electric field *E* can either raise or lower it to $G_s \pm eE\Delta x$ ($\Delta x$ the distance between adjacent free sites along *E*) depending on whether the ion diffuses along or against the applied *E*. Similarly, the energy barriers for ion hopping from a free site to the paired trapping site and its escaping backward are defined as $G_t$ and $G_e$, respectively. According to our MD simulations (Fig. S6), we have $G_e > G_t$, and, also, it is reasonable to assume that the diffusion and trapping steps share a similar energy barrier, $G_t \approx G_s$. At every specific site (either free or trapping), the ion can transport to its adjacent site at a rate $k = v_0 \exp(-G/k_BT)$ with $v_0$ the attempt rate.

To calculate the total time for ion transport, we employed the well-documented Gillespie algorithm[39] to implement procedures detailed in Algorithm S1. We chose the following parameters that considered both the authenticity for physics and convenience for simulation: $v_0 = 10^{12}$ Hz; $T = 300$ K; $\Delta x = 1$ nm; number of free or trapping sites, 100; number of transported ions, $10^5$; $G_t = G_s = k_BT$; $E = 0.05k_BT/(e\Delta x) \approx 10^6$ V/m; $G_e$ was taken from a uniform distribution $U$ ($k_BT$, $20k_BT$). The upper limit of *U* is determined according to our estimations described in section "Atomistic simulations", whereas the lower limit corresponds to the situation where no confinement is imposed by the graphene ripples, leading to $G_e = G_t = G_s$. This distribution fully takes into account the randomness in the formation of rippled edge structure.

Working with the described algorithm and parameters, our simulations for the transport of all ions show that the characteristic time falls into a distribution that spreads over one order of magnitude and from about $6 \times 10^{-3}$ s to $6 \times 10^{-2}$ s (Fig. S7b). This suggests that the studied ion transport processes could exhibit multiple timescales. The observed wide time distribution is a direct consequence of random diffusion along the transport pathway connecting the source and the drain. In particular, it can be attributed to different combinations of free and trapping sites that are experienced by the transported ions during their bidirectional hopping (resulting in different diffusion lengths), variations in the residential time at each site and in $G$ at different sites (resulting in different hopping rates). The result was obtained by simulating the rippled edge structure as a 1D transport pathway. In practice, the structure extends along the entire pore periphery, which is expected to allow for more assessable transport pathways, and therefore, a wider time distribution. Further statistics over all the ion transport processes shows that an ion repeatedly gets into every trapping site 9 times on average during its bidirectional random diffusion from the source to the drain. Furthermore, we intentionally cut off the backward pathways and only allowed the ions to transport unidirectionally. Comparisons between the results for unidirectional and bidirectional diffusions show that the latter takes notably longer time (Fig. S7b). These results echo our analysis detailed in section "Atomistic simulations", further proving that the ions can cross an individual ripple multiple times and bidirectionally during their random diffusion within the entire edge structure.

The above simulations adopted a fixed number of trapping sites. In practice, the number of ripples along the ion transport pathway $n$ should affect the ion transportation time and hence, the memory timescale $\tau$. Because both the electrical field-driven ion drift and diffusion contribute to the observed memory effect whereas the time for the former process is proportional to $n$ and that for the latter scales as a function of $n^2$, the expected $\tau$ should increase with $n$ at a rate between proportional and quadratic. This is further supported by our kMC simulations shown in Fig. S7c. The stronger memory effect at a higher $n$ along the transport pathway is also supported by our measurements for devices having different lengths $L$ (lower inset of Fig. 2e and Fig. S9b). The observed linear-in-$L$ dependences for both the absolute and normalized values of $A_{\text{loop}}$ corroborate that the memory effect is enhanced with increasing the length of rippled structure decorating the pore rim.

In addition to ripples, abundant defects on graphene surface could also serve as trapping sites and potentially enhance the memory effect, but the enhancement should depend on the number of defects $n$ along the ion transport pathway and the associated free energy barrier height $G$. The above simulations (Figs. S6, S7) suggest that the memory effect is closely related to the overall residence time $t_{\text{tot}}$ of transported ions at all traps along the pathway. Specifically, the residence time at a given type of trap is determined by its $n$ and $G$ and can be expressed as $t \propto n \exp(G/k_BT)$. In the presence of surface defects acting as additional adsorption sites, the overall residence time is described as $t_{\text{tot}} = t_1 + t_2 \propto n_1 \exp(G_1/k_BT) + n_2 \exp(G_2/k_BT)$, where subscripts "1" and "2" denote ripples and surface defects, respectively. Assuming that ripples and surface defects are not coupled, the residence time is expected to roughly increase by a ratio of $n_2/n_1 \exp(\Delta G/k_BT)$ with $\Delta G = G_2 - G_1$. Accordingly, the enhancement of memory effect by additional surface defects should depend on their number ratio with respect to ripples, $n_2/n_1$, and the difference in free energy barriers, $\Delta G$, which is expected to be pronounced for sufficiently large defect densities and high barriers.

**Concentration effects**

We know from Fig. S2d that the rectification observed in the *I-V* curves originated from the concentration difference across the pore. Trying to understand this effect, we performed the following qualitative analysis under quasi-static conditions. Because the ion current $I$ is mostly confined near the pore edge (Fig. S5), this edge region is simplified as a 1D transport channel. Its length is $L$ and the coordinate within the channel is assigned as $x$ ($0 \le x \le L$). Accordingly, the ion current through the pore is expressed as

$$I = E(x)c^+(x)P^+ + E(x)c^-(x)P^- \quad \text{(S6)}$$

where $P^+$ and $P^-$ are permeabilities of cations and anions, respectively, and $E$ is the electric field. In eq. S6, $c^+$ and $c^-$ can differ because of ion selectivity as observed in the inset of Fig. 1d.

On the other hand, the potential difference $V$ across the pore can be described as

$$V = \int_0^L E(x)\mathrm{d}x = I\int_0^L \frac{1}{c^+(x)P^+ + c^-(x)P^-}\mathrm{d}x \quad \text{(S7)}$$

Reorganizing eq. S7 yields the quasi-static conductance $G_{qs}$ as

$$G_{\mathrm{qs}} = \frac{I}{V} = \left(\int_0^L \frac{1}{c^+(x)P^+ + c^-(x)P^-}\mathrm{d}x\right)^{-1} \quad \text{(S8)}$$

Assuming $P^+ >> P^-$, namely, a strong cation selectivity, a positive $V$ applied to the $C_{high}$ side would drive cations from $C_{high}$ to $C_{low}$, accumulate more cations in the pore (especially in the rippled graphene edges) and according to eq. S8, increase $G_{qs}$. Similarly, a reverse $V$ would deplete cations and decrease $G_{qs}$. The dynamic accumulation and depletion of cations would lead to a rectified *I-V* curve exhibiting systematically higher $I$ or $G$ at $V > 0$ than that at $V < 0$. On the other hand, for $P^+ << P^-$, that is, a strong anion selectivity, eq. S8 predicts that the rectification direction would reverse. The analysis is fully consistent with the experimental *I-V* curves for KCl/NaCl (Fig. 1b) and $MgCl_2/CaCl_2$ (Fig. 1d), where cations and anions are preferentially transported, respectively. In the absence of a concentration difference, no such ion dynamics would occur and the concentration in the pore would remain essentially unchanged regardless of the polarity of applied $V$, resulting in a constant $G_{qs}$ according to eq. S8 and a linear *I-V* response as evidenced in Fig. 1b. The described processes are schematically illustrated in Fig. S8.

**Influence of spike parameters on the current modulation**

To find out how the spike parameters including amplitude $A$, duration $t_d$ and frequency $f$ influenced the observed conductance or current modulation, we performed the following measurements (Fig. S10) using devices of $D \approx 2$ µm, $L \approx 500$ nm, and KCl solutions of $C_{high} = 1$ M, $C_{low} = 1$ mM. First, to measure the $t_d$ dependences, we used a series of positive and negative spikes with fixed $A = 0.6$ V and time interval $\Delta t = 1$ s but varying $t_d$ from 1 s to 9 s (inset of Fig. S10b). Next, to measure the $f$ dependences, we fixed the applied spikes at $A = 0.6$ V and $t_d = 0.5$ s but varied $\Delta t$ so that the resulting $f$ was in the range from 0.1 Hz to 1 Hz (Fig. S10c, top). Finally, to measure the $A$ dependences, we applied a series of positive and negative spikes having $t_d = \Delta t = 1$ s with increasing $A$ from 0.6 V to 3 V (inset of Fig. S10e). Following the programmed spike trains, the current evolutions were recorded (Figs. S10a, c, e) and the current ratios after the last and first spikes, $I_n/I_1$ ($n$, the number of the last spike) were plotted as a function of $t_d$, $f$ or $A$ (Figs. S10b, d, f). Results show that the changes in $I$ increase with $t_d$, $f$ and $A$, and then tend to saturate.

The above measurements employed voltage spikes up to a few volts and lasting for seconds so that the evolutions in $I$ could be clearly discerned. It is instructive to estimate the lower limits for $A$, $t_d$ and $\Delta t$ that our devices could successively respond to. This test was performed using the same devices and conditions as described above. The setup yielded zero-current voltage $V_0 = -(218 \pm 12)$ mV according to continuous *I-V* measurements for several devices. The found $V_0$ served as a DC power source (section "Equivalent circuit model", Fig. S3) and with respect to $V_0$ as determined for a specific device, we increased $A$ of the applied spikes at a step of 2 mV. Also, we decreased $t_d$ from 0.5 s at a step of 10 ms and fixed $\Delta t = t_d$. Figure S11 shows results for one of our devices. Successive increase and decrease in $I$ were detected in response to minimum voltage spikes $V_{min}$ of +62 mV and -8 mV, respectively. The minimum discernable $t_d$ and $\Delta t$ for both spikes were about 10 ms (Fig. S11). The resulting difference in $I$ after two successive spikes was a few tens of pA. This is just slightly above our lower detection limit (~10 pA). Similar results were also observed for other tested devices. Taking the errors in determining $V_0$ into account, the observed $V_{min}$ then translate into an energy input of a few tens of meV to the permeating $K^+$ and $Cl^-$ ions, which is comparable to that of thermal fluctuations, $k_BT$ (~26 meV at room $T$). The capability of distinguishing effective current responses to minute voltage stimuli from thermal fluctuations highlights the devices' high sensitivity. We would like to note that the determined amplitudes and timescales (or frequencies) are relevant to biological systems[40,41] and represent the lower limits that our devices could successively respond to. Furthermore, the measurement allows us to estimate the minimum energy consumption for an individual spike, which is in the range from <1 pJ to a few tens of pJ. This energy cost is comparable to that in biological systems (0.01 pJ – 10 pJ)[42,43] and could be further reduced by, for example, using smaller pores.

**Image identification**

Image identification has been well-documented for fluidic memristors and typically implemented by reservoir computing[44]. The latter utilizes the short-term memory and nonlinear mapping of reported devices so that temporal signals can be mapped into manageable reservoir status. Different from past demonstrations, our experiment utilizes the devices' long-term memory and plasticity, and takes advantage of their long-term endurance (Fig. 3). The ultimate goal is to assess the fidelity and reliability of our devices for information storage and transmission.

The workflow for our experiment and data processing procedures is illustrated in Fig. S12a. Two well-known image datasets[45,46], MNIST and CIFAR-10, were employed in our experiments. MNIST[45] is a dataset for the grayscale and handwritten digits of 0 to 9. Each image is 28 × 28 pixels in size and in total, 60,000 training and 10,000 testing images constitute the full dataset. CIFAR-10 is a color image dataset containing 50,000 training and 10,000 testing images[46]. They are 32 × 32 pixels in size and across 10 different classes. We intended to encode these image datasets into voltage spike trains, namely, sequentially applied spikes with programmable parameters such as those in Figs. 3 and S10. In principle, at least $10^7$ – $10^8$ spike trains are required to encode all pixels' intensities in these large datasets. Considering that each spike train should consist of many individual spikes, processing such a huge number of spikes is, however, impractical for our devices. To simplify the problem, we note that the grayscale or RGB (red, green, blue) intensities can be universally measured by 256 integers (from 0 to 255), with '0' and '255' the lower and upper intensity bounds, respectively, whereas the total number of pixels in a full dataset is orders of magnitude larger. Accordingly, the same intensity could repetitively appear across different pixels. For this reason, we only focused on the 256 intensity values and encoded them into spike trains (Figs. S12b and S13a) following the latency encoding approach, which is also known as time-to-first-spike (TTFS) encoding[47]. The approach aims to encode data into spike trains with varying timings of the first spike, namely, a larger data value spikes earlier than a smaller one. We first converted each intensity into a series of discrete and single-bit events, either '0' or '1', spanning over 1,024 time-steps. This was implemented using the "spiken.latency" function in the Python package "snntorch". For the lowest intensity '0', we filled all the 1,024 positions with bit '0', and for intensities from '1' to '255', only one specific position in the series was replaced by bit '1' and its position was shifted from right to left with increasing the intensity. To translate the number series into a spike train, we dictated bit '1' as a write spike $V_{write}$ of $A$ = 3 V and '0' as a read spike $V_{read}$ of $A$ = 0.02 V. Both $V_{write}$ and $V_{read}$ lasted for 1 s and in between adjacent spikes, a 1-s resting period was inserted. Accordingly, the time span for such a spike train was 2,047 s, and transmitting all the 256 spike trains required at least $5 \times 10^5$ s, which is accessible by our devices (Fig. 3e).

The spike trains were applied to the fluidic circuit shown in Fig. S12c, where the constituting devices transmitted and stored the carried information as time evolutions of $G$ (Fig. S12d). For intensity '0', all the $V_{read}$ had negligible influence on $G$, which remained approximately constant as $G_1 \approx G_0$. On the other hand, for intensities '1' to '255', $V_{write}$ either increased or decreased $G$ depending on its polarity and the electrolytes (Fig. 3). The time evolution of $G$ (normalized by $G_0$) as recorded by $V_{read}$ displayed an abrupt change (either a peak or a valley) that separated two long-term states, $G_1$ ($\approx G_0$) and $G_2$ (Fig. S12d). Their time lengths were exclusively determined by the position of $V_{write}$ and hence, the encoded intensity. To recover $G_0$ before the second spike train, we normally applied a few $V_{erase}$ of the same $A$ but opposite polarities as $V_{write}$ (Fig. S12c) to trigger an inverse modification of $G$ (insets of Fig. 3c). We found that the mean value of all the recorded $G$ after transmitting a spike train could be used as a feature to characterize the corresponding intensity and following this way, different intensities could be effectively distinguished (see below). All the pixels' intensities in an image could be expressed as the extracted features from a single device, forming a feature map (Figs. S12e and S13b). On the other hand, thanks to the rectified and ion-selective characteristics of observed memory effect (Figs. 1b, d), four sets of features corresponding to a single intensity could be obtained simultaneously from the circuit in Fig. S12c where the constituting devices worked with $V_{write}$ of opposite polarities and different electrolytes. Accordingly, multiple feature maps could also be combined for image identification. The effectiveness of our devices in transmission and storage of a large amount of data ($10^5$ – $10^6$ samples) over a long time ($10^5$ – $10^6$ seconds) could be assessed by comparing the testing results using the extracted feature maps with those directly using the original data.

Taking the greyscale digits (28 × 28 pixels in size) in the MNIST dataset as an example, we employed two neural network models to identify these simple images. One of them was a one-layer, fully connected neural network with 784 input and 10 output neurons (Fig. S13c). We fed one of the four feature maps obtained from the circuit (Fig. S12c) into this model. The other one was a convolutional neural network (CNN), as illustrated in Fig. S13e. It was constructed by a 3×3 convolutional layer with 95 filters and a stride of 1, which were followed by a 2×2 max pooling layer and a fully connected layer with 16,055 neurons connecting to 10 output neurons (Fig. S13e). This CNN model allows multiple feature maps (typically ≤ 3) so that we could combine, for example, two of the four sets of features obtained from two outputs to form two feature maps. This allows us to test if different electrolytes could help improve the accuracies in image identification. We tuned the optimiser, activation function and hyperparameters, including filters, learning rate, batch size, and epochs, using Bayesian optimization combined with 5-fold cross-validation (CV).

Results for identification of the greyscale and handwritten digits are summarized in Table S3. It shows that both the training and testing accuracies obtained using our feature maps are quite close to those obtained directly using the original data. Comparing with the training accuracies, the testing ones are more instructive because they essentially assess the performance of the model on the unseen data. To be specific, the testing accuracies using the one-layer model and one of the four sets of features are in the range of 91.8% – 92.9%, while that for the original data is 92.9%. Same features but using the CNN model yield systematically higher testing accuracies, all about 98.3%, and close to that for the original data, 98.4%. By contrast, different combinations of two sets of features obtained from KCl and $MgCl_2$ solutions, respectively, yield testing accuracies of 98.3% – 98.4%, which are only marginally improved with respect to those using a single feature set.

To move a step further, we sought to process color images, such as those in the CIFAR-10 dataset. Identification of such images requires three feature maps to express the RGB intensities of all pixels in a single image. To analyze these feature maps, we employed a more complex model of residual neural network with 18 layers (ResNet18) (Fig. S12f). In principle, we could either combine three of the four sets of features or simply use a single set to express the RGB intensities. However, our preliminary attempt using both methods showed no noticeable difference in the resulting accuracies (Table S3). Accordingly, the RGB intensities were mainly described using one of the four sets of features for simplicity. Furthermore, because it was computationally expensive and time-consuming to train a ResNet18 from scratch, we instead started with a pretrained model and tuned the parameters using our feature maps under Bayesian optimization combined with 5-fold CV. The parameters included optimiser and hyperparameters, such as learning rate, batch size, epochs and dropout rate. Results for identification of the color images using the described feature maps are summarized in Table S3. The testing accuracies are 92.6% – 94.1%, which are only slightly lower than that of the original data, 94.6%. Summarizing all the numerical results for three neural networks and two image datasets, we conclude that the testing accuracies after processing by our fluidic circuit are quite close to that directly using the original data. The closeness is further detailed in the confusion matrices for our best results and their comparisons with those for the original data (Figs. S12g, S13d and f), which in turn proves the high fidelity and reliability of our devices and their integrated circuit for storing and transmitting largescale and event-based information over a long time.

**Real-time analysis of emulated neural signals**

We constructed voltage spike trains using write spikes $V_{\text{write}}$ having physiologically relevant $A$ = 100 mV and $t_{\text{d}}$ = 10 ms to emulate their real neural counterparts. They were arranged along a single time line to form three characteristic firing pattern segments, namely, tonic, bursting and adapting, as illustrated in the top panels of Figs. 4a-c. Specifically, the tonic segments were constructed by $V_{\text{write}}$ of equal interval $\Delta t$ = 140 ms. In the bursting segments, three successive $V_{\text{write}}$ with $\Delta t$ = 10 ms first formed a unit and then such units were placed with equal inter-unit interval of $\Delta T$ = 450 ms. Finally, the adapting ones were constructed by repeating a nine-$V_{\text{write}}$ unit at zero $\Delta T$. The nine successive $V_{\text{write}}$ in the unit were arranged with $\Delta t$ starting from 20 ms, which was then increased stepwise by a factor of 50% in between subsequent spikes. This resulted in a total time span of about 1.5 s for transmitting such a unit. For all the three firing patterns, $V_{\text{read}}$ were incorporated uniformly at $\Delta t$ = 40 ms throughout the entire time line to record time evolutions of $G$. It is evident from the bottom panels in Figs. 4a-c that three different firing patterns could be clearly discerned from corresponding $G$ responses.

Such a spike train was employed to emulate signals transmitted from a neuron. To further assess the ability of our devices for identifying the synchronization states between two neurons, namely, in-phase, anti-phase and no-phase, a similar spike train with different $\Delta t$ or $\Delta T$ in its tonic, bursting and adapting segments, respectively, was used to emulate signals from the other neuron. In brief, the in-phase state was represented by two identical spike trains with the same time lines and the anti-phase was emulated by imposing a relative shift of 100 ms in their time lines. To construct no-phase state, the time intervals between adjacent spikes ($\Delta t$) or units ($\Delta T$) were intentionally modified in the second spike train. To be specific, in the tonic segments, we used $\Delta t$ = 190 ms (instead of 150 ms) in between adjacent spikes and in the bursting ones, we kept the same $\Delta t$ = 10 ms in each unit but increased $\Delta T$ by 100 ms (from 450 ms). For the adapting segments, each unit was still constructed by nine successive spikes and the same $\Delta t$ = 20 ms was imposed between the first two spikes. However, $\Delta t$ was increased progressively by a factor of 80% (instead of 50%) in between subsequent spikes. This resulted in a prolonged time span of about 5 s (rather than 1.5 s) for transmitting a single unit.

To generate sufficiently large training and testing datasets for simultaneous recognition of neural firing patterns and synchronization states, we constructed two pairs of spike trains of different lengths (7.5 h- and 2.5 h-long, respectively) using the parameters as described above. Each pair was composed of three different firing patterns (tonic, bursting and adapting) of equal time length and within each pattern, three different synchronization states (in-phase, anti-phase and no-phase) also spanned the same length. Two spike trains in the same pair were first divided, respectively, into three segments of the same pattern. The resulting three pairs of shorter segments were then fed into the six inputs of our fluidic circuit (Figs. 4d, S14a). Three devices were connected in parallel to the same input and each of them was used to successively process 1/3 of the transmitted spike train. Using such a fluidic circuit, the three different synchronization states within the three pairs of shorter spike trains and their firing patterns could be identified in an in-parallel and simultaneous manner. The resulting $G$ responses collected from all the outputs were used as training (7.5 h-long) and testing (2.5 h-long) datasets, respectively. Thanks to the long-term stability of our devices, processing these two long spike train pairs allowed us to have a total number of $1.1 \times 10^6$ and $0.4 \times 10^6$ conductance samples in the training and testing datasets, respectively. Figures 4f and g display the programmed 7.5 h-long spike train pair and corresponding time evolutions of $G$. Because relatively weak voltage spikes were used, the raw data were inevitable to have low signal-to-noise ratios and slight baseline drifts. To mitigate these issues, we applied a first-order difference to the raw data of $G$ evolutions and smoothed the outliers in the differentiated data using an interpolation approach. Then we applied a global Z-score normalization to each time series to remove any scale dependences. This ensured that variations across different recorded time series in both the training and testing datasets could be directly compared.

To assess the quality of the entire training and testing datasets, we employed the different $G$ responses to the three neuron firing patterns (Figs. 4a-c) as features to implement pattern classification. For the training dataset (that is, time evolutions of $G$ in response to the 7.5 h-long spike train pair), we implemented an 80/20 temporal split to partition the six shorter segments of $G$ evolutions into training and validation sets. Each segment corresponded to a certain firing pattern that was fed into one of the six inputs of the circuit in Fig. 4d. To prevent data leakage in the training stage, the input features were generated using a non-overlapping sliding window (width and step size of 100), which partitioned the data into 100 × 1 vectors. Each vector was assigned a ground-truth label (namely, adapting, bursting or tonic). To implement online evaluation, we processed the testing dataset (that is, $G$ responses to the 2.5 h-long spike train pair) using a sliding window (width of 100 and step length of 1). The described data processing resulted in a continuous stream of conductance samples, which enabled high-resolution and real-time prediction.

A simple CNN model containing only one 1D convolution layer with a Gaussian noise layer for data augmentation, and a fully connected layer (Fig. S14b), was used for the classification task. The model was trained on the prepared training and validation sets as described above and then online predictions were made on the testing dataset, yielding a high raw prediction accuracy of 96.7%. The hyperparameters of the model, including filters, neuron numbers in the fully connected layer, learning rate, batch size, dropout rate, and epochs, were tuned using Bayesian optimization that was integrated with a 5-fold CV. To mitigate prediction glitches (namely, temporal flickering), a median filter (window size of 30) was applied to the output stream, which further refined the final accuracy to 96.8%. Details about the prediction results and their comparison with the real data are shown in the confusion matrix in Fig. S14c.

To implement simultaneous identification of neuron firing patterns and synchronization states, we extracted shared features from correlated $G$ responses to a pair of spike trains. Same as recognition of neural firing patterns alone, the described 80/20 stratified temporal split was also implemented and three pairs of temporal $G$ responses corresponding to shorter spike train pairs featuring the same pattern but different synchronization states were partitioned into training and validation sets. We then used a sliding window (width and step length of 100) to separate each pair of $G$ evolutions into 100 × 2 non-overlapping matrices to prevent data leakage. Each matrix was dual-labelled by the corresponding neuron firing pattern and synchronization state. Specially, transitions between different synchronization states were labelled as no-phase to maintain explicit boundaries. To prepare the testing set for online predictions, we applied a sliding window (width of 100 but step length of 1) and separated the testing dataset into 100 × 2 matrices. These matrices enabled the model to make online predictions of both the neuron firing patterns and synchronization states in every 100 × 2 segment of all the $G$ evolution pairs. Furthermore, because effective analysis of the synchronization states relies on identifying correlated temporal trends rather than the absolute values of $G$, we employed a local Z-score normalization within every 100 × 2 matrix. Unlike the global Z-score normalization used above, the described local scaling allowed the model to prioritize local temporal trends.

A multi-scale CNN model architecture constructed by three parallel 1D convolutional branches with kernel sizes $k$ = 5, 20, and 50, respectively, was developed for the multi-task predictions (Fig. S14d). After concatenation, a Squeeze-and-Excitation (SE) attention module was integrated to perform dynamic channel-wise feature recalibration and adaptively weigh the importance of each scale. The features were then aggregated by a dual-pooling strategy, combining both the global max pooling and global average pooling, so that the model could capture the prominent peaks and general trend simultaneously. These features were mapped to a shared dense representation layer before being passed to a multi-task output layer. This dual-head configuration in the output layer enabled the model to leverage shared features for simultaneous classification of synchronization states and neuron firing patterns. The hyperparameters, including filters, learning rate, batch size, dropout rate, rate for the Spatial Dropout layer, reduction ratio for the SE block, weight of the pattern output loss, and epochs, were tuned using Bayesian optimization and a 5-fold CV. The multi-scale CNN model was trained using the prepared training and validation sets and online multi-task predictions were made on the testing set. After applying median filtering to smooth the prediction stream, high overall accuracies of 98% and 95% were achieved for predicting neural firing patterns and synchronization states, respectively. Details are shown in Figs. 4h and i.

## Supplementary figures and tables

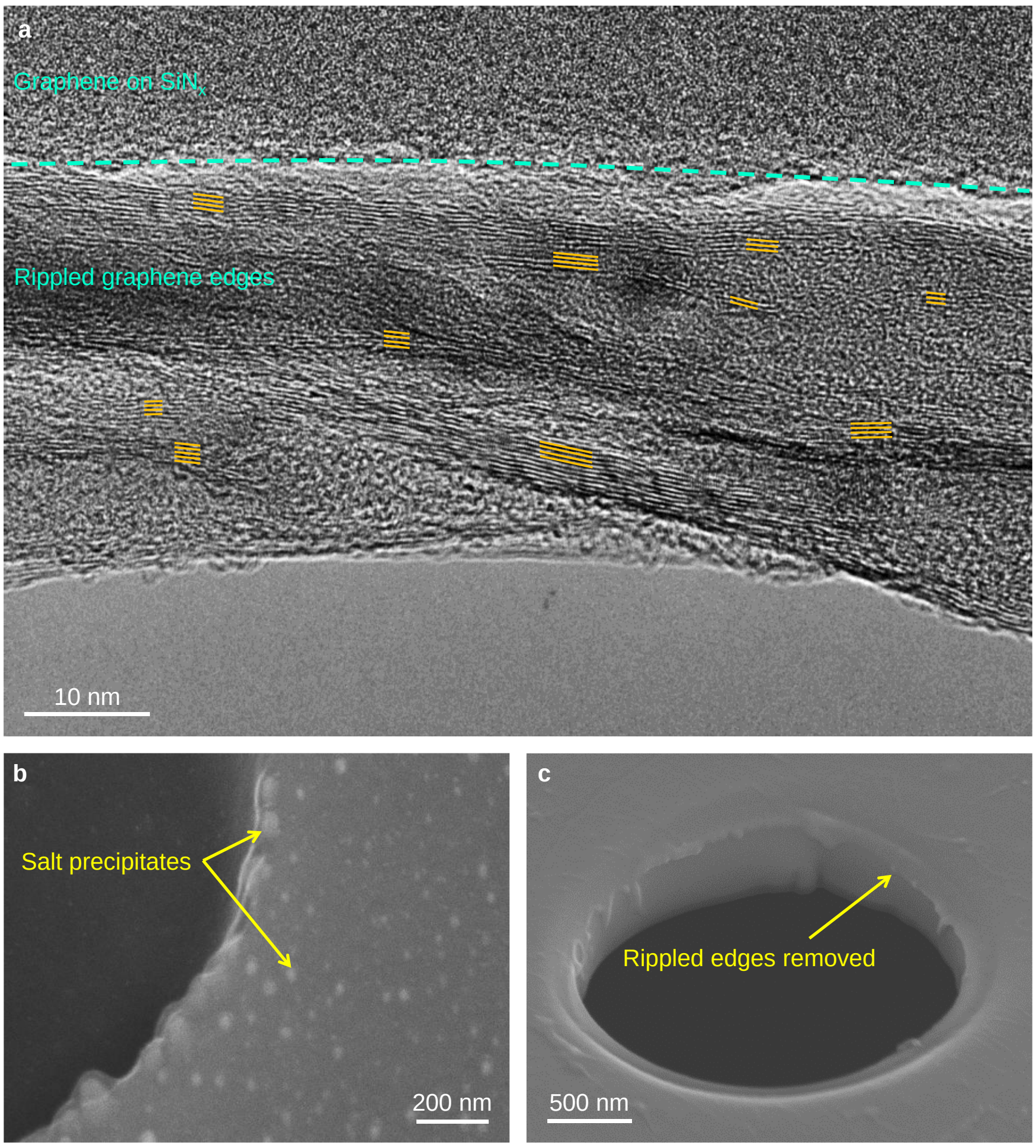

**Figure S1. Additional electron micrographs. a**, A representative TEM image of the rippled graphene edges. Dashed curve marks the position of $SiN_x$ aperture rim. Short yellow lines mark some of the parallel fringes seen in the rippled graphene edges. Statistics on a few of them yields interlayer spacing of 0.36 ± 0.02 nm. **b**, SEM image for one of our devices after long-term measurements in KCl solutions. One can see many nanoscale particles (salt precipitates) accumulating inside the rippled edges and on the graphene surface. **c**, SEM image for a control sample with its rippled edges removed by dry etching.

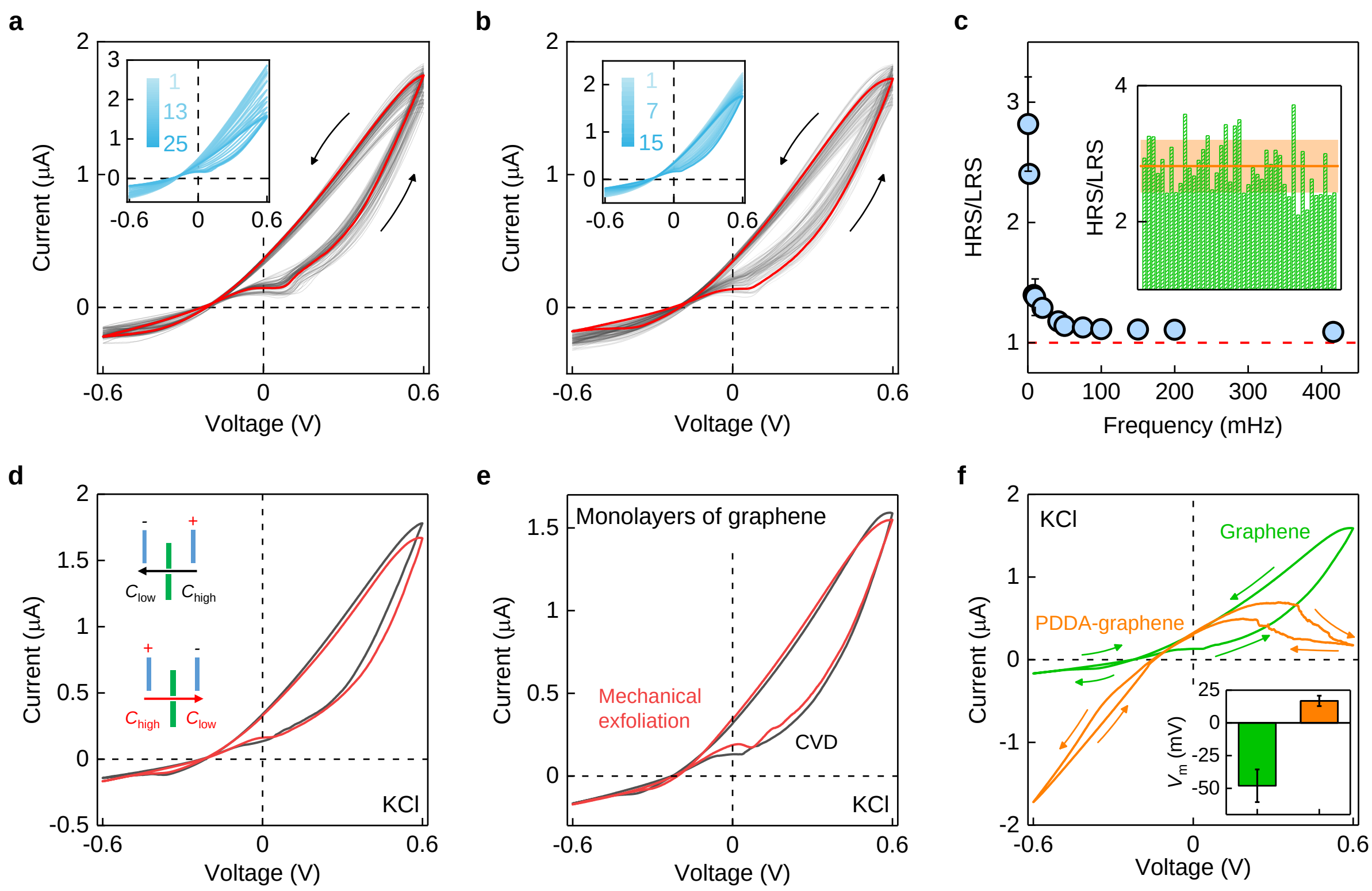

**Figure S2. Additional *I-V* measurements. a, b,** Full sets of cyclic *I-V* measurements for two additional devices. Main figures, *I-V* curves after reaching stabilization (coded in grey). Red curves, representatives. The total number of stable *I-V* curves shown in (a, b) are 90 and 130, respectively. Insets, evolutions of *I-V* curves before stabilization (color coded). **c,** Frequency dependence of HRS/LRS ratio. Symbols, average data based on multiple devices. Error bars, SD (shown if larger than the symbols). Inset, HRS/LRS ratios at $f$ = 0.4 mHz for 43 devices. Each bar represents a different device. Solid orange line, average HRS/LRS with the shaded area indicating SD. **d**, Representative stable *I-V* curves acquired for one of our devices under opposite concentr ation directions (color coded), as illustrated in the insets. The arrows indicate directions from $C_{high}$ to $C_{low}$ (same color coding as the *I-V* curves). **e**, Monolayer graphene samples made from mechanical exfoliation (red) and CVD (black). **f**, Comparison between a control device with its graphene surface functionalized by PDDA and an untreated device (color coded). Inset, extracted $V_m$ with error bars indicating SD for 3 devices (same color coding). The arrows in (a, b, f) indicate the directions of hysteresis loops. Black dashed lines mark zero $I$ and $V$. All the shown *I-V* curves were acquired in KCl solutions ($C_{high}$ = 1 M, $C_{low}$ = 1 mM) under a triangular voltage wave with $A$ = 0.6 V and $r_s$ = 1 mV/s.

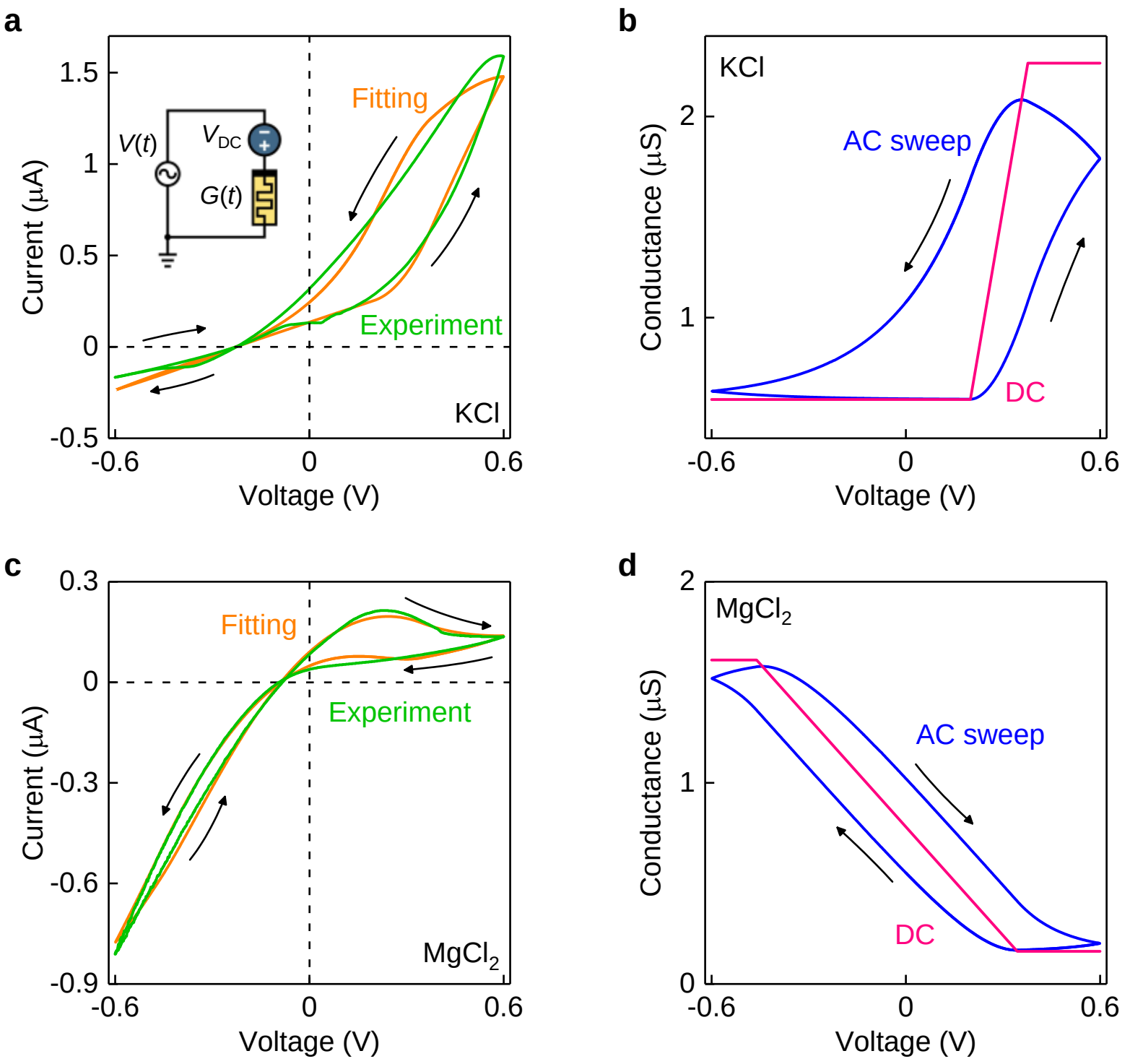

**Figure S3. Equivalent circuit model. a**, **c**, Experimental and fitted *I*-*V* curves (color coded) for KCl and $MgCl_2$ solutions, respectively. The experimental *I*-*V* curves were taken from Fig. 1. Inset in (a), equivalent circuit model of a memristor in series with a DC voltage source. **b**, **d**, Simulated $G(V)$ under the same AC voltage sweep as in (a, c) and compared with the DC quasi-static conductance $G_{qs}$ (color coded). The arrows indicate loop directions.

**Table S1.** Fitting parameters in eqs. S1-S3 for KCl and $MgCl_2$ solutions, respectively.

| **Electrolyte** | **KCl** | **$MgCl_2$** |
|---|---|---|
| $\tau_m$ (s) | 243 | 139 |
| $k$ (μS/V) | 9.42 | -1.80 |
| $b$ (μS) | -3.42 | 0.940 |
| $G_{on}$ (μS) | 2.26 | 1.61 |
| $G_{off}$ (μS) | 0.593 | 0.162 |

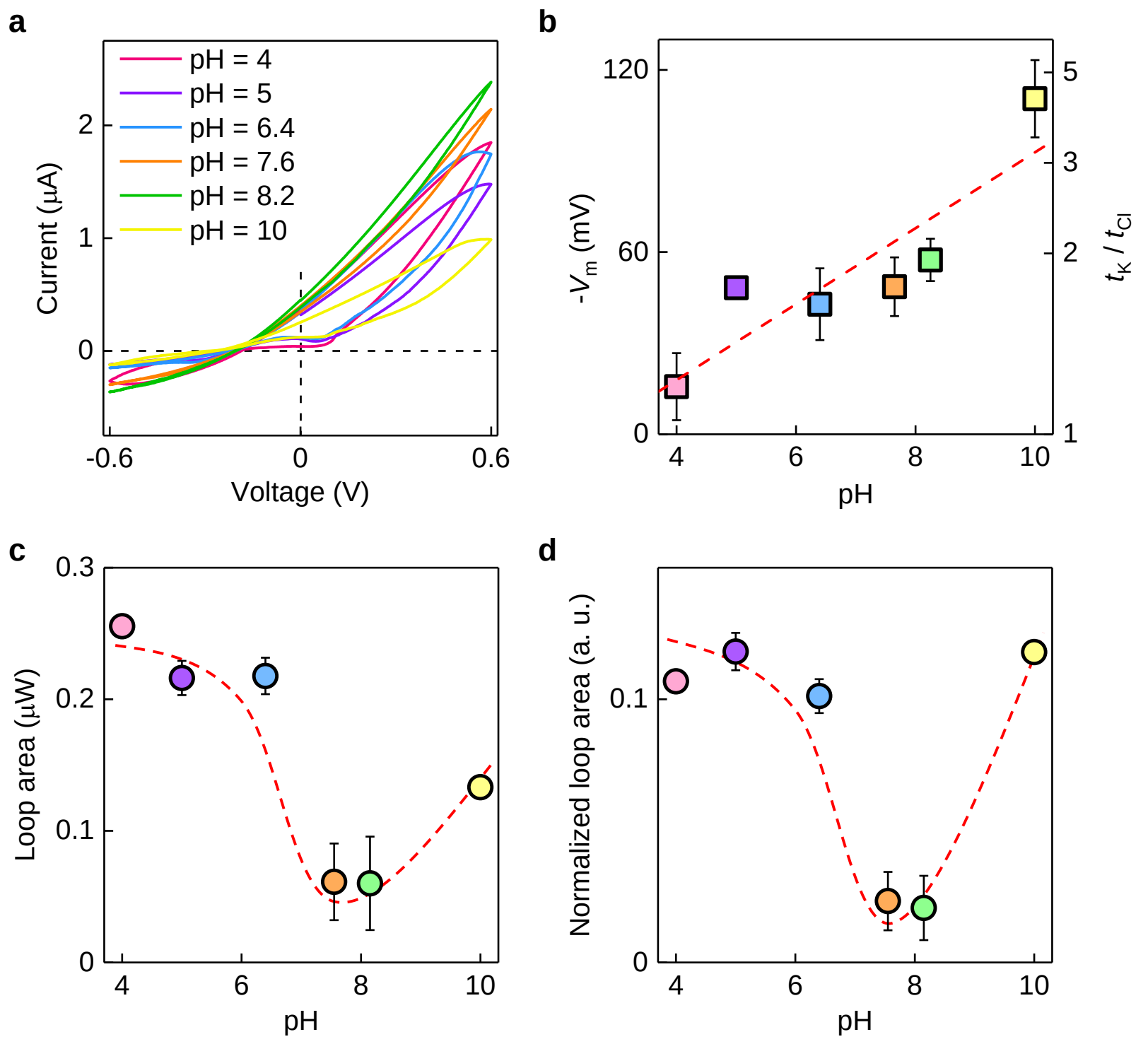

**Figure S4. PH dependence. a**, Representative *I*-*V* curves after reaching stabilization at different pH (color coded). Dashed black lines mark *I* = 0 and *V* = 0, respectively. **b**, $t_K/t_{Cl}$ (right-*y*) calculated from $V_m$ (left-*y*) at different pH. Note the nonlinear scale on the right-*y* axis. **c**, $A_{loop}$ as a function of pH. **d**, Same as in (c) but for normalized $A_{loop}$ (pH). Symbols in (b-d), experimental data averaged by 10 independent *I*-*V* measurements with error bars indicating SD; same color coding as in (a). Red dashed curves in (b-d), guides to the eyes.

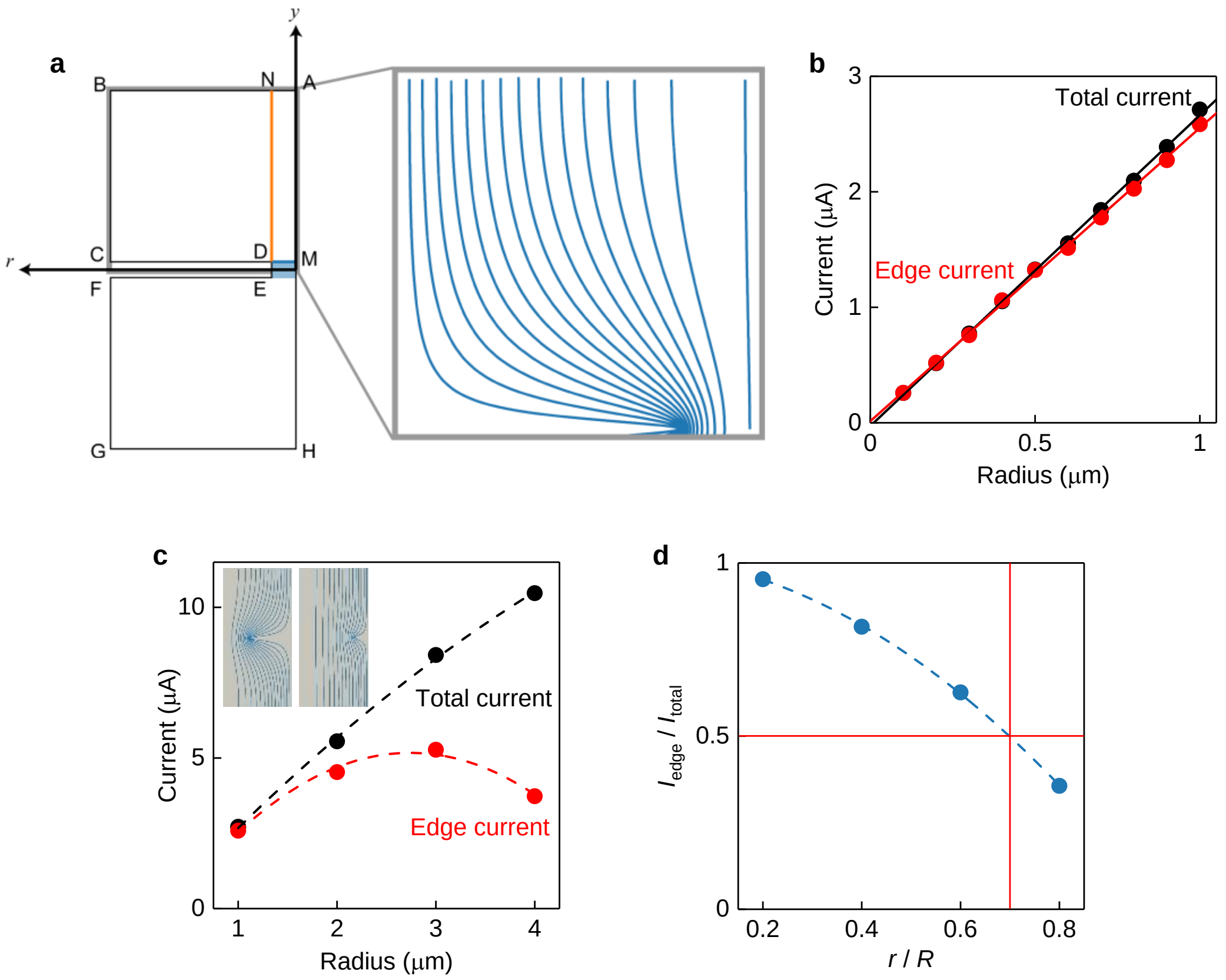

**Figure S5. Spatial distribution of ion current. a**, Left, schematic diagram for the axisymmetric and cylindrical simulation domain with *y* the axis of symmetry. AB and GH are boundaries to bulk KCl solutions with $C_{high}$ = 1 M and $C_{low}$ = 1 mM. CDEF is the device's surface. The pore region is shaded in blue with an effective length DE varying from 2 nm to 500 nm. Right, an example of the simulated distribution map for the current streamlines. Pore radius, *r* = 1 μm and length, *L* = 2 nm. **b**, Comparison of the edge current with the total current (color coded) at *r* = 0.1 – 1 μm and fixed *L* = 2 nm. The edge current is defined by its streamlines crossing ND in (a). **c**, Same comparison as in (b) but for *r* = 1 – 4 μm (same color coding) at a fixed half bulk size of *R* = 5 μm. Insets, current streamlines through pores of *r* = 2 μm (left) and 4 μm (right), respectively. The current fluxes between two adjacent streamlines in (a, c) are equal. **d**, Evolution of $I_{edge}/I_{total}$ as a function of $r/R$. Red solid lines mark the position for $I_{edge}/I_{total}$ = 0.5 and $r/R \approx 0.7$. Symbols in (b-d), calculations; solid lines in (b), best linear fits; dashed curves in (c, d), guides to the eyes.

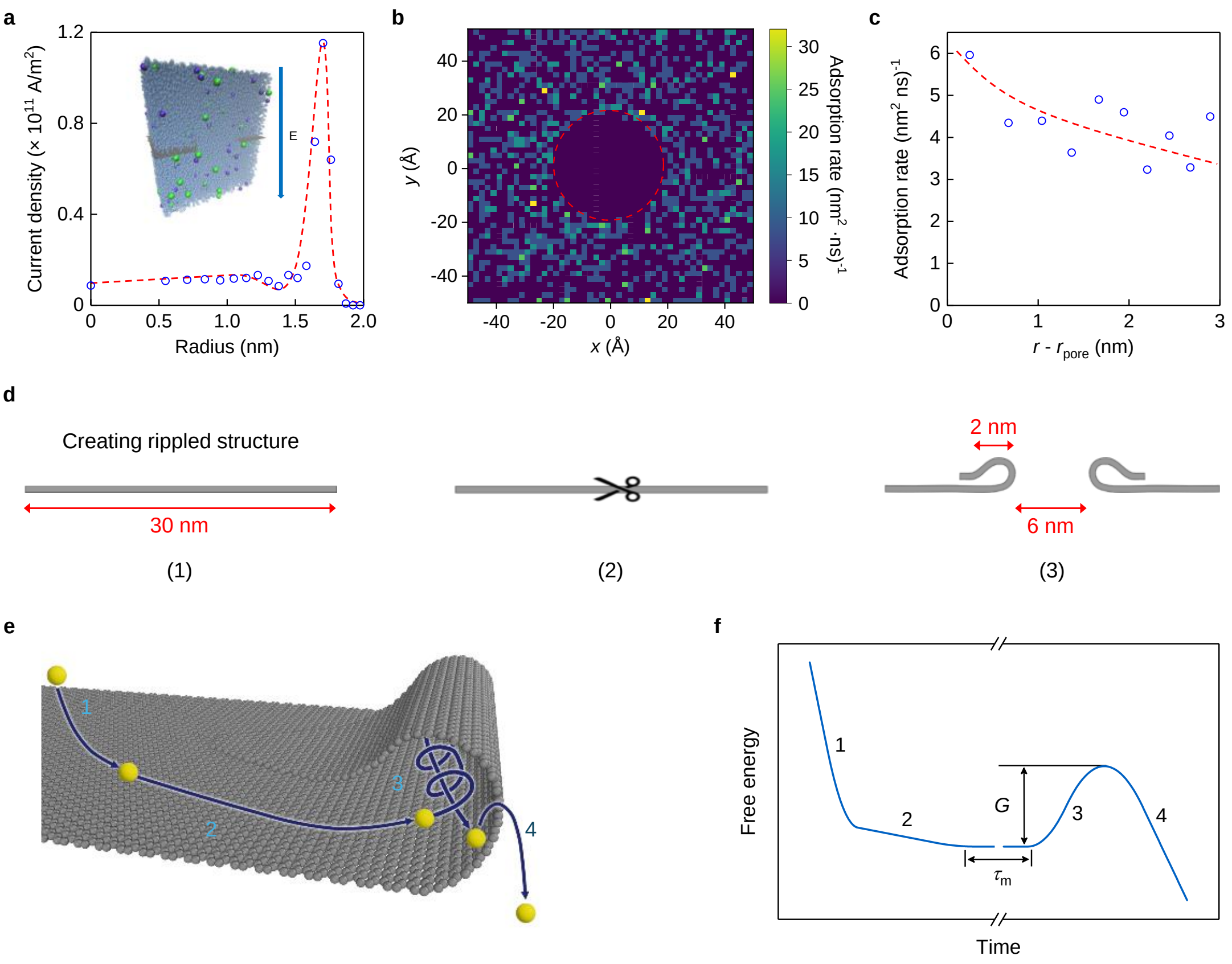

**Figure S6. MD simulations. a**, Radial distribution of current density across a 4-nm graphene pore. Concentration of KCl solutions, 0.4 M. Electric field, 0.5 V/nm. Inset, schematic illustration of the MD model. **b**, A snapshot for mapping the ion adsorption rate over the surface of graphene. **c**, Ion adsorption rate as a function of the distance away from the pore edge. Symbols in (a, c), simulation results; red dashed curves, guide to the eyes. **d**, Creating the model for the rippled structure. A monolayer graphene sheet with length of 30 nm and width of about 10 nm (1) was cut off in the middle along the width direction (2). Then the two freshly cleaved edges were scrolled along the length direction and away from each other, resulting in a 6-nm wide trench in the middle and two graphene ripples with curvature of 1 $nm^{-1}$ (3). The empty space inside the ripples is about 2 nm in size. **e**, The resulting model for an individual graphene ripple and the ion transport trajectory with numbers denoting different steps. Practically, the entire edge structure could be simplified as stacks of multiple ripples. **f**, Free energy profile along the trajectory; same number labelled as in (e).

**Algorithm S1.** Gillespie algorithm to calculate the total time for ion transport.

```
Input: starting site S, destination site D, adjacency site lists and energy barriers for all sites
Output: total time for ion transport t
Initialize total time t = 0 and ion position P = S
while P ≠ D
    Set state transition rate R_t = 0
    Create a list L to store partial sum of rates
    for each site Q in getAdjacentSites (P)
        Get energy barrier G = getEnergyBarrierBetween (P, Q)
        Calculate transition rate for the path k = v_0exp(-G/k_BT)
        Update state transition rate R_t = R_t + k
        Append R_t to L
    end for
    Generate random numbers r_1 and r_2 from uniform distribution U(0, 1)
    Set the transition site index n = 0
    for i = 0 to length(L) – 1
        if L[i] ≥ R_tr_1
            n = i
            break
        end if
    end for
    Move the ion to site P = getAdjacentSites(P)[n]
    Calculate transition time dt = -ln(r_2/R_t)
    Update total time t = t + dt
end while
return t
```

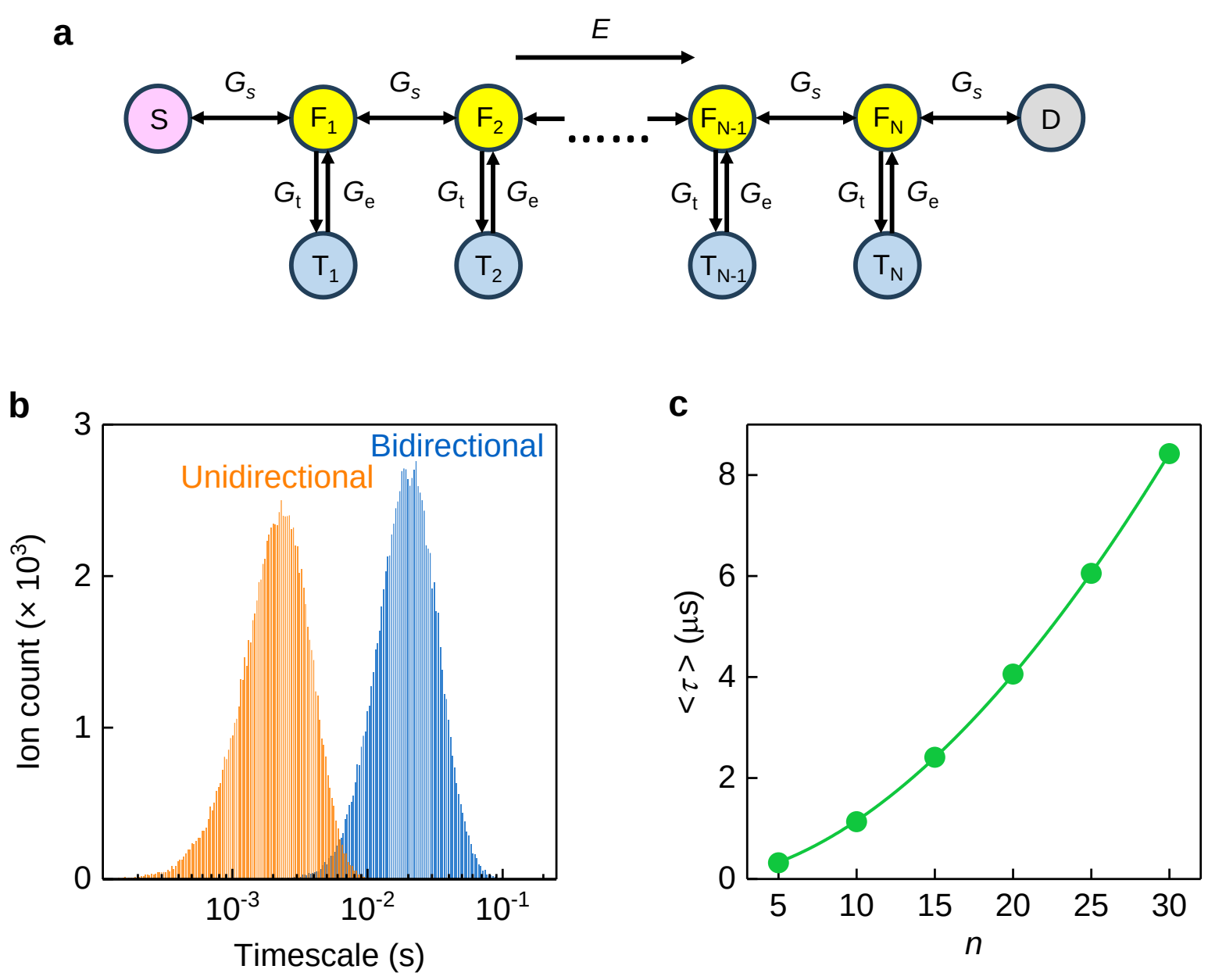

**Figure S7. Kinetic Monte Carlo simulations. a**, Schematic of the simulation system. The ion transport starts at source (S) and stops at drain (D). A series of free (F) and trapping (T) sites that are connected in pairs constitutes the transport pathway between source and drain. **b**, Histograms for the transport time of all ions along a bidirectional and a unidirectional pathway (color coded), respectively. **c**,

Evolution of the mean ion transport time ⟨$\tau$⟩ as a function of the number of energy barriers $n$. In this specific example, the energy barrier is set at $10k_BT$; the voltage difference between adjacent sites is 0.01 $k_BT/e$, but can be varied depending on the relative contribution between ion drift and diffusion. The solid curve is a power law fit, $\langle\tau\rangle = 0.0168\ n^{1.83}$.

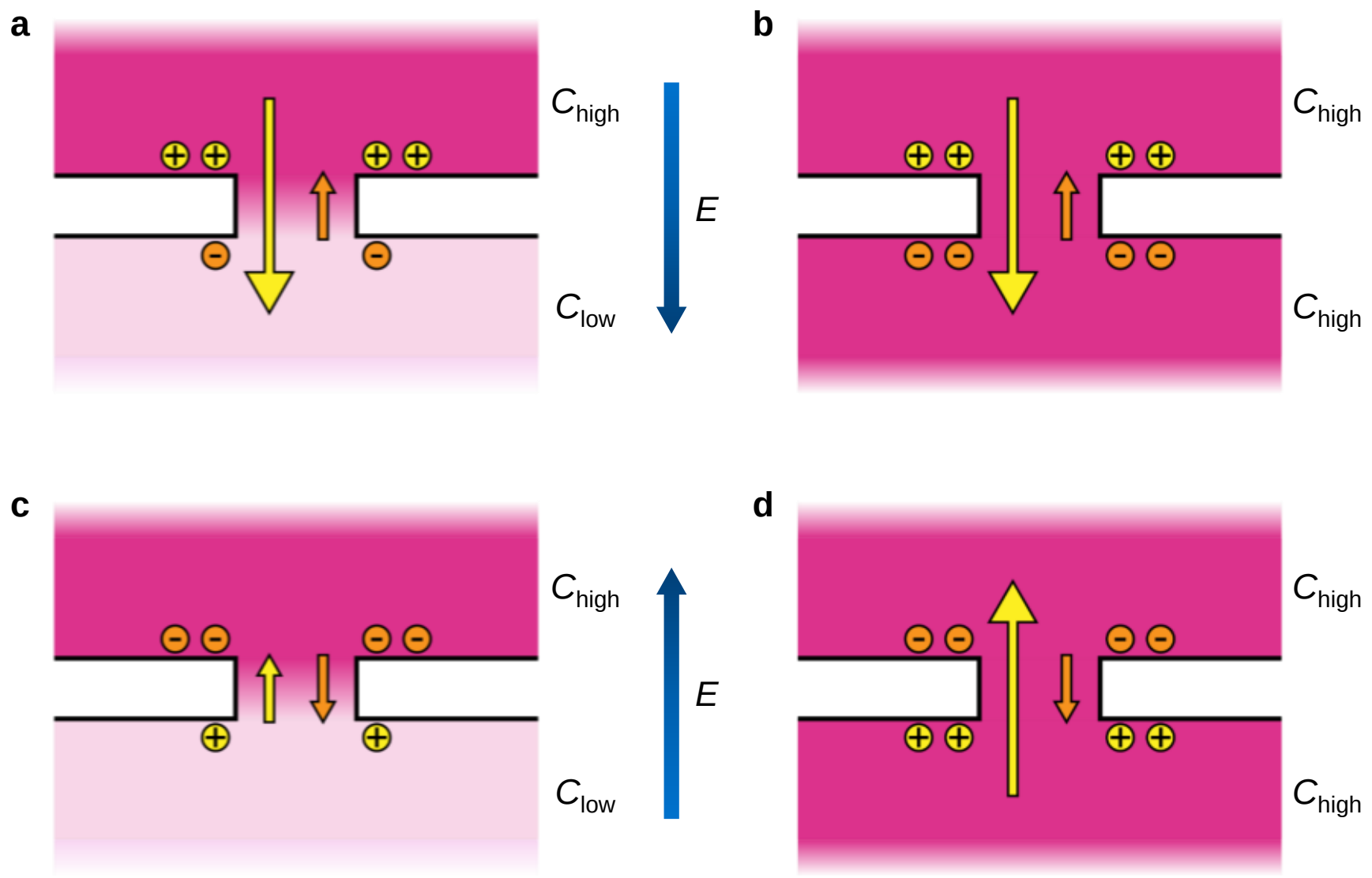

**Figure S8. Concentration effects.** Schematic diagrams for the transport of cations and anions under (**a**, **c**) different $C$ and (**b**, **d**) the same $C$ across the pore and driven by either a forward or a backward electric field. The permeability of cations is assumed to be much larger than that of anions.

**Table S2.** Detailed measurements performed on all our working devices (89 in total).

| | Experiment | Parameters | Material |
|---|---|---|---|
| Device 1 | Cyclic *I-V* measurements, Frequency dependence | 1 mM – 3 M KCl | CVD graphene |
| Device 3 | Cyclic *I-V* measurements, Frequency dependence | 1, 10 mM – 0.1 M KCl, 1 mM – 1 M KCl | CVD graphene |
| Device 4 | Cyclic *I-V* measurements, Frequency dependence | 0.1 M – 0.1 M, 1 mM – 1 M KCl | CVD graphene |
| Device 5 | Cyclic *I-V* measurements, Frequency dependence | 1 mM – 3 M KCl | CVD graphene |
| Device 6 | Cyclic *I-V* measurements, Frequency dependence, Concentration dependence | 1 mM – 0.1, 0.5, 1, 3 M KCl | CVD graphene |
| Device 7 | Cyclic *I-V* measurements, Frequency dependence, Concentration dependence | 1 mM – 0.1, 0.5, 1, 3 M KCl | CVD graphene |
| Device 8 | Cyclic *I-V* measurements, Frequency dependence | 1 mM – 3 M KCl | CVD graphene |
| Device 9 | Cyclic *I-V* measurements | 1 mM – 3 M KCl | CVD graphene |
| Device 10 | Cyclic *I-V* measurements, Frequency dependence | 1 mM – 3 M KCl | CVD graphene |
| Device 11 | Cyclic *I-V* measurements, Frequency dependence | 1 mM – 3 M KCl | CVD graphene |
| Device 13 | Cyclic *I-V* measurements | 1 mM – 1, 3 M KCl | CVD graphene |
| Device 15 | Concentration dependence, pH dependence | 0.1 mM – 0.1 M KCl; 1 mM – 1, 3 M KCl; 0.3 mM – 3 M KCl | CVD graphene |
| Device 17 | Concentration dependence, pH dependence | 1mM – 1 M KCl; 3 mM – 3 M KCl; 0.3 mM – 3 M KCl | CVD graphene |
| Device 18 | Pore geometry dependence, pH dependence | 1 mM – 1 M KCl; 1 mM – 3 M KCl | CVD graphene |
| Device 19 | Pore geometry dependence, pH dependence | 1 mM – 1 M KCl; 1 mM – 3 M KCl | CVD graphene |
| Device 20 | Cyclic *I-V* measurements, Concentration dependence | 1 mM – 0.001, 0.01, 0.1, 1, 3 M KCl | CVD graphene |
| Device 21 | Pore geometry dependence, Spikes measurements | 1 mM – 1, 3 M KCl | CVD graphene |
| Device 22 | Spikes measurements | 1 mM – 3 M KCl | CVD graphene |
| Device 23 | Concentration dependence, *I-V* measurements for different salts, Spikes measurements | 1 mM – 1, 3 M KCl; 1 mM – 0.001, 0.01, 0.1, 1 M $CaCl_2$ | CVD graphene |
| Device 25 | Integrated circuit, *I-V* measurements | 1 mM – 3 M KCl | CVD graphene |
| Device 26 | Integrated circuit, *I-V* measurements | 1 mM – 3 M KCl | CVD graphene |
| Device 27 | Spikes measurements | 1 mM – 1, 3 M KCl | CVD graphene |
| Device 28 | Integrated circuit, *I-V* measurements | 1 mM – 3 M KCl | CVD graphene |
| Device 29 | Cyclic *I-V* measurements, *I-V* measurements for different salts, Spikes measurements | 1 mM – 1 M KCl, NaCl | CVD graphene |
| Device 30 | *I-V* measurements for different salts, Spikes measurements | 1 mM – 0.001, 0.01, 0.1, 1, 3 M NaCl, $AlCl_3$; 1 mM – 1, 3 M $CaCl_2$; 1 mM – 3 M KCl | CVD graphene |
| Device 31 | Pore geometry dependence, *I-V* measurements for different salts, Spikes measurements | 1 mM – 1 M KCl; 1 mM – 0.1, 1 M $CaCl_2$ | CVD graphene |
| Device 32 | *I-V* measurements for different salts, Pore geometry dependence | 1 mM – 0.001, 0.01, 0.1, 1, 3 M $CaCl_2$ | CVD graphene |
| Device 33 | Spikes measurements | 1 mM – 3 M KCl | CVD graphene |
| Device 35 | Ripples removed, *I-V* measurements | 1 mM – 3 M KCl | CVD graphene |
| Device 36 | Ripples removed, *I-V* measurements | 1 mM – 3 M KCl | CVD graphene |
| Device 37 | Ripples removed, *I-V* measurements | 1 mM – 3 M KCl | CVD graphene |
| Device 38 | *I-V* measurements for different salts | 1 mM – 0.001, 0.01, 0.1, 1, 3 M $MgCl_2$ | CVD graphene |
| Device 39 | Cyclic *I-V* measurements | 1 mM – 1 M KCl | CVD graphene |
| Device 40 | Cyclic *I-V* measurements, *I-V* measurements for different salts, Frequency dependence | 1 mM – 1 M KCl, $AlCl_3$ (Frequency dependence); 1 mM – 3 M $MgCl_2$ | CVD graphene |
| Device 41 | Ripples removed, *I-V* measurements | 1 mM – 1 M KCl | CVD graphene |
| Device 42 | *I-V* measurements for different salts, Spikes measurements | 1 mM – 1 M KCl, $AlCl_3$ | CVD graphene |
| Device 43 | Spikes measurements | 1 mM – 1 M KCl, $MgCl_2$ | CVD graphene |
| Device 44 | Pore geometry dependence | 1 mM – 1 M KCl | CVD graphene |
| Device 45 | Ripples removed, *I-V* measurements | 1 mM – 1 M KCl | CVD graphene |
| Device 46 | Concentration gradient direction dependence, Spikes measurements | 1 mM – 1 M KCl | CVD graphene |
| Device 47 | *I-V* measurements for different salts, Spikes measurements | 1 mM – 1 M KCl, $AlCl_3$ | CVD graphene |
| Device 48 | Spikes measurements | 1 mM – 1 M KCl | CVD graphene |
| Device 49 | Spikes measurements | 1 mM – 1 M KCl | CVD graphene |
| Device 50 | Spikes measurements | 1 mM – 1 M KCl | CVD graphene |
| Device 51 | Pore geometry dependence | 1 mM – 1 M KCl | CVD graphene |
| Device 52 | Integrated circuit, *I-V* measurements | 1 mM – 1 M KCl | CVD graphene |
| Device 53 | Integrated circuit, *I-V* measurements | 1 mM – 1 M KCl | CVD graphene |
| Device 54 | Pore geometry dependence | 1 mM – 1 M KCl | CVD graphene |
| Device 55 | Pore geometry dependence | 1 mM – 1 M KCl | CVD graphene |
| Device 56 | Pore geometry dependence | 1 mM – 1 M KCl | CVD graphene |
| Device 57 | Pore geometry dependence | 1 mM – 1 M KCl | CVD graphene |
| Device 58 | Pore geometry dependence | 1 mM – 1 M KCl | CVD graphene |
| Device 60 | *I-V* measurements for different salts, Frequency dependence, Spikes measurements | 1 M – 1 M KCl; 1 mM – 1 M $MgCl_2$, $AlCl_3$ | CVD graphene |
| Device 61 | Pore geometry dependence | 1 mM – 1 M KCl | CVD graphene |
| Device 62 | Spikes measurements | 1 mM – 1 M KCl | CVD graphene |
| Device 63 | Ripples removed, *I-V* measurements | 1 mM – 1 mM KCl | CVD graphene |
| Device 64 | Cyclic *I-V* measurements, Frequency dependence, Spikes measurements, pH dependence | 1 mM – 1 M KCl | CVD graphene |
| Device 65 | PDDA-functionalized graphene | 1 mM – 1 M KCl | CVD graphene |
| Device 66 | Pore geometry dependence, Frequency dependence | 1 mM – 1 M KCl, 0.1 mM – 0.1 M KCl (Voltage) | CVD graphene |
| Device 67 | Spikes measurements | 1 mM – 1 M KCl | CVD graphene |
| Device 68 | Thickness dependence | 1 mM – 1 M KCl | CVD graphene |
| Device 69 | Thickness dependence | 1 mM – 1 M KCl | CVD graphene |
| Device 70 | Thickness dependence | 1 mM – 1 M KCl | CVD graphene |
| Device 71 | Thickness dependence | 1 mM – 1 M KCl | CVD graphene |
| Device 72 | Pore geometry dependence | 1 mM – 1 M KCl | CVD graphene |
| Device 73 | Pore geometry dependence | 1 mM – 1 M KCl | CVD graphene |
| Device 75 | Concentration dependence, *I-V* measurements for different salts, Spikes measurements, PDDA-functionalized graphene | 1 mM – 1 M KCl, $MgCl_2$; 0.01 M – 1 M KCl | CVD graphene |
| Device 76 | Concentration dependence, Spikes measurements, *I-V* measurements for different salts, PDDA-functionalized graphene | 1 mM – 1 M KCl, $MgCl_2$; 0.01 M – 1 M KCl | CVD graphene |
| Device 77 | *I-V* measurements for different salts, Spikes measurements | 1 mM – 1 M KCl, $MgCl_2$ | CVD graphene |
| Device 78 | Cyclic *I-V* measurements | 1 mM – 1 M KCl | Single layer exfoliated graphene |
| Device 79 | Cyclic *I-V* measurements, Spikes measurements | 1 mM – 1 M KCl | Single layer exfoliated graphene |
| Device 80 | Integrated circuit, *I-V* measurements, Frequency dependence | 1 mM – 1 M KCl | CVD graphene |
| Device 81 | Integrated circuit, *I-V* measurements | 1 mM – 1 M KCl | CVD graphene |
| Device 82 | Integrated circuit, *I-V* measurements | 1 mM – 1 M KCl | CVD graphene |
| Device 83 | Integrated circuit, *I-V* measurements | 1 mM – 1 M KCl | CVD graphene |
| Device 84 | Cyclic *I-V* measurements | 1 mM – 1 M KCl | Single layer exfoliated graphene |
| Device 85 | Cyclic *I-V* measurements, Spikes measurements | 1 mM – 1 M KCl | CVD graphene |
| Device 86 | Cyclic *I-V* measurements, Spikes measurements | 1 mM – 1 M KCl | CVD graphene |
| Device 87 | Cyclic *I-V* measurements, Spikes measurements | 1 mM – 1 M KCl | CVD graphene |
| Device 88 | Cyclic *I-V* measurements, Spikes measurements | 1 mM – 1 M KCl | CVD graphene |
| Device 89 | Cyclic *I-V* measurements, Spikes measurements | 1 mM – 1 M KCl | CVD graphene |
| Device 90 | Cyclic *I-V* measurements, Spikes measurements | 1 mM – 1 M KCl | CVD graphene |
| Device 91 | Cyclic *I-V* measurements, Spikes measurements | 1 mM – 1 M KCl | CVD graphene |
| Device 92 | Cyclic *I-V* measurements, Spikes measurements | 1 mM – 1 M KCl | CVD graphene |
| Device 93 | Cyclic *I-V* measurements, Spikes measurements | 1 mM – 1 M KCl | CVD graphene |
| Device 94 | Cyclic *I-V* measurements, Spikes measurements | 1 mM – 1 M KCl | CVD graphene |
| Device 95 | Cyclic *I-V* measurements, Spikes measurements | 1 mM – 1 M KCl | CVD graphene |
| Device 96 | Cyclic *I-V* measurements, Spikes measurements | 1 mM – 1 M KCl | CVD graphene |
| Device 97 | Cyclic *I-V* measurements, Spikes measurements | 1 mM – 1 M KCl | CVD graphene |
| Device 98 | Cyclic *I-V* measurements, Spikes measurements | 1 mM – 1 M KCl | CVD graphene |
| Device 99 | Cyclic *I-V* measurements, Spikes measurements | 1 mM – 1 M KCl | CVD graphene |
| Device 100 | Cyclic *I-V* measurements, Spikes measurements | 1 mM – 1 M KCl | CVD graphene |
| Device 101 | Cyclic *I-V* measurements, Spikes measurements | 1 mM – 1 M KCl | CVD graphene |
| Device 102 | Cyclic *I-V* measurements, Spikes measurements | 1 mM – 1 M KCl | CVD graphene |
| Device 103 | Cyclic *I-V* measurements, Spikes measurements | 1 mM – 1 M KCl | CVD graphene |

Note: Devices 2, 12, 14, 16, 24, 34, 59, 74 were failed devices; Devices 35, 36, 37, 41, 45, 63 were used for control experiments by intentionally removing their rippled edges.

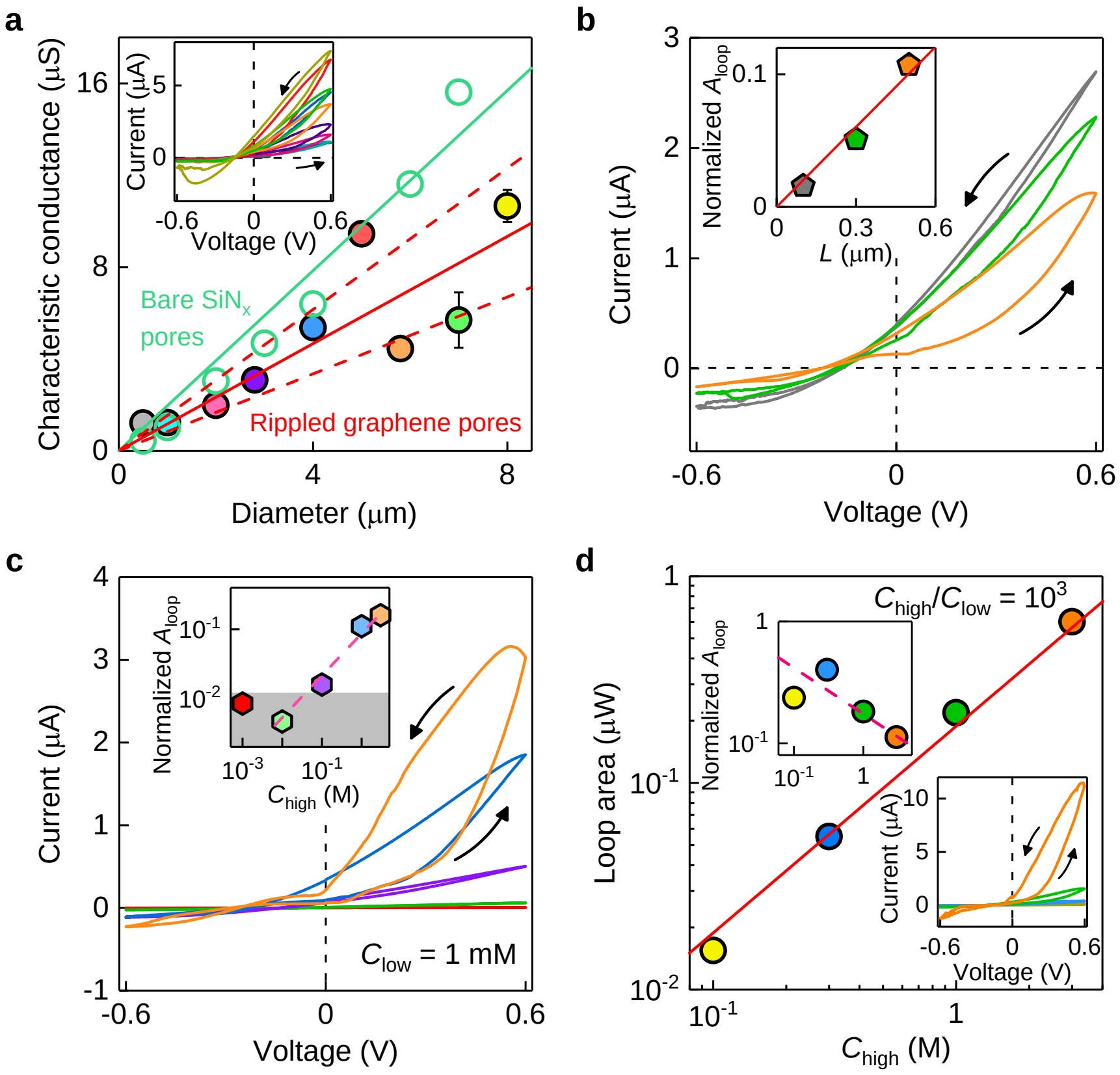

**Figure S9. Pore geometry and concentration dependent *I-V* measurements.** **a**, Main figure, *D* dependence of *G** for bare $SiN_x$ pores (empty symbols) and rippled graphene pores (solid), respectively. Solid lines, best linear fits (color coded). Red dashed lines, upper and lower limits for the linear fit coded in red, which together indicate a scatter of about 30%. Inset, representative *I-V* curves (same color coding as the solid symbols in the main figure) after reaching stabilization at different *D*. **b**, Main figure, stable *I-V* curves for rippled graphene pores with different *L* (same color coding as in the inset). Inset, *L* dependence of normalized $A_{loop}$. Red solid line, best linear fit. **c**, Main, stable *I-V* curves at different $C_{high}$ but fixed $C_{low}$ = 1 mM (same color coding as the inset). Inset, same data as in the lower inset of Fig. 2f but color coded for different $C_{high}$ Dashed pink line is a guide to the eyes and the grey shaded area marks normalized $A_{loop}$ smaller than that for $C_{high}$ = $C_{low}$ = 1mM. **d**, Main figure, $A_{loop}$ ($C_{high}$) at fixed $C_{high}/C_{low}$ = $10^3$. Solid red line, linear fit at a slope of 1. Upper inset, normalized $A_{loop}$ ($C_{high}$); same color coding as in the main figure. Pink dashed line, guide to the eyes. Lower inset, representative *I-V* curves after reaching stabilization (same color coding). For all *I-V* curves, black dashed lines mark zero *I* and *V*; the arrows indicate the same directions of hysteretic loops.

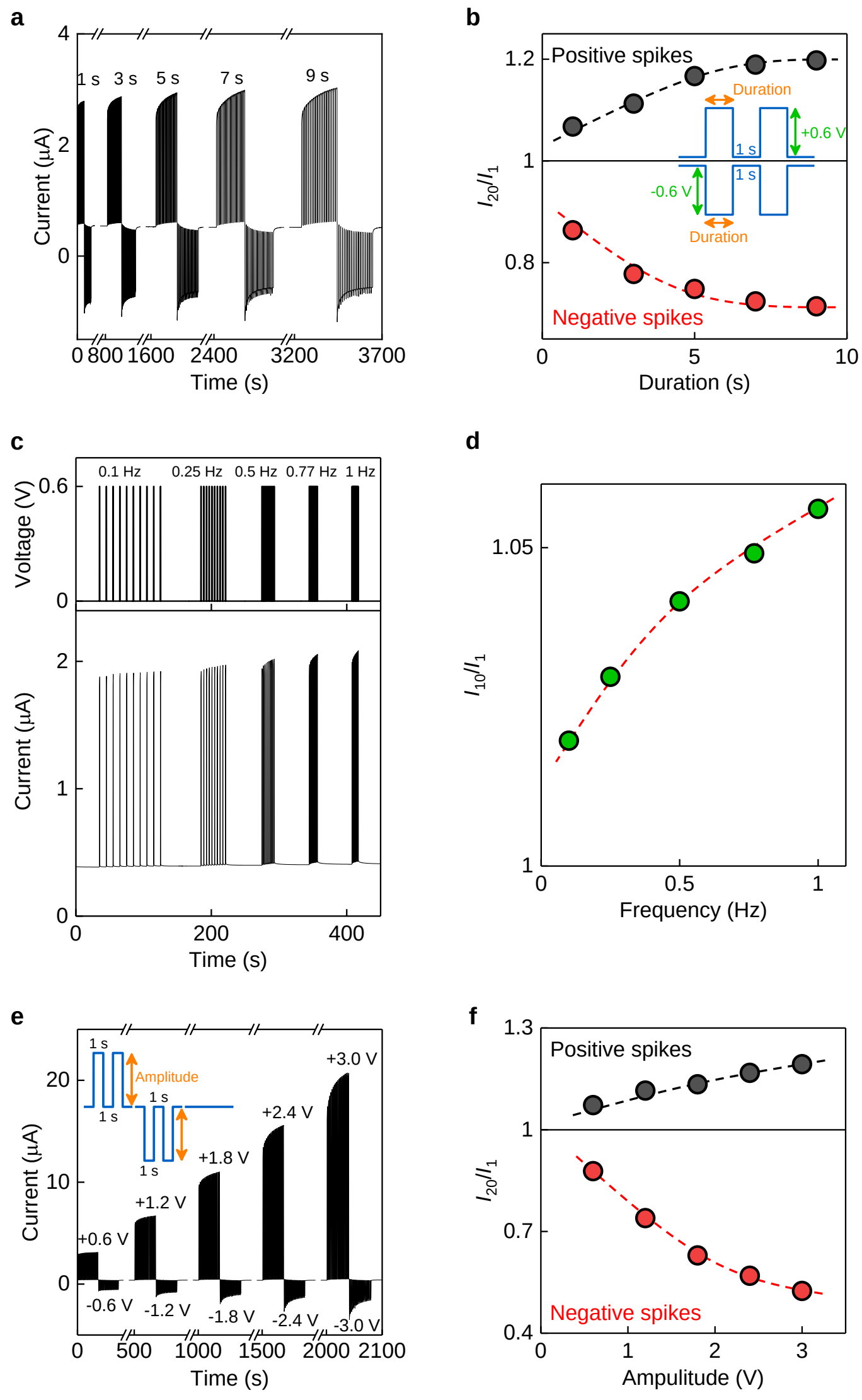

**Figure S10. Dependence of plasticity on spikes' parameters.** **a**, Evolution of ion current $I$ under successive positive and negative voltage spikes with increasing duration $t_d$ from 1 s to 9 s. For each $t_d$, 20 positive spikes were followed by 20 negative spikes. Amplitude, $A$ = 0.6 V; time interval, $\Delta t$ = 1 s. **b**, Ion current ratio after the 20$^{th}$ and 1$^{st}$ spikes, $I_{20}/I_1$, as a function of $t_d$ for positive and negative spikes, respectively (color coded). Insets, details of the applied spikes. **c**, Voltage spikes of $A$ = 0.6 V and $t_d$ = 0.5 s with increasing frequency $f$ (top) and the resulting $I$ (bottom). For each $f$, 10 successive spikes were applied. **d**, $I_{10}/I_1$ as a function of $f$. **e**, Changes in $I$ under successive positive and negative spikes with $A$ from 0.6 V to 3 V. For each $A$, 20 positive and negative spikes were tested, respectively. Inset, details of the spikes, $t_d = \Delta t$ = 1 s. **f**, $I_{20}/I_1$ as a function of $A$ for positive and negative spikes (color coded), respectively. Dashed curves in (b, d, f), guides to the eyes. Solid black lines in (b, f) mark the positions of $I_{20}/I_1$ = 1. All measurements were done with devices of $D \approx$ 2 µm, $L \approx$ 500 nm, and in KCl solutions of $C_{high}$ = 1 M, $C_{low}$ = 1 mM.

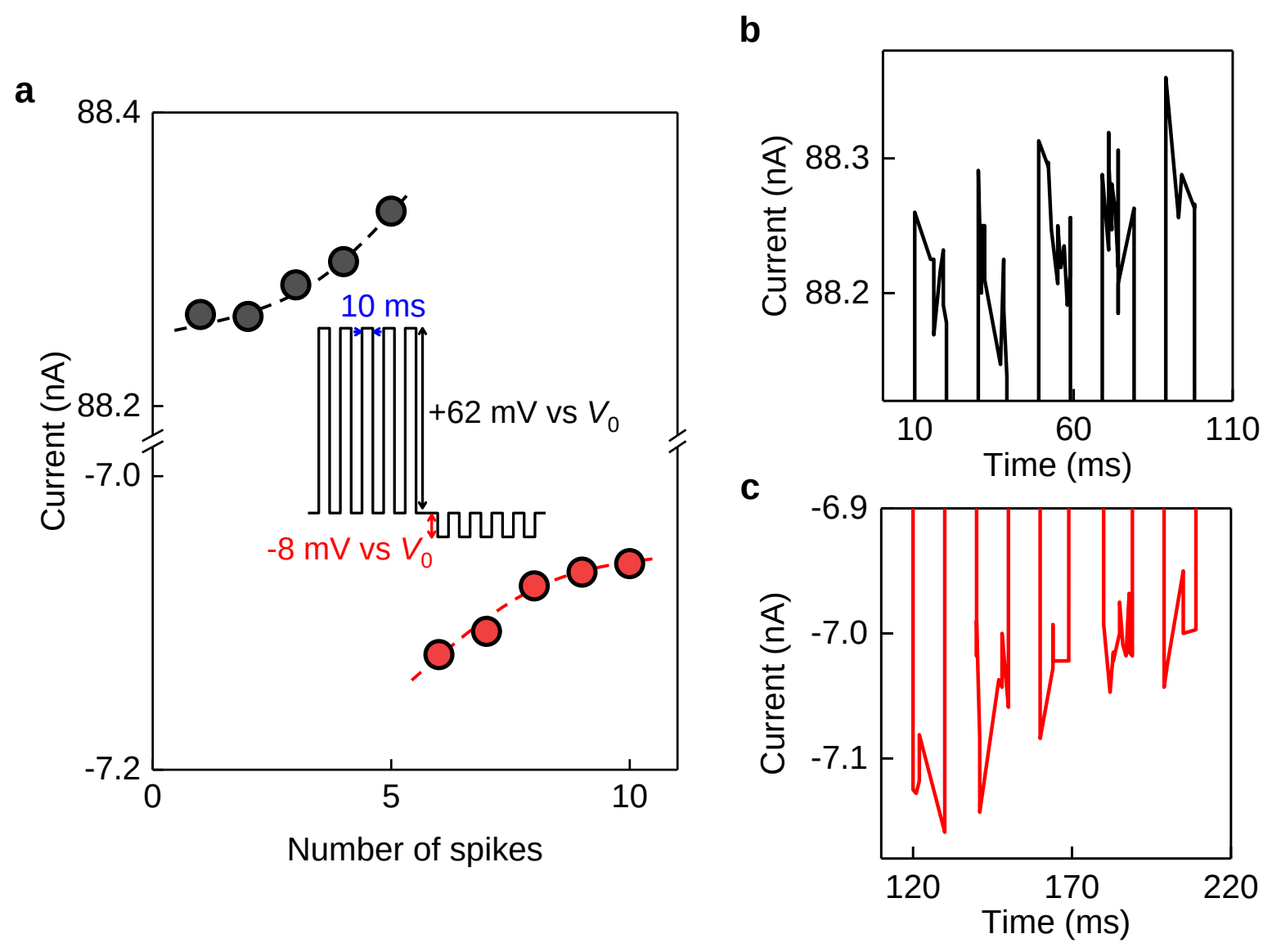

**Figure S11. Lower response limits. a**, Evolution of $I$ after 5 spikes of +62 mV (vs $V_0$) followed by another 5 spikes of -8 mV (vs $V_0$) (color coded), for one of our tested devices ($D \approx 2$ µm, $L \approx 500$ nm). Inset, details of the applied spikes, $t_d = \Delta t = 10$ ms. These parameters represent the lower limits that the tested device could successively respond to. Symbols, experimental data averaged over 5 independent measurements with error bars indicating SD. Dashed curves, guides to the eyes. **b**, **c**, Zoom-in views for the responses of $I$ to the spikes shown in the inset of (a), same color coding.

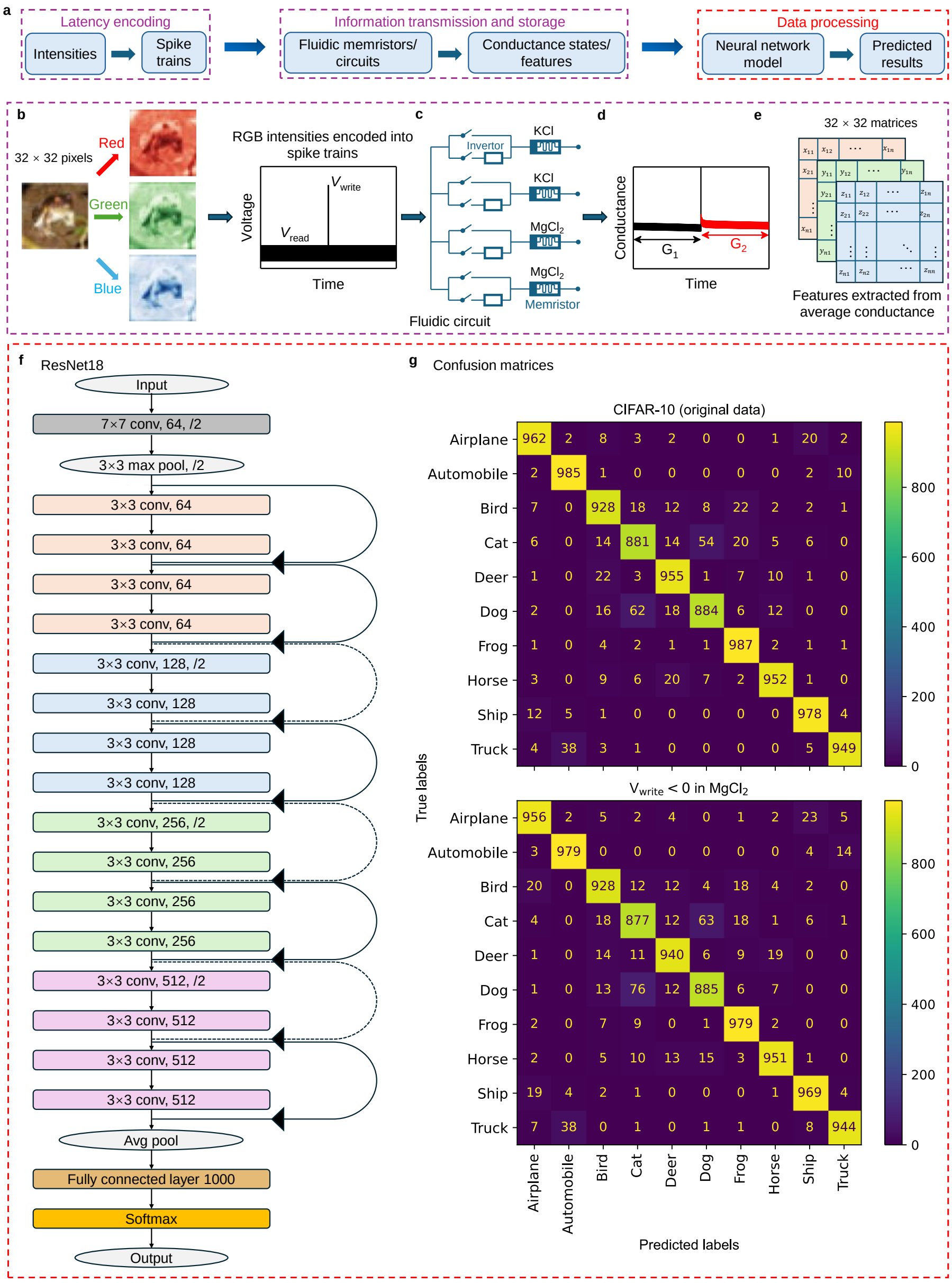

**Figure S12. Identifying color images. a**, Workflow of the experiment and data processing procedures. **b**, The RGB intensities extracted from a color image (left) were encoded into spike trains (right). **c**, The spike trains were then input into a fluidic circuit constructed by four devices ($D \approx 2$ µm, $L \approx 500$ nm) working with KCl or $MgCl_2$ solutions. $C_{high}$ = 1 M, $C_{low}$ = 1 mM. **d**, The resulting time evolution of $G$ displayed a peak separating two long-term states $G_1$ and $G_2$. Their time lengths were determined by the position of $V_{write}$. **e**, The mean values of recorded $G(t)$ such as in (d) were extracted as features and filled into three 32 × 32 matrices as feature maps. **f**, Architecture of the ResNet18 model used for identification of color images. **g**, The obtained confusion matrix for our highest testing accuracy (~94.1%, right) using features extracted at $V_{write}$ = -3 V in $MgCl_2$ solutions was compared with that directly using the original data (~94.6%, left).

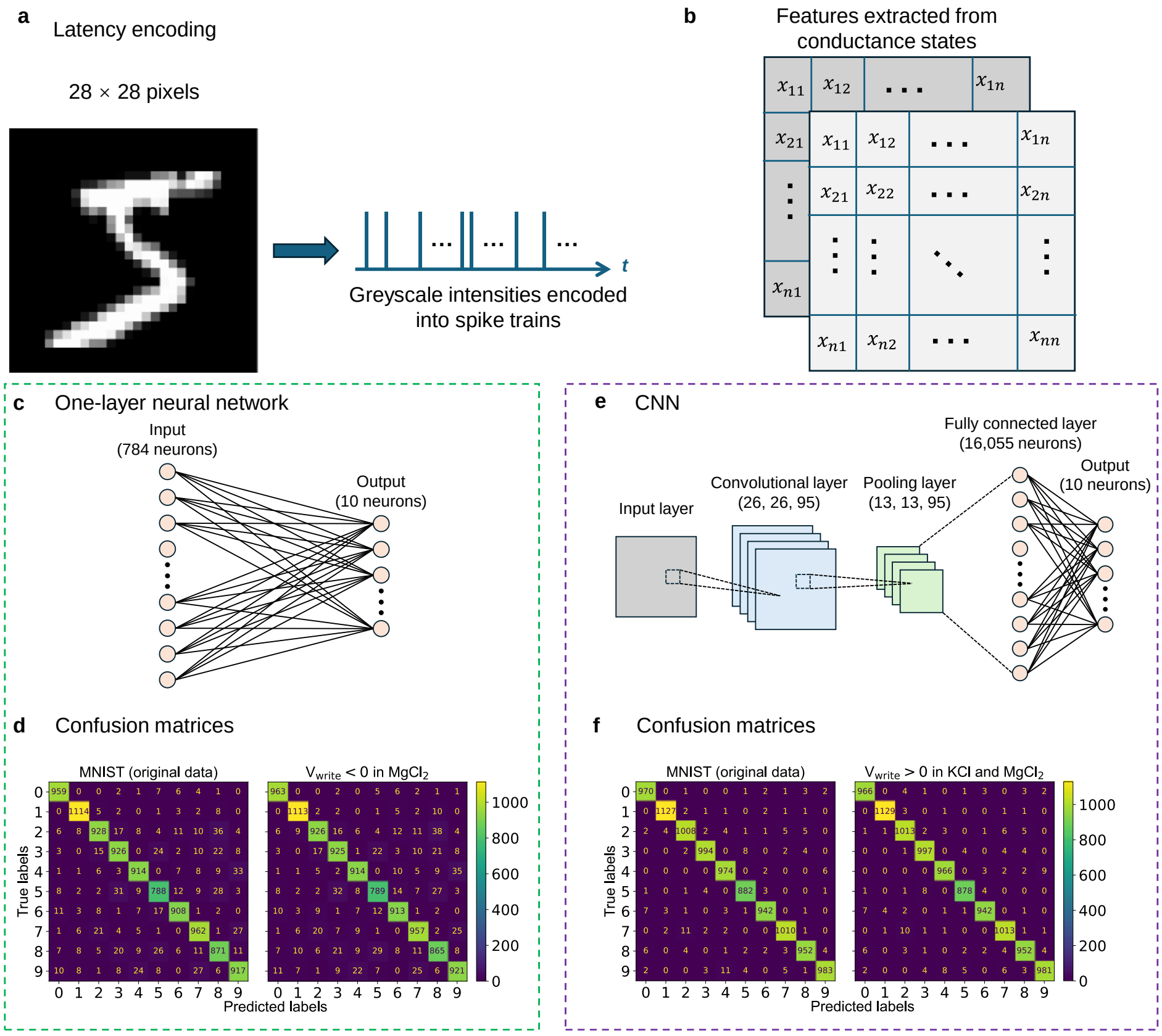

**Figure S13. Identifying greyscale images.** **a**, The greyscale intensities were encoded as voltage spike trains. **b**, The spike trains were input into the fluidic circuit in Fig. 12c and feature values were extracted from time evolutions of *G*. They were used to represent the pixels' intensities in the tested images and filled into 28 × 28 matrices as feature maps for image identification. **c**, **e**, Architectures of the one-layer fully connected neural network and the CNN model. **d**, **f**, Confusion matrices for the original data and for our best results obtained using the two neural network models, respectively. Details for $V_{write}$ and electrolytes are specified in the figures.

**Table S3.** Training and testing results for MNIST and CIFAR-10 datasets using three neural network models.

| **One-layer neural network** | | |
|---|---|---|
| **Feature maps** | **Training accuracy** | **Testing accuracy** |
| Original dataset for MNIST | 0.9369 | 0.9287 |
| $V_{write} > 0$ in KCl | 0.9350 | 0.9270 |
| $V_{write} < 0$ in $MgCl_2$ | 0.9333 | 0.9286 |
| $V_{write} < 0$ in KCl | 0.9267 | 0.9203 |
| $V_{write} > 0$ in $MgCl_2$ | 0.9224 | 0.9177 |
| **CNN** | | |
| **Feature maps** | **Training accuracy** | **Testing accuracy** |
| Original dataset for MNIST | 0.9998 | 0.9842 |
| $V_{write} > 0$ in KCl | 0.9992 | 0.9825 |
| $V_{write} < 0$ in $MgCl_2$ | 0.9993 | 0.9834 |
| $V_{write} < 0$ in KCl | 0.9995 | 0.9833 |
| $V_{write} > 0$ in $MgCl_2$ | 0.9991 | 0.9834 |
| $V_{write} > 0$ in KCl and $V_{write} < 0$ in $MgCl_2$ | 0.9999 | 0.9836 |
| $V_{write} < 0$ in KCl and $V_{write} > 0$ in $MgCl_2$ | 0.9989 | 0.9826 |
| $V_{write}$ < 0 in KCl and $V_{write}$ < 0 in $MgCl_2$ | 0.9996 | 0.9833 |
| $V_{write}$ > 0 in KCl and $V_{write}$ > 0 in $MgCl_2$ | 0.9990 | 0.9837 |
| **ResNet18** | | |
| **Feature maps** | **Training accuracy** | **Testing accuracy** |
| Original dataset for CIFAR-10 | 1.0000 | 0.9461 |
| $V_{write}$ > 0 in KCl (red)<br>$V_{write}$ < 0 in KCl (green)<br>$V_{write}$ < 0 in $MgCl_2$ (blue) | 0.9976 | 0.9357 |
| $V_{write}$ > 0 in KCl (RGB) | 1.0000 | 0.9398 |
| $V_{write}$ < 0 in $MgCl_2$ (RGB) | 1.0000 | 0.9408 |
| $V_{write}$ < 0 in KCl (RGB) | 0.9946 | 0.9260 |
| $V_{write}$ > 0 in $MgCl_2$ (RGB) | 0.9947 | 0.9278 |

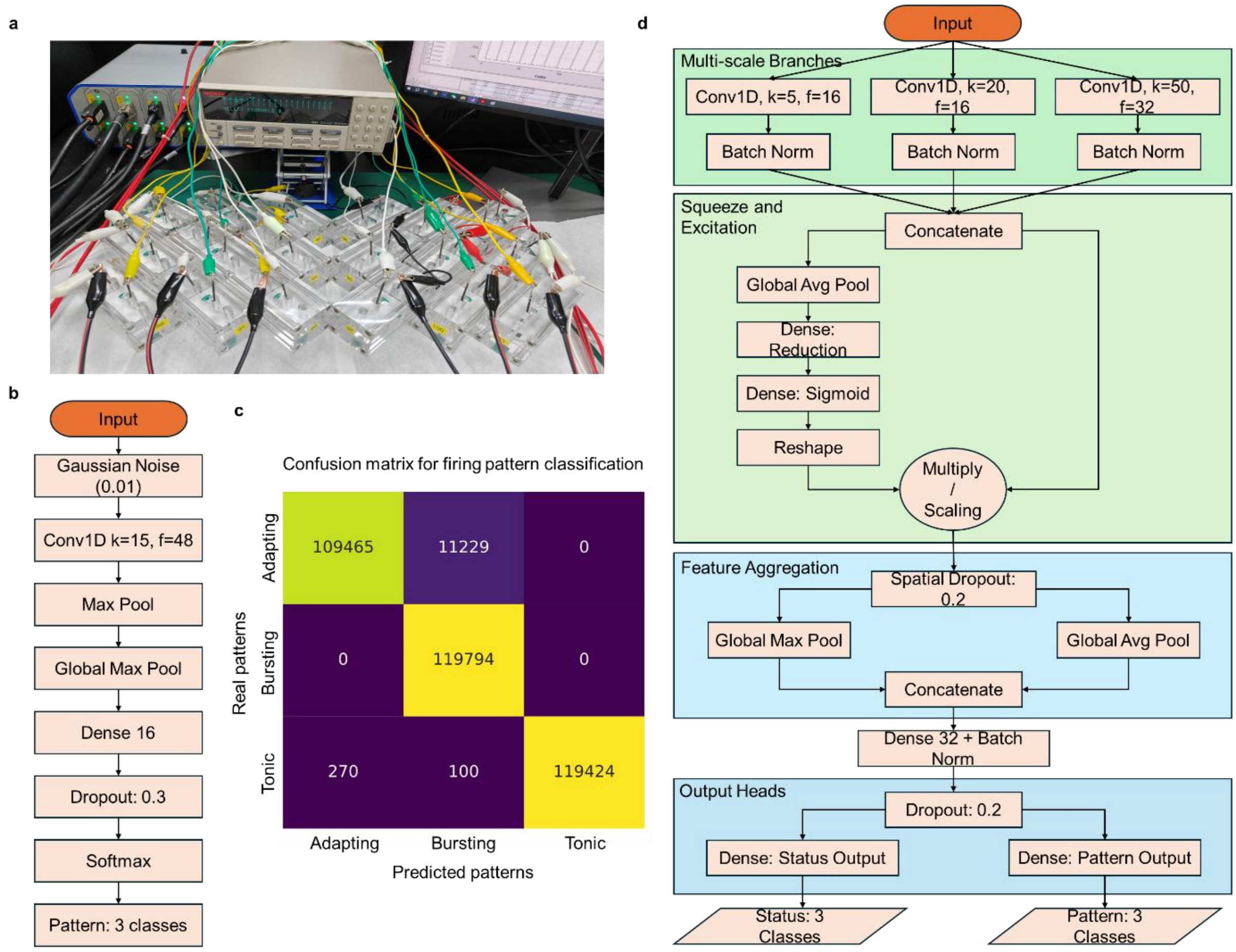

**Figure S14. Additional results for real-time analysis of emulated neural signals. a,** Image of the fluidic circuit used for the experiment. **b**, Architecture of the one-layer CNN model used for classification of neural firing patterns alone. **c**, Confusion matrix showing an overall accuracy of 96.8%. **d**, Architecture of the multi-scale CNN model used for simultaneous identification of neuron firing patterns and synchronization states.

**Supplementary reference**

1. Wahab OJ, Daviddi E, Xin B *et al.* Proton transport through nanoscale corrugations in two-dimensional crystals. *Nature* 2023;**620**:782–6. https://doi.org/10.1038/s41586-023-06247-6.
2. Wu ZF, Sun PZ, Wahab OJ *et al.* Proton and molecular permeation through the basal plane of monolayer graphene oxide. *Nat Commun* 2023;**14**:7756. https://doi.org/10.1038/s41467-023-43637-w.
3. Ji Y, Hao GP, Tan YT *et al.* High proton conductivity through angstrom-porous titania. *Nat Commun* 2024;**15**:10546. https://doi.org/10.1038/s41467-024-54544-z.
4. Geim AK, Grigorieva IV. Van der Waals heterostructures. *Nature* 2013;**499**:419–25. https://doi.org/10.1038/nature12385.
5. Wang L, Meric I, Huang PY *et al.* One-Dimensional Electrical Contact to a Two-Dimensional Material. *Science* 2013;**342**:614–7. https://doi.org/10.1126/science.1244358.
6. Cheng LJ, Guo LJ. Rectified Ion Transport through Concentration Gradient in Homogeneous Silica Nanochannels. *Nano Lett* 2007;**7**:3165–71. https://doi.org/10.1021/nl071770c.
7. Poggioli AR, Siria A, Bocquet L. Beyond the Tradeoff: Dynamic Selectivity in Ionic Transport and Current Rectification. *J Phys Chem B* 2019;**123**:1171–85. https://doi.org/10.1021/acs.jpcb.8b11202.
8. Robin P, Emmerich T, Ismail A *et al.* Long-term memory and synapse-like dynamics in two-dimensional nanofluidic channels. *Science* 2023;**379**:161–7. https://doi.org/10.1126/science.adc9931.
9. Ismail A, Nam GH, Lokhandwala A *et al.* Programmable memristors with two-dimensional nanofluidic channels. *Nat Commun* 2025;**16**:7008. https://doi.org/10.1038/s41467-025-61649-6.
10. Rollings RC, Kuan AT, Golovchenko JA. Ion selectivity of graphene nanopores. *Nat Commun* 2016;**7**:11408. https://doi.org/10.1038/ncomms11408.
11. Meyer JC, Girit CO, Crommie MF *et al.* Imaging and dynamics of light atoms and molecules on graphene. *Nature* 2008;**454**:319–22. https://doi.org/10.1038/nature07094.
12. Sun P, Zheng F, Zhu M *et al.* Selective Trans-Membrane Transport of Alkali and Alkaline Earth Cations through Graphene Oxide Membranes Based on Cation−π Interactions. *ACS Nano* 2014;**8**:850–9. https://doi.org/10.1021/nn4055682.
13. Chen L, Shi G, Shen J *et al.* Ion sieving in graphene oxide membranes via cationic control of interlayer spacing. *Nature* 2017;**550**:380–3. https://doi.org/10.1038/nature24044.
14. Song R, Wang P, Zeng H *et al.* Nanofluidic Memristive Transition and Synaptic Emulation in Atomically Thin Pores. *Nano Lett* 2025;**25**:5646–55. https://doi.org/10.1021/acs.nanolett.4c06297.
15. Lin K, Lin CY, Polster JW *et al.* Charge Inversion and Calcium Gating in Mixtures of Ions in Nanopores. *J Am Chem Soc* 2020;**142**:2925–34. https://doi.org/10.1021/jacs.9b11537.
16. Marcus Y. Thermodynamics of solvation of ions. Part 5.—Gibbs free energy of hydration at 298.15 K. *J Chem Soc, Faraday Trans* 1991;**87**:2995–9. https://doi.org/10.1039/FT9918702995.
17. Sah MP, Yang C, Kim H *et al.* A Generic Model of Memristors With Parasitic Components. *IEEE Trans Circuits Syst I* 2015;**62**:891–8. https://doi.org/10.1109/TCSI.2014.2373674.
18. Xiao Y, Sun W, Gao C *et al.* Neural Functions Enabled by a Polarity-Switchable Nanofluidic Memristor. *Nano Lett* 2024;**24**:12515–21. https://doi.org/10.1021/acs.nanolett.4c03449.
19. Nagel LW, Pederson DO. *SPICE (Simulation Program with Integrated Circuit Emphasis)*. 1973. http://www2.eecs.berkeley.edu/Pubs/TechRpts/1973/22871.html.
20. Chua LO, Sirakoulis GC, Adamatzky A (eds). *Handbook of Memristor Networks*. Cham: Springer, 2019.
21. Xu Z. Soft Nanofluidic Machinery. *ACS Nano* 2024;**18**:9765–72.

https://doi.org/10.1021/acsnano.3c10760.
22. Lee C, Joly L, Siria A *et al.* Large Apparent Electric Size of Solid-State Nanopores Due to Spatially Extended Surface Conduction. *Nano Lett* 2012;**12**:4037–44. https://doi.org/10.1021/nl301412b.
23. Logg A, Mardal KA, Wells G (eds). *Automated Solution of Differential Equations by the Finite Element Method: The FEniCS Book*. Heidelberg: Springer, 2012.
24. Hall JE. Access resistance of a small circular pore. *J Gen Physiol* 1975;**66**:531–2. https://doi.org/10.1085/jgp.66.4.531.
25. Plimpton S. Fast Parallel Algorithms for Short-Range Molecular Dynamics. *J Comput Phys* 1995;**117**:1–19. https://doi.org/10.1006/jcph.1995.1039.
26. Berendsen HJC, Grigera JR, Straatsma TP. The missing term in effective pair potentials. *J Phys Chem* 1987;**91**:6269–71. https://doi.org/10.1021/j100308a038.
27. Ryckaert JP, Ciccotti G, Berendsen HJC. Numerical integration of the cartesian equations of motion of a system with constraints: molecular dynamics of n-alkanes. *Journal of Computational Physics* 1977;**23**:327–41. https://doi.org/10.1016/0021-9991(77)90098-5.
28. Jorgensen WL, Tirado-Rives J. The OPLS [optimized potentials for liquid simulations] potential functions for proteins, energy minimizations for crystals of cyclic peptides and crambin. *J Am Chem Soc* 1988;**110**:1657–66. https://doi.org/10.1021/ja00214a001.
29. Zhou K, Xu Z. Deciphering the nature of ion-graphene interaction. *Phys Rev Research* 2020;**2**:042034. https://doi.org/10.1103/PhysRevResearch.2.042034.
30. Hockney RW, Eastwood JW. *Computer Simulation Using Particles*. CRC Press, 2021. https://doi.org/10.1201/9780367806934.
31. Luo J, Jang HD, Sun T *et al.* Compression and Aggregation-Resistant Particles of Crumpled Soft Sheets. *ACS Nano* 2011;**5**:8943–9. https://doi.org/10.1021/nn203115u.
32. Cranford SW, Buehler MJ. Packing efficiency and accessible surface area of crumpled graphene. *Phys Rev B* 2011;**84**:205451. https://doi.org/10.1103/PhysRevB.84.205451.
33. Xie B, Yu P. Ion-shuttling memristor: towards ionic computing and neuromorphic sensing. *Iontronics* 2026;**2**:2. https://doi.org/10.20517/iontronics.2026.02.
34. Li X, Li R, Li S *et al.* Triboiontronics with temporal control of electrical double layer formation. *Nat Commun* 2024;**15**:6182. https://doi.org/10.1038/s41467-024-50518-3.
35. Li X, Wei Y, Gao X *et al.* Harnessing triboiontronic Maxwell's demon by triboelectric-induced polarization for efficient energy-information flow. *Joule* 2025;**9**:101888. https://doi.org/10.1016/j.joule.2025.101888.
36. Sint K, Wang B, Král P. Selective Ion Passage through Functionalized Graphene Nanopores. *J Am Chem Soc* 2008;**130**:16448–9. https://doi.org/10.1021/ja804409f.
37. Zhou K, Xu Z. Field-enhanced selectivity in nanoconfined ionic transport. *Nanoscale* 2020;**12**:6512–21. https://doi.org/10.1039/C9NR10731B.
38. Dočkal J, Moučka F, Lísal M. Molecular Dynamics of Graphene–Electrolyte Interface: Interfacial Solution Structure and Molecular Diffusion. *J Phys Chem C* 2019;**123**:26379–96. https://doi.org/10.1021/acs.jpcc.9b07487.
39. Gillespie DT. Exact stochastic simulation of coupled chemical reactions. *J Phys Chem* 1977;**81**:2340–61. https://doi.org/10.1021/j100540a008.
40. Laughlin SB, de Ruyter van Steveninck RR, Anderson JC. The metabolic cost of neural information. *Nat Neurosci* 1998;**1**:36–41. https://doi.org/10.1038/236.
41. Fu S, Gao H, Wang S *et al.* Constructing artificial neurons with functional parameters comprehensively matching biological values. *Nat Commun* 2025;**16**:8599. https://doi.org/10.1038/s41467-025-63640-7.
42. Kuzum D, Yu S, Philip Wong HS. Synaptic electronics: materials, devices and applications. *Nanotechnology* 2013;**24**:382001. https://doi.org/10.1088/0957-4484/24/38/382001.

43. Paulo G, Sun K, Di Muccio G *et al.* Hydrophobically gated memristive nanopores for neuromorphic applications. *Nat Commun* 2023;**14**:8390. https://doi.org/10.1038/s41467-023-44019-y.
44. Kamsma TM, Kim J, Kim K *et al.* Brain-inspired computing with fluidic iontronic nanochannels. *Proc Natl Acad Sci USA* 2024;**121**:e2320242121. https://doi.org/10.1073/pnas.2320242121.
45. LeCun Y, Cortes C, Burges C. The MNIST database of handwritten digits. 1998. http://yann.lecun.com/exdb/mnist/.
46. Krizhevsky A, Hinton G. *Learning Multiple Layers of Features from Tiny Images.* University of Toronto Technical Report, 2009.
47. Eshraghian JK, Ward M, Neftci EO *et al.* Training Spiking Neural Networks Using Lessons From Deep Learning. *Proc IEEE* 2023;**111**:1016–54. https://doi.org/10.1109/JPROC.2023.3308088.